\documentclass[longauth]{aa}  
\usepackage{ulem}
\usepackage{multirow}
\usepackage{graphicx}
\usepackage{txfonts}
\usepackage[colorlinks=true,linkcolor=red,anchorcolor=blue,citecolor=blue,filecolor=black,menucolor=black,runcolor=black,urlcolor=blue]{hyperref}

\usepackage[dvipsnames]{xcolor}

\begin{document}

   \title{PRISMS. UNCOVER-26185, a metal-poor SFG at $z$\,$=$\,10.05 with no evidence for a X-ray-luminous AGN}


   \author{J. \'Alvarez-M\'arquez\inst{\ref{inst:CAB}}\thanks{jalvarez@cab.inta-csic.es} 
   \and L.~Colina\inst{\ref{inst:CAB}} 
   \and A.~Crespo-Gomez\inst{\ref{inst:Baltimore}}
   \and S.~Kendrew\inst{\ref{inst:ESA-BALTIMORE}} 
   \and J.~Zavala\inst{\ref{inst:Amherst}} 
   \and R.~Marques-Chaves\inst{\ref{inst:Geneva}}
   \and C.~Prieto-Jiménez\inst{\ref{inst:CAB}} 
   \and Abdurro’uf\inst{\ref{inst:Indiana}}
   \and C.~Blanco-Prieto\inst{\ref{inst:CAB}}
   \and L.~A.~Boogaard\inst{\ref{inst:Leiden}}
   \and M.~Castellano\inst{\ref{inst:INAF-ROME}}
   \and A.~Fontana\inst{\ref{inst:INAF-ROME}}
   \and Y.~Fudamoto\inst{\ref{inst:CHIBA}}
   \and S.~Fujimoto\inst{\ref{inst:Toronto},\ref{inst:Dunlap}}
   \and M.~García-Marín\inst{\ref{inst:ESA-BALTIMORE}}
   \and Y.~Harikane\inst{\ref{inst:Tokyo}}
   \and S.~Harish\inst{\ref{inst:Baltimore}}
   \and T.~Hashimoto\inst{\ref{inst:Tsukuba},\ref{inst:TCHoU}}
   \and T.~Hsiao\inst{\ref{inst:Austin}}
   \and E.~Iani\inst{\ref{inst:ISTA}}
   \and A.~K.~Inoue\inst{\ref{inst:SASE},\ref{inst:Waseda}}
   \and D.~Langeroodi\inst{\ref{inst:DARK}}
   \and R.~Lin\inst{\ref{inst:Amherst}}
   \and J.~Melinder\inst{\ref{inst:Stockholm}}
   \and L.~Napolitano\inst{\ref{inst:INAF-ROME}}
   \and G.~${\rm \ddot{O}}$stlin\inst{\ref{inst:Stockholm}}
   \and P.~G.~Pérez-González\inst{\ref{inst:CAB}} 
   \and P.~Rinaldi\inst{\ref{inst:AURA}}
   \and B.~Rodríguez~Del~Pino\inst{\ref{inst:CAB}}
   \and P.~Santini\inst{\ref{inst:INAF-ROME}}
   \and Y.~Sugahara\inst{\ref{inst:SASE},\ref{inst:Waseda}}
   \and T.~Treu\inst{\ref{inst:California}} 
   \and A.~Varo-O'ferral\inst{\ref{inst:CAB}} 
   \and G.~Wright\inst{\ref{inst:Edinburgh}} }

   \institute{Centro de Astrobiolog\'{\i}a (CAB), CSIC-INTA, Ctra. de Ajalvir km 4, Torrej\'on de Ardoz, E-28850, Madrid, Spain\label{inst:CAB}
   \and Space Telescope Science Institute (STScI), 3700 San martin Drive, Baltimore, MD 21218, USA\label{inst:Baltimore}
   \and European Space Agency (ESA), ESA Office, Space Telescope Science Institute, 3700 San Martin Drive, Baltimore, MD 21218, USA\label{inst:ESA-BALTIMORE}
   \and University of Massachusetts Amherst, 710 North Pleasant Street, Amherst, MA 01003-9305, USA\label{inst:Amherst}
   \and Geneva Observatory, Department of Astronomy, University of Geneva, Chemin Pegasi 51, CH-1290 Versoix, Switzerland \label{inst:Geneva}
   \and Department of Astronomy, Indiana University,727 East Third Street, Bloomington, IN 47405, USA\label{inst:Indiana}
   \and Leiden Observatory, Leiden University, PO Box 9513, NL-2300 RA Leiden, The Netherlands\label{inst:Leiden}
   \and INAF – Osservatorio Astronomico di Roma, via Frascati 33, 00078, Monteporzio Catone, Italy\label{inst:INAF-ROME}  
   \and Center for Frontier Science, Chiba University, 1-33 Yayoi-cho, Inage-ku, Chiba 263-8522, Japan\label{inst:CHIBA}
   \and David A. Dunlap Department of Astronomy and Astrophysics, University of Toronto, 50 St. George Street, Toronto, Ontario, M5S 3H4, Canada\label{inst:Toronto}
   \and Dunlap Institute for Astronomy and Astrophysics, 50 St. George Street, Toronto, Ontario, M5S 3H4, Canada\label{inst:Dunlap}
   \and Institute for Cosmic Ray Research, The University of Tokyo, 5-1-5 Kashiwanoha, Kashiwa, Chiba 277-8582, Japan\label{inst:Tokyo}
   \and Division of Physics, Faculty of Pure and Applied Sciences, University of Tsukuba, Tsukuba, Ibaraki 305-8571, Japan\label{inst:Tsukuba}
   \and Tomonaga Center for the History of the Universe (TCHoU), Faculty of Pure and Applied Sciences, University of Tsukuba, Tsukuba, Ibaraki 305-8571, Japan\label{inst:TCHoU}   
   \and Department of Astronomy, University of Texas, Austin, TX 78712, USA\label{inst:Austin}
   \and Institute of Science and Technology Austria (ISTA), Am Campus 1, 3400 Klosterneuburg, Austria\label{inst:ISTA}
   \and Department of Physics, School of Advanced Science and Engineering, Faculty of Science and Engineering, Waseda University, 3-4-1 Okubo, Shinjuku, Tokyo 169-8555, Japan \label{inst:SASE}
   \and Waseda Research Institute for Science and Engineering, Faculty of Science and Engineering, Waseda University, 3-4-1 Okubo, Shinjuku, Tokyo 169-8555, Japan \label{inst:Waseda}
   \and DARK, Niels Bohr Institute, University of Copenhagen, Jagtvej 155A, 2200 Copenhagen, Denmark\label{inst:DARK}
   \and Department of Astronomy, Stockholm University, Oscar Klein Centre, AlbaNova University Centre, 106 91 Stockholm, Sweden\label{inst:Stockholm}
   \and AURA for the European Space Agency (ESA), Space Telescope Science Institute, 3700 San Martin Dr., Baltimore, MD 21218, USA \label{inst:AURA}
   \and Physics and Astronomy Department University of California Los Angeles CA 90095\label{inst:California}
   \and UK Astronomy Technology Centre, Royal Observatory Edinburgh, Blackford Hill, Edinburgh EH9 3HJ, UK\label{inst:Edinburgh}
   }

   \date{Received ; accepted}

\abstract
{This work presents the first results of the PRImordial galaxy Survey with MIRI Spectroscopy (PRISMS), a JWST cycle 4 program (PID\,8051) aimed at the characterization of a relatively large sample of ten primordial galaxies about 500\,Myr after the Big Bang. Here, we present deep (13.9\,hours) spectroscopy with the MIRI Low-Resolution Spectrograph (LRS) of the lensed galaxy UNCOVER-26185 at a redshift of $z$\,$=$\,10.054\,$\pm$\,0.011. It is a faint UV galaxy (M$_{\mathrm{UV}}$\,$=$\,-18.83$\pm$0.07\,mag) previously identified as a X-ray luminous active galactic nuclei. MIRI LRS detects the H$\beta$+[OIII]$\lambda$$\lambda$4960,5008$\AA$ complex and H$\alpha$ emission line with a significance of 10$\sigma$ and 8$\sigma$, respectively, as well as the optical continuum emission at rest-frame 0.45\,$\mu$m and 0.57\,$\mu$m with a signal-to-noise ratio of 6$-$7. The UV-to-optical spectral energy distribution, combining continuum and emission lines, is compatible with: (i) a low stellar (A$_V$= 0.2$\pm$0.09) and nebular (A$_V$=0$^{+0.4}_{-0.0}$) extinction, (ii) a star-formation history composed by a young (7$\pm$3\,Myr) starburst and an intermediate-age (65$\pm$20\,Myr) stellar population, and (iii) a total stellar mass of (1.7$\pm$0.3)\,$\times$\,10$^{8}$\,M$_{\odot}$. The H$\alpha$-derived star-formation rate is SFR=1.3$\pm$0.3\,M$_{\odot}$\,yr$^{-1}$ yielding a specific star formation rate of 7.6$\pm$1.2\,Gyr$^{-1}$. The low optical emission line ratios (e.g., $\log$(R23)\,$=$\,0.59$\pm$0.10 and $\log$(O32)\,$=$\,0.65$\pm$0.10) locate UNCOVER-26185 as the most metal-poor galaxy (Z\,$=$\,0.04$\pm$0.01\,Z$_{\odot}$), and as outlier with the lowest ionization ($\log$U\,$=$\,-2.5) identified so far at redshifts above 9. The measured low internal extinction, low ionization, and high H$\alpha$ equivalent width (1399$\pm$271$\AA$) do not support an scenario where the AGN is obscured by dust (host and torus) or by the BLR gas clouds, and alternative scenarios have to be considered. With no evidence of an active galactic nuclei in the rest-frame UV-to-optical spectrum, UNCOVER-26185 has the properties of a low-mass, metal-poor, main-sequence star-forming galaxy at redshift 10, with interestellar medium and ionization properties very different than those of the already studied UV-bright (M$_{\mathrm{UV}}$\,$<$\,-20 mag) galaxies at redshifts beyond 10 (e.g., GNz11, GHz2, or JADES-GS-z14-0). 
PRISMS is starting to explore the population of intermediate-UV luminosity galaxies at $z$\,$\sim$\,10, covering UV absolute magnitudes in the range of $-$17.9\,$\leq$\,M$_{\mathrm{UV}}$\,[mag]\,$\leq$\,$-$20.5, fainter than those of UV-bright galaxies studied so far.}

\titlerunning{UNCOVER-26185, a Main-Sequence, Metal-Poor SFG at $z$\,$=$\,10.05}
\maketitle

\section{Introduction}\label{Sect:intro}

JWST is revolutionizing our understanding of galaxy formation in the Epoch of Reionization (EoR) and beyond, pushing the spectroscopic detection frontier to redshifts slightly above 14, with galaxies such as JADES-GS-z14-0 \citep{Carniani+24, Schouws+2025, Helton2025} and MoM-z14 \citep{Naidu+2025}. NIRCam has identified several hundred photometric candidates at $z$\,$\gtrsim$\,10 \citep{Finkelstein+23, Perez-Gonzalez+23b, Robertson+23}, with tentative ones extending to $z$\,$\sim$\,17 and even $z$\,$\sim$\,25$-$30 (e.g., \citealt{Perez-Gonzalez+2025, Castellano2025}), awaiting spectroscopic confirmation. To date, roughly fifty galaxies have been spectroscopically confirmed with NIRSpec in the first 500\,Myr of the Universe (e.g., \citealt{Curtis-Lake+23, Harikane+23a, Arrabal-Haro+23, Bunker+23, Roberts-Borsani+23Natur, Hsiao+23-NIRSpec, Fujimoto2024_UNCOVER, Castellano2024, Carniani+24, Witstok2025Natur_z13, Napolitano2025_glass_sample, Tang2025_z9-14, Roberts-Borsani+2025}. 

Some of these galaxies show signatures consistent with active black holes (BH, e.g., \citealt{Kovacs24-GHz9, Bogdan24-UNCOVER, Maiolino2024_BH, Napolitano25-GHz9}), while others resemble extreme, young, compact, low-metallicity starbursts (e.g., \citealt{Hsiao+2024_MIRI, Zavala2024, Alvarez-Marquez_2025, Helton2025}). The physical nature of these primordial galaxies, and the mechanisms driving their rapid stellar build-up and BH growth, remain the subject of intense debate. Proposed explanations span a wide range of scenarios, ranging from high star-formation efficiency in dense, metal-poor, feedback-free environments \citep{Dekel2023}, to the coevolution of active galaxy nuclei (AGN) and star-formation \citep{Ji2025_gnz11}. Others include dust-free conditions from strong radiation-driven outflows \citep{Ferrara23}, top-heavy initial mass functions \citep{Hutter2025}, and stochastic star-forming histories (SFHs; \citealt{Ciesla2024}). 

The limited spectral coverage of NIRCam and NIRSpec ($<$\,0.4\,$\mu$m) at $z$\,$\geq$\,9.7 biasses our ability to reveal the nature of these galaxies. At such redshifts, key optical features$-$H$\alpha$ \& [OIII]5008Å, and their underlying continua$-$are redshifted into the Mid-Infrared Instrument (MIRI; \citealt{Rieke+15,Wright+15,Wright+23}) spectral range. To date, only four galaxies have reported detections of these lines$-$MACS0647-JD (z=10.2; \citealt{Hsiao+2024_MIRI}), GNz11 ($z$\,$=$\,10.6; \citealt{Alvarez-Marquez_2025}), GHZ2 ($z$\,$=$\,12.3; \citealt{Zavala+2024}), and JADES-GS-z14-0 ($z$\,$=$\,14.2; \citealt{Helton2025})$-$revealing: (i) the first direct-T$_{\rm e}$ metallicity (Z) measurements at $z$\,$>$\,10, showing relatively high metallicities of 10$-$20\%\,Z$_{\odot}$, (ii) the lack of significant dust attenuation based on their Balmer decrements, (iii) elevated electron densities (n$_e$\,$>$\,250\,cm$^{-3}$) and temperatures (T$_{e}\sim1.5\times10^4$\,K), (iv) high photon production efficiencies ($\log$($\xi_\mathrm{ion}$\,[$\mathrm{Hz\,erg^{-1}}$])\,$\geq$\, 25.2), and (v) the dominant starburst nature with high specific star formation (sSFR). These findings highlight the MIRI's pivotal role in constraining the nature and physical properties of these primordial galaxies. However, these studies have been limited so far to ultraviolet (UV) bright galaxies (M$_{UV}$< $-$21).

A more comprehensive study into the nature and physical properties of these sources requires to enlarge the sample of z$\sim$10 galaxies with MIRI spectroscopy covering a wider range in M$_{\rm UV}$ to investigate their properties as a function of UV luminosity. The PRImordial galaxy Survey with MIRI Spectroscopy (PRISMS) program (PID\,8051, PIs: J. \'Alvarez-M\'arquez and L. Colina) has selected a sample of ten spectroscopically confirmed galaxies ranging redshifts of 9.7\,$<$\,$z$\,$<$\,10.4 and UV absolute magnitudes of $-$17.9\,$\leq$\,M$_{\mathrm{UV}}$\,[mag]\,$\leq$\,$-$20.5, including lensed galaxies and AGN candidates. 

The galaxy UNCOVER-26185 or UHz1 (hereafter U26185) is a member of the PRISMS galaxy sample that was first identified using deep NIRCam imaging as a high-redshift candidate galaxy with a photo-$z$ of 10.32 \citep{Castellano23-UNCOVER}. NIRSpec spectroscopy follow-ups confirmed its high redshift at 10.073$\pm$0.002 based on the detection of several UV and optical emission lines \citep{Goulding23-UNCOVER, Fujimoto2024_UNCOVER}. U26185 was detected as a strong X-ray source with the \textit{Chandra} X-ray Observatory \citep{Goulding23-UNCOVER, Bogdan24-UNCOVER}. The lack of evidence of AGN features in the rest-frame UV has been interpreted as a heavily obscured luminous AGN hosting a 10$^7$\,$-$\,10$^8$\,M$_{\odot}$ supermassive BH (SMBH) accreting at the Eddington rate. If confirmed, U26185 would represent one of the earliest SMBH detected so far, when the Universe was just about 500\,Myr old favoring formation scenarios based on heavy seeds. 

This paper presents the first results of the PRISMS project, where we analyze new MIRI spectroscopy, detecting the optical emission lines and underline continua, together with existing ancillary JWST and ALMA data for U26185. The paper is organized as follows. Section \ref{Sect2:data_calibration_lines} introduces the MIRI observations and calibrations, together with the ancillary data. Section \ref{Sect:analysis} presents the analysis of the MIRI observations, where fluxes of the emission lines and continuum are derived, together with the overall UV to far-infrared (FIR) spectral energy distribution (SED) fitting analysis. Section \ref{Sect:results_dis} shows the calculation of all physical properties, such as dust attenuation, SFH, ionizing photon production efficiency, ionization and emission line ratios, and metallicity. Section \ref{Sect4:disc} discusses the nature of the ionized source. Section \ref{Sect:conclusion_Summary} concludes and summarized the main results and discussions of the paper. Throughout this paper, we assume a Chabrier initial mass function (IMF, \citealt{Chabrier+03}), vacuum emission line wavelengths, and a flat $\Lambda$CDM cosmology with $\Omega_\mathrm{m}$\,=\,0.310, and H$_0$\,=\,67.7\,km\,s$^{-1}$\,Mpc$^{-1}$ \citep{PlanckCollaboration18VI}.

\section{Data and calibrations}\label{Sect2:data_calibration_lines}

\subsection{MIRI LRS observations and calibrations}\label{Sect:obs_cal}

U26185 was observed with the Low Resolution Spectrograph (LRS, \citealt{Kendrew2015}) of MIRI on 19th and 20th of November of 2025 as part of the cycle 4 JWST program ID\,8051 (PIs: J. \'Alvarez-M\'arquez \& L. Colina). The MIRI LRS spectral coverage ranges from 4.85\,$\mu$m to 14\,$\mu$m, and uses a slit with a size of 0.51\,$\times$\,4.7 arcsec$^2$. It covers the optical rest-frame spectrum of U26185, including bright optical emission lines, such as H$\beta$, [O\,III]$\lambda\lambda$4960,5008$\AA$, and H$\alpha$, and their underlined continuum.  

The MIRI LRS total on-source integration time corresponds to 49917 seconds (13.9\,hours), distributed in 12 dither positions. Each dither uses 149 groups and 10 integrations in FASTR1 readout mode. We implement a non-standard four-point dither pattern. We combine a two-column mosaic and a two-point dither pattern. This strategy locates the target in four equidistant positions along the LRS slit instead of only two offered by the standard two-point dither pattern, and it is repeated 3 times to get the total number of dithers. This strategy allows a better mitigation of bad/hot pixels and pixels affected by cosmic rays, as well as an improvement on the background subtraction. A GAIA DR3 star is used to perform an offset target acquisition (TA) and locate our target in the center of the slit. The complementary verification LRS image confirms that the TA strategy works within subpixel uncertainties.     

The LRS observations are processed with version 1.20.2 of the JWST calibration pipeline and context 1464 of the Calibration Reference Data System. We follow the standard MIRI LRS pipeline procedure \citep{bushouse_2025_17515973}, with additional customized steps to improve the quality of the final LRS calibrated products. We run the first and second stage of the pipeline following the default configuration, except for the \texttt{jump} step where we activate the \texttt{find$\_$showers} keyword to search and correct the cosmic ray showers events. Before stage 3 and in the fully calibrated detector images, we implement four customized steps: 
\begin{itemize}

\item Wavelength masking. We mask the calibrated detector image by assigning \texttt{NaN} values to pixels corresponding to wavelengths lower than 4.85\,$\mu$m and greater than 8\,$\mu$m, where no relevant spectral information is expected for this source.

\item Master background correction. We generate a master background by performing a sigma-clipped median of all the calibrated detector masked images, independently of the pointing. That master background is individually subtracted from each of the detector images. The decision to subtract the background at the end of the second stage by the median of all the pointings is driven by the fact that the main source is not detected in individual exposures, and the noise is dominated by background and detector effects. Therefore, this pixel-by-pixel background subtraction would correct any optical or detector level residuals as well as any features and noise injected by the reference files (e.g. photometry and flat-field correction). A similar methodology has been already implemented in the calibration of deep MRS observations (e.g. \citealt{Alvarez-Marquez+23-SPT,Alvarez-Marquez+23-MACS}).

\item Residual background subtraction. Over the long exposure time (13.9\,hours) of the LRS observation, the total background emission is seen to drift, causing changes in the background level between exposures. Also, scattering from the dispersed background in the MIRI Imager field causes spatial gradients in the slit region. The master background correction takes care of the mean background emission, but background residuals in the order of 1\% compared to the total background are found between different observations. That residuals are in the range of the integrated fluxes of the H$\beta$+[O\,III]$\lambda\lambda$4960,5008$\AA$ complex (hereafter H$\beta$+[O\,III]) and H$\alpha$ emission line, and a factor of about 10 higher than the continuum emission, being extremely important to correct it. Then, we perform a second order correction row by row (i.e., wavelength by wavelength), where we model the residual background emission in the spatial direction of the slit. We take each of the row of the detector image and: (i) perform sigma clipping over the full row to remove outliers and pixels where the source emission is larger than 3$\sigma$, (ii) smooth the flux values of the row (spatial direction of the slit) with a Gaussian filter, (iii) fit the smoothed row with a polynomial function, and (iv) subtract the derived polynomial function to the original row fluxes. 

\item Sigma clipping. We perform a sigma clipping over the masked and background subtracted calibrated image to remove any residual from bright cosmic rays, cosmic ray showers, and detector artifact. We have carefully checked that this step does not remove any flux from our target. In the case of U26185, the emission of the source is not detected over 3$\sigma$ in individual pointings. 

\end{itemize}

\begin{figure*}
\centering
   \includegraphics[width=\linewidth]{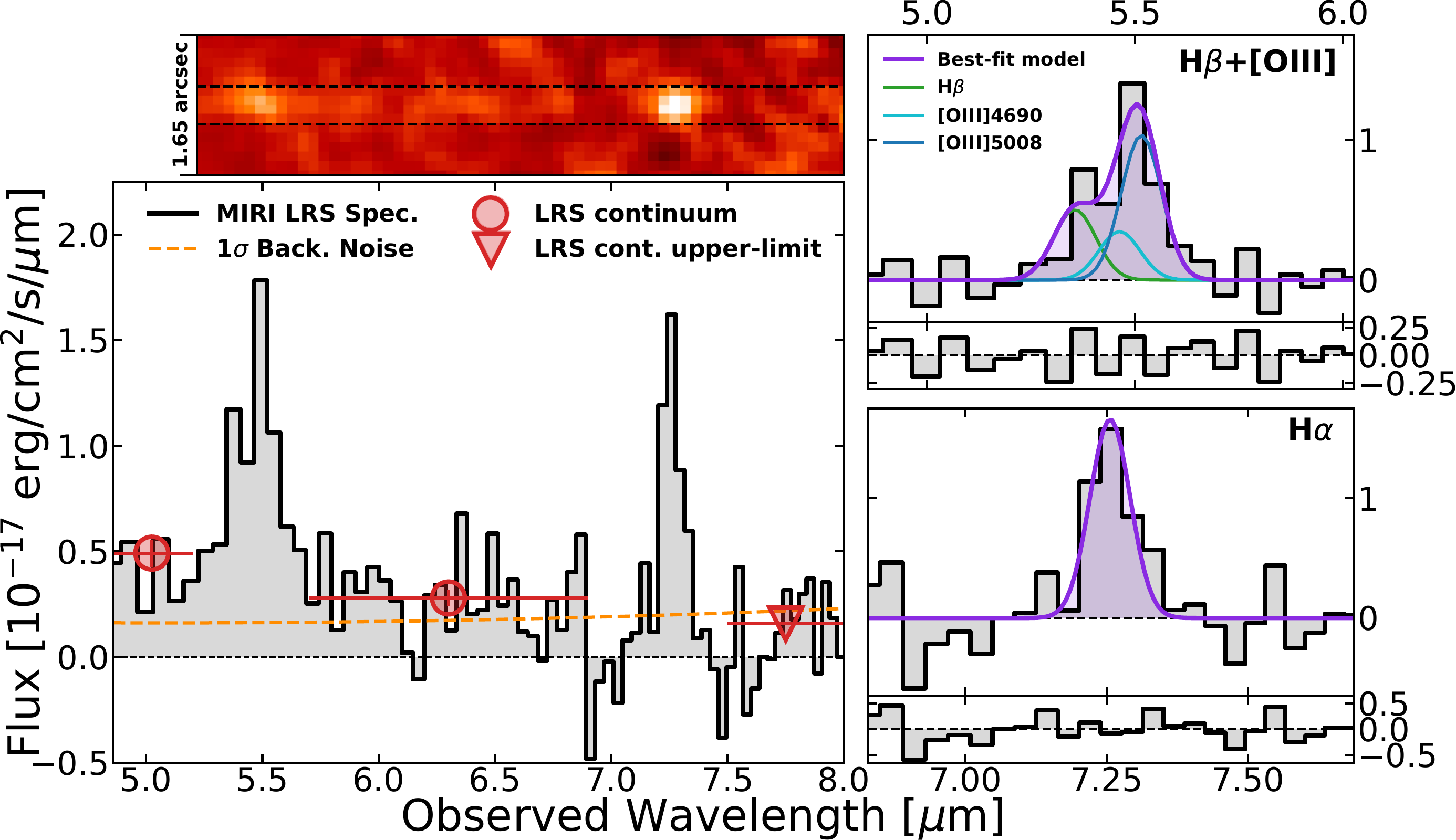}
      \caption{MIRI LRS spectrum and emission line fitting of U28165 galaxy at a redshift of 10.054\,$\pm$\,0.011. Upper left panel: area of the MIRI LRS 2D detector combined image with a total exposure time of 13.9\, hours. The dashed black lines define the aperture of 0.44\,arcsec used to extract the 1D spectrum. Bottom left panel: MIRI LRS 1D extracted spectrum and continuum fluxes. Black and gray area: 1D extracted spectrum showing the detection of the H$\beta$+[OIII] complex and the H$\alpha$ emission line. Red dots and triangle: stacked MIRI LRS continuum fluxes and 3$\sigma$ upper-limit at 5.0, 6.3, and 7.8\,$\mu$m. Orange dashed line: 1$\sigma$ error of the MIRI LRS spectrum. Upper right panel: three-component Gaussian fit of the H$\beta$+[OIII] complex (purple line and area). Green, cyan, and blue lines represent the H$\beta$, [OIII]4960$\AA$, and [OIII]5008$\AA$, respectively. Bottom right panel: one-component Gaussian fit of H$\alpha$ (purple line and area). Bottom panels of right panels are the residual of the line fitting.}
         \label{fig:LRS_spec}
\end{figure*}

Finally, we run the stage 3 of the pipeline to generate the combined 2D detector image and the 1D extracted spectrum (see left panels of Figure \ref{fig:LRS_spec}). The 1D extracted spectrum is calculated in an aperture with a diameter of 0.44\,arcsec (dashed lines in left-upper panel of Figure \ref{fig:LRS_spec}). We implement aperture correction following the reference file in the JWST pipeline (e.g. \texttt{jwst\_miri\_apcorr\_0017.fits}) and assuming a point-like source. Additionally, we extract two 1D background spectra from the full exposure time coverage of the 2D detector image to estimate the noise level (see orange dashed line in left-bottom panel of Figure \ref{fig:LRS_spec}).  

\subsection{Ancillary observations and absolute cross-instrument flux calibration}\label{Sect:ancillary}

NIRCam imaging and NIRSpec spectroscopy data are publicly available in the UNCOVER project\footnote{UNCOVER webpage: \url{https://jwst-uncover.github.io/}}. The NIRSpec R100 calibrated observations are obtained from the UNCOVER data release 4 (DR4; \citealt{Bezanson2024}). We use the NIRSpec 1D calibrated spectrum and emission lines fluxes derived by \cite{Price2025}\footnote{UNCOVER DR4 and DJA V4 NIRSpec spectra are compatible, and the emission lines fluxes agree within 1$\sigma$. DJA v4 include an extension of the wavelength up to 5.5\,$\mu$m, which covers the H$\beta$ emission line. The 3$\sigma$ upper-limit of the H$\beta$ flux is $<$\,66\,$\times$\,10$^{-20}$ erg\,s$^{-1}$\,cm$^{-2}$. We decide not to include that upper-limit in this work as the extension to longer wavelengths uses an extrapolated calibration. However, the value obtained is compatible with the H$\beta_{\mathrm{\,Case\,B}}$ used as fiducial in this work (see Sections \ref{Sect:LRS_fluxes} and \ref{Sect:extinction}).}, together with the redshift ($z$\,$=$\,10.061$^{+0.011}_{-0.013}$). The NIRCam photometry, combining the medium and wide band filters, is obtained from the SUPER catalog of the UNCOVER DR3 \citep{Suess2024}. We also use the updated gravitational magnification from the UNCOVER DR4 ($\mu$\,$=$\,4.07$^{+0.03}_{-0.11}$; \citealt{Furtak2023,Weaver2024,Suess2024,Price2025}). Finally, we include in the analysis the ALMA [OIII]88$\mu$m line and 90\,$\mu$m continuum emissions from \cite{Algera25-ALMA}.

To correct unknown uncertainties in the calculation of the absolute slit path-losses of NIRSpec and MIRI spectra, we normalize both spectra to the NIRCam photometry. We first normalized NIRSpec to NIRCam photometry. We use the broad and medium band fluxes presented in Table \ref{tab:fluxes} from the UNCOVER DR3 SUPER catalog, and generate synthetic NIRSpec photometry by integrating through the same filter transmission functions. We match both multi-band photometries, and derive a normalization factor of 0.75$\pm$0.09 for the NIRSpec spectra. NIRSpec emission lines fluxes from \cite{Price2025} are also normalized by the same factor, and presented in Table \ref{tab:fluxes}. MIRI LRS spectrum is normalized to the normalized NIRSpec spectrum as NIRCam or MIRI photometry is not available in the MIRI spectral range. We generate a synthetic filter in the spectral range 4.85-5.30\,$\mu$m, common to both NIRSpec and MIRI LRS, and then normalize MIRI LRS to NIRSpec. We obtain a factor of 1.24$\pm$0.27. Despite the relatively large uncertainty, we renormalize the MIRI LRS spectrum by this factor to have all spectroscopic measurements (e.g. continuum and lines) in the NIRCam photometric scale. 

This work only uses the relative uncertainties associated with the specific observations. The additional absolute uncertainty due to the current calibration of the NIRCam, NIRSpec and MIRI absolute spectrophotometry is not taken into account. In general, this is estimated to be at the 1-2\%, 15\% and 3-8\% level\footnote{Additional information on the absolute flux calibrations in the JWST instruments: \url{https://jwst-docs.stsci.edu/jwst-calibration-status}}, respectively.

\begin{table}
\caption{Magnified, scaled fluxes of all detected lines and continuum}\label{tab:fluxes}

\begin{tabular}{lclc}
\hline
\hline
\multicolumn{4}{l}{MIRI LRS line fluxes [10$^{-20}$ erg\,s$^{-1}$\,cm$^{-2}$]}  \\
\hline
H$\beta_{\mathrm{\,Gaussian-fit}}$ & 78$\pm$25 & {[OIII]}5008 & 158$\pm$17 \\
H$\beta_{\mathrm{\,Case\,B}}^{(a)}$ & 66$\pm$8 & H$\alpha$ & 186$\pm$23\\
\hline
\multicolumn{4}{l}{MIRI LRS cont. fluxes  [nJy]}  \\
\hline
LRS$_{5.0\mu m}$ & 41$\pm$6 & LRS$_{7.8\mu m}$ & $<$\,40$^{(f)}$\\
LRS$_{6.3\mu m}$ & 46$\pm$7 & & \\
\hline
\multicolumn{4}{l}{NIRSpec R100 line fluxes [10$^{-20}$ erg\,s$^{-1}$\,cm$^{-2}$]$^{(b)}$}  \\
\hline
{[OII]}3727,3730 & 46.7$\pm$9.1 & H$\gamma$ &  $<$\,29$^{(f)}$\\
{[NeIII]}3870 & 32.4$\pm$9.1 & & \\
\hline
\multicolumn{4}{l}{NIRCam band fluxes [nJy]$^{(c)}$}  \\
\hline
F115W & $<$\,8.9 & F300M & 44$\pm$4\\
F140M & 36$\pm$7 & F335M & 47$\pm$4\\
F150W & 48$\pm$3 & F356W & 40$\pm$2\\
F162M & 65$\pm$6 & F360M & 40$\pm$3\\
F182M & 50$\pm$5 & F410M & 48$\pm$3\\
F200W & 50$\pm$3 & F430M & 45$\pm$7\\
F210M & 54$\pm$5 & F444W & 49$\pm$3\\
F250M & 50$\pm$6 & F460M & 41$\pm$9\\
F277W & 48$\pm$2 & F480M & 55$\pm$8\\
\hline
\multicolumn{4}{l}{ALMA line and continuum fluxes$^{(d)}$}\\
\hline
\multicolumn{2}{l}{[OIII]88$\mu$m} & \multicolumn{2}{c}{$<$\,17$\times$10$^{-20}$\,erg\,s$^{-1}$\,cm$^{-2}$$^{(f)}$}\\ 
\multicolumn{2}{l}{Band 7 cont.} & \multicolumn{2}{c}{$<$\,28.8\,$\mu$Jy$^{(f)}$}\\
\hline
\hline
\end{tabular}
\tablefoot{$^{(a)}$\,H$\beta$ is derived from the H$\alpha$ assuming H$\alpha$/H$\beta$\,=\,2.80. $^{(b)}$ NIRSpec fluxes are from the UNCOVER DR4 \citep{Bezanson2024}. $^{(c)}$ NIRCam fluxes are from the SUPER catalog of the UNCOVER DR3 \citep{Suess2024}. $^{(d)}$ ALMA fluxes are from \cite{Algera25-ALMA}. $^{(f)}$ 3$\sigma$ upper-limits. All emission line and continuum fluxes are scaled to NIRCam photometry following the description of Section \ref{Sect:ancillary}.}
\end{table}

\section{Analysis}\label{Sect:analysis}

\subsection{MIRI LRS emission line and continuum fluxes}\label{Sect:LRS_fluxes}

Figure \ref{fig:LRS_spec} shows the MIRI LRS 2D detector image and the 1D extracted spectrum of U26185 galaxy. The H$\beta$+[OIII] complex and H$\alpha$ emission line are detected with integrated signal-to-noise (SNR) of 10 and 8, respectively. A faint continuum emission is detected at 5.0 and 6.3\,$\mu$m, with a SNR of 6-7, when multiple spectral elements are stacked. At 7.8\,$\mu$m, where there is no significant detection above 3$\sigma$, a strong upper-limit is estimated.  

MIRI LRS 1D spectrum has a resolving power ranging from 40 to 100 (i.e. FWHMs of 7500\,-\,3000\,km/s) at wavelengths between 5 and 8\,$\mu$m \citep{Kendrew2015}, respectively. At these spectral resolutions, the H$\beta$+[OIII] complex is spectrally unresolved, and all emission lines, including H$\alpha$, are also kinematically unresolved. We perform a three and one-component Gaussian fit to calculate the emission line fluxes and redshift of H$\beta$+[OIII] and H$\alpha$, respectively. We assume: (i) all emission lines to be unresolved fixing the FWHM to the corresponding spectral resolution of the instrument, (ii) independent redshift for each set of lines as uncertainties on the wavelengths are on average $\pm$\,20\,nm, and larger at 5\,$\mu$m where the H$\beta$+[OIII] complex is located (see also \citealt{Zavala+2024}), (iii) the theoretical value of 2.98 for the [OIII] line ratio ([OIII]$\lambda$5008$\AA$/[OIII]$\lambda$4960$\AA$; \citealt{Storey2000}), and (iv) an one-order polynomial function for the dependency of the continuum on wavelength. The uncertainties on the best-fit parameters, such as flux and redshift, are obtained as the standard deviation of all the individual measurements of 1000 bootstrapped spectra after adding a random Gaussian noise equal to the RMS of the original spectrum. The Gaussian fits are presented in the right panels of Figure \ref{fig:LRS_spec}, while the H$\beta$, [OIII]$\lambda$5008$\AA$, and H$\alpha$ emission line fluxes are given in Table \ref{tab:fluxes}. We provide two H$\beta$ flux estimates: one obtained directly from the fit to the H$\beta$+[OIII] complex, and another inferred from the H$\alpha$ line assuming Case B recombination (H$\alpha$/H$\beta$\,=\,2.8; for T$_{\mathrm{e}}$\,=\,15000\,K and n$_{\mathrm{e}}$\,=\,1000\,cm$^{-3}$). Both H$\beta$ fluxes are consistent within uncertainties (see Section \ref{Sect:extinction} for additional details).

We derive a redshift equal to $z$\,$=$\,10.010\,$\pm$\,0.020 and $z$\,$=$\,10.054\,$\pm$\,0.011 for the H$\beta$+[OIII] complex and the H$\alpha$ line, respectively. The H$\beta$+[OIII] redshift is less reliable due to the larger uncertainties in the LRS wavelength calibration around 5\,$\mu$m and because the lines are spectrally unresolved.
We therefore consider the H$\alpha$-based value as the fiducial redshift of U26185 and adopt it for all subsequent analysis. The H$\alpha$–based redshift is in agreement within uncertainties with the NIRSpec-based redshift value ($z$\,$=$\,10.061$^{+0.011}_{-0.013}$). 

We also calculate the continuum emission at 5.0, 6.3, and 7.8\,$\mu$m by stacking multiple spectral elements. We define three spectral windows clean of emission lines: 4.85\,-\,5.2\,$\mu$m, 5.7\,-\,6.9\,$\mu$m, and 7.5\,-\,8\,$\mu$m. We obtain fluxes of 41$\pm$6 and 46$\pm$7\,nJy at 5.0 and 6.3\,$\mu$m, and a 3$\sigma$ upper-limit of 40\,nJy at 7.8\,$\mu$m. Values are given in Table \ref{tab:fluxes} and represented in Figure \ref{fig:LRS_spec}. 

We combine the measured emission-line fluxes with the stacked continuum measurements to derive the equivalent widths (EWs) of H$\beta$, [OIII]$\lambda\lambda$4960,5008$\AA$, and H$\alpha$ emission lines. The continuum flux density underlying each line is estimated by fitting a linear relation to the two MIRI continuum fluxes and extrapolating to the wavelength of each line. The EWs are then computed as the ratio of the line flux to the corresponding continuum flux density. The resulting values are reported in Table~\ref{tab:properties}. 

\subsection{SED-fitting analysis}\label{sect:sed}

We perform a SED fitting analysis using the SpectroPhotometric version of \texttt{CIGALE} \citep{Boquien2019,Burgarella2025}, and the combination of NIRCam photometry, NIRSpec spectrum, and MIRI LRS and ALMA emission lines and continuum fluxes (see Table \ref{tab:fluxes}). NIRCam wide- and medium-band filters, together with synthetic continuum bands derived from MIRI LRS, probe the rest-frame UV to optical wavelengths up to 0.7\,$\mu$m, while ALMA provides access to the rest-frame far-IR continuum at 90\,$\mu$m. NIRSpec spectra cover the rest-frame UV up to 0.45\,$\mu$m, revealing both the continuum and lines, including CIII]$\lambda$1909\AA, {[OII]}$\lambda\lambda$3727,3730\AA\ and {[NeIII]}$\lambda$3870\AA. In addition to the continuum fluxes, MIRI LRS provides the {[OIII]}$\lambda$5008\AA\ and H$\alpha$ , while ALMA provides the {[OIII]}\,$88\mu$m emission line fluxes. Altogether, this constitutes one of the most comprehensive multiwavelength datasets available for a galaxy at $z$\,$\sim$\,10, simultaneously constraining the continuum and lines from the rest-frame UV to the FIR (see also \citealt{Zavala2024,Carniani+24_ALMA,Castellano2025_GHz2}).

The SFH is modeled as the combination of a delayed-$\tau$ component, representing the bulk of the stellar mass, and a constant star-formation burst accounting for the young stellar population. The delayed-$\tau$ component is characterized by e-folding times ($\tau$) ranging from 1 to 250\,Myr and ages between 25 and 200\,Myr. The constant young burst spans ages from 1 to 10\,Myr. We adopted the stellar population models from \cite{Bruzual&Charlot+03} and the \cite{Chabrier+03} IMF. We included nebular continuum and lines, using an electron density of 1000\,cm$^{-3}$ following the electron density-redshift relation derived for high-z galaxies up to $z$\,$\sim$\,10 \citep{Abdurrouf+24}, ionised parameter, $\log(U)$, values from $-$3.2 to $-$2.0, and the fraction of Lyman continuum photons escaping the galaxy set to zero. Both stellar populations and nebular emission use metallicities ranging from 2\% to 20\%\,Z$_{\odot}$.

\begin{figure*}
\centering
   \includegraphics[width=\linewidth]{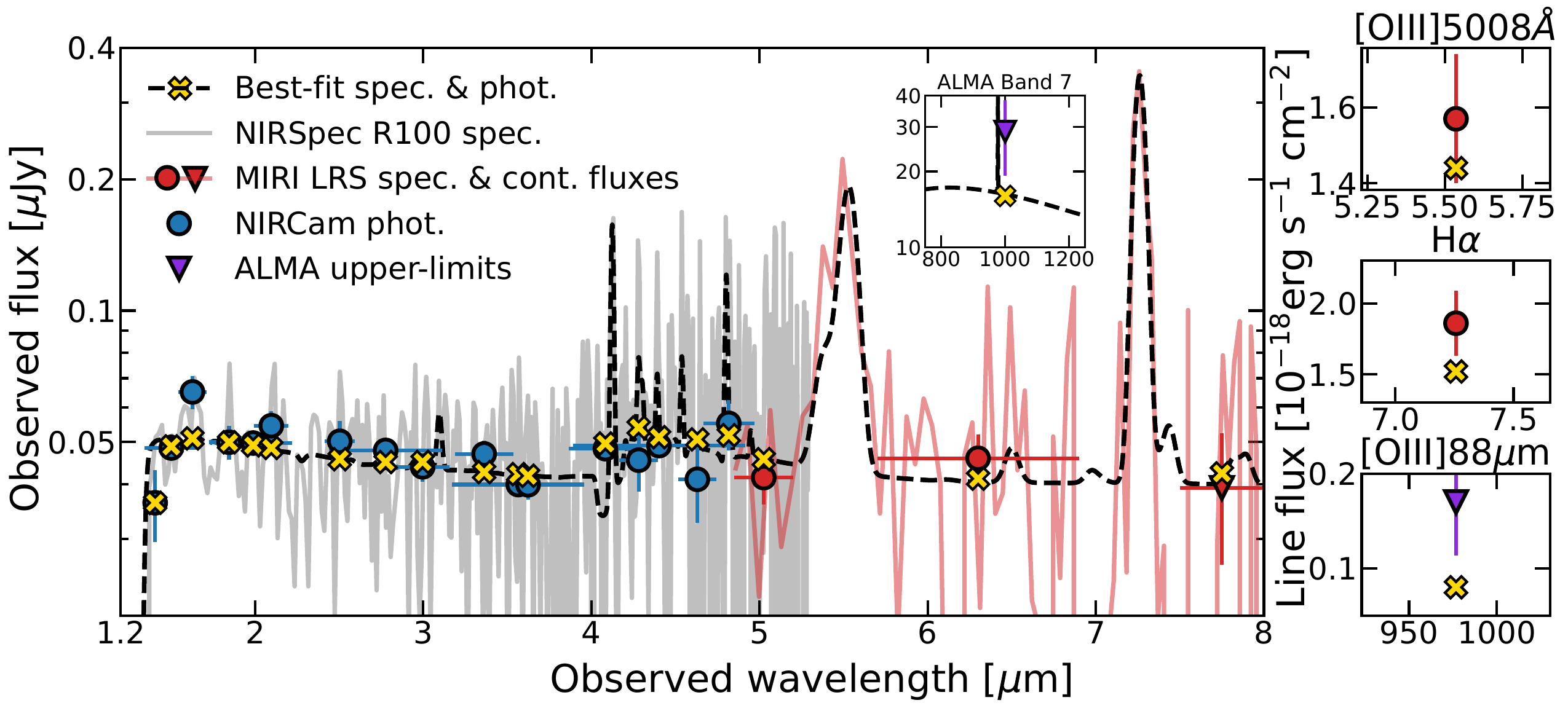}
      \caption{SED-fitting analysis using the magnified, scaled fluxes of all detected lines and continuum emission of U26185 at redshift of 10.05. The analysis combine the NIRCam photometry (Blue dots), NIRSpec R100 spectrum (gray line), MIRI LRS derived continuum photometry and [OIII]$\lambda$5008$\AA$ and H$\alpha$ fluxes (Red dots for detections and red triangles for 3$\sigma$ upper-limits), and the ALMA Band 7 continuum and [OIII]$\lambda$88$\mu$m 3$\sigma$ upper-limits (Purple triangles). The best-fit CIGALE model is represented with dashed back line, and its synthetic NIRCam photometry, and MIRI LRS and ALMA continuum fluxes and emission lines fluxes are represented by yellow crosses.}
         \label{fig:SED}
\end{figure*}

Dust attenuation is modeled using the module of \texttt{CIGALE} (\texttt{dustatt\_modified\_starburst}). Stellar attenuation follows the modified Calzetti prescription of \citet{Noll2009}, based on the Calzetti law \citep{Calzetti+00}. We vary the attenuation curve slope $\delta$ between 0 and $-0.6$, and fix the UV bump amplitude to zero. Nebular dust attenuation is modeled using the Small Magellanic Cloud (SMC) extinction curve \citep{Pei1992}. The visible nebular dust attenuation ranges from 0 to 0.4 mag, as derived from the Balmer decrements (see Section \ref{Sect:extinction}), and the stellar-to-nebular dust attenuation ratio goes from 0.33 to 1. This configuration provides the flexibility to explore dust attenuation curves steeper than the Calzetti law, as commonly observed in galaxies with low stellar masses and metallicities (e.g. \citealt{Alvarez-Marquez+19_SED,Shivaei2020,Fudamoto2020}). Infrared emission is modeled using the dust templates of \citet{Draine2007}, with parameters fixed to those derived for the average population of high$-z$ ALPINE galaxies \citep{Burgarella+22}.

\begin{table}
\caption{Intrinsic, de-lensed physical properties of U26185 assuming a star-forming galaxy scenario.}\label{tab:properties}

\begin{tabular}{lll}
\hline
\hline
 \multicolumn{3}{c}{Properties derived from emission lines and continuum} \\
\hline
Redshift & & 10.054\,$\pm$\,0.011\\
M$_{\mathrm{UV}}$ & [mag] & -18.83$\pm$0.07 \\
A$_{\mathrm{V-lines}}$ & [mag] & 0.0$_{-0.0}^{+0.4}$ \\
SFR$_{\mathrm{H}\alpha}$ & [M$_{\odot}$\,yr$^{-1}$] & 1.3\,$\pm$\,0.2 \\
$\log(\xi_\mathrm{ion})$ & [$\mathrm{Hz\,erg^{-1}}$] & 25.50\,$\pm$\,0.06 \\
EW(H$\beta$) & [$\AA$] &  138$\pm$26 \\
EW([OIII]4960,5008) & [$\AA$] & 467$\pm$81 \\
EW(H$\alpha$) & [$\AA$] & 1399$\pm$271 \\
$\log($U$)$ & & -2.5$\pm$0.1 \\
Metallicity & [Z$_{\odot}$] & 0.04$\pm$0.01  \\
\hline
 \multicolumn{3}{c}{Properties derived from SED-fitting analysis} \\
\hline
A$_{\mathrm{V-SED}}$ & [mag] & 0.20$\pm$0.09\\
$\beta_{\mathrm{UV}}$ & & -2.29$\pm$0.06 \\
SFR$_{\mathrm{SED-10\,Myr}}$ & [M$_{\odot}$\,yr$^{-1}$] & 1.7$\pm$0.4\\
M$_{\mathrm{Stellar}}$ & [$\times$10$^{8}$\,M$_{\odot}$] & 1.7$\pm$0.3\\
Mass weighted Age & [Myr] & 65$\pm$20\\
sSFR & [Gyr$^{-1}$] & 7.6\,$\pm$\,1.2 \\
M$_{\mathrm{Stellar-burst}}$ & [$\times$10$^{8}$\,M$_{\odot}$] & 0.17$\pm$0.04\\
Age$_{\mathrm{burst}}$ & [Myr] & 7$\pm$3 \\
D4000 & & 1.33$\pm$0.03 \\
$\log($U$)$ & & -2.6$\pm$0.2 \\
Metallicity & [Z$_{\odot}$] & 0.05$\pm$0.01 \\
\hline
\hline
\end{tabular}
\end{table}

The results of the \texttt{CIGALE} Bayesian analysis are presented in Table~\ref{tab:properties}, while the best-fit SED model is shown in Figure~\ref{fig:SED}. Assuming a two-component SFH, we derive a total stellar mass of (1.7\,$\pm$\,0.3)\,$\times$\,10$^{8}$\,$\mathrm{M}_{\odot}$ and a mass-weighted stellar age of 65\,$\pm$\,20\,Myr. The ongoing burst, with an age of 7\,$\pm$\,3\,Myr, is characterized by a star formation rate (SFR) of 1.7\,$\pm$\,0.4\,$\mathrm{M}_{\odot}$\,$\mathrm{yr}^{-1}$ and stellar mass of (1.7\,$\pm$\,0.4)\,$\times$\,10$^{7}$\,$\mathrm{M}_{\odot}$, i.e. 10\% of the stellar mass of the galaxy. The Balmer break strength (D4000; \citealt{Balogh1999}), defined as the ratio of the flux in the red continuum to that in the blue continuum, is 1.33$\pm$0.03 supporting the presence of a more mature stellar population. The inferred dust attenuation favors a curve steeper than the standard Calzetti law, with a $\delta$\,$=$\,$-$0.3\,$\pm$\,0.2 and a A$_{\mathrm{V}}$\,$=$\,0.20\,$\pm$\,0.09\,mag. The nebular metallicity is constrained to 0.05\,$\pm$\,0.01\,Z$_{\odot}$, with an ionization parameter of $\log$(U)\,$=$\,$-$2.6\,$\pm$\,0.2. While the A$_{\mathrm{V}}$ values derived independently from emission lines and SED fitting are consistent within the uncertainties (see Section~\ref{Sect:extinction}), the fit is affected by limitations in the nebular model grid at very low metallicities ($<$\,0.05\,Z$_{\odot}$). In particular, for U26185 the observed [O III]$\lambda5008$-to-H$\alpha$ line ratio of $\sim$\,0.85 implies a metallicity below 0.05\,Z$_{\odot}$ according to the nebular models implemented in \texttt{CIGALE}. As the available models do not have a grid to fully reproduce such low-metallicity line ratios, the fitting procedure could compensate by invoking differential dust attenuation between the two lines (e.g., a steeper attenuation slope or higher overall dust attenuation), or may even fail to reproduce within 1$\sigma$ the observed line fluxes of H$\alpha$ (see Figure \ref{fig:SED}). This degeneracy may lead to a modest overestimation of the total dust attenuation and, consequently, affect the inferred metallicity. Nevertheless, the combined continuum and line model provides a good match to the MIRI LRS spectrum (see the left panel of Figure~\ref{fig:SED}), and the dust attenuation and metallicity are in agreement with the ones derived using only emission lines (see Table \ref{tab:properties}).  

\section{Results}\label{Sect:results_dis}

\subsection{Dust attenuation}\label{Sect:extinction}

We estimate the nebular dust attenuation using Balmer decrements derived from the hydrogen recombination lines covered by the combined MIRI LRS and NIRSpec observations. The observed H$\alpha$/H$\beta$, H$\alpha$/H$\gamma$, and H$\beta$/H$\gamma$ line ratios are $2.4 \pm 0.8$, >6.4, and >2.69, respectively. Within the uncertainties, these Balmer decrements are consistent with the Case B recombination values expected for ISM inferred in previously observed $z>10$ galaxies (e.g., $T_{\mathrm{e}} = 15{,}000$ K and $n_{\mathrm{e}} = 1000\,\mathrm{cm}^{-3}$; \citealt{Hsiao+2024_MIRI, Abdurrouf+24, Alvarez-Marquez_2025}): H$\alpha$/H$\beta$ = 2.80, H$\alpha$/H$\gamma$ = 5.93, and H$\beta$/H$\gamma$ = 2.12. Considering \cite{Cardelli+89}, these results are compatible with no dust attenuation, but the 1$\sigma$ uncertainty in the H$\alpha$/H$\beta$ provides an upper-limit of 0.4 mag (A$_{\mathrm{V-lines}}$\,$=$\,0$^{+0.4}_{-0.0}$\,mag).

\texttt{CIGALE} SED-fitting analysis, combining rest-frame UV$-$optical spectra, emission lines, and photometry together with the ALMA continuum upper-limits, provides an absolute (nebular + stellar) visible dust attenuation of A$_{\mathrm{V-SED}}$\,$=$\,0.20\,$\pm$\,0.09\,mag. These values are consistent, within uncertainties, with those derived from \texttt{Bagpipes} SED-fitting analysis combining only NIRSpec spectra and NIRCam wide-band photometry (A$_{\mathrm{V-SED}} = 0.08^{+0.08}_{-0.05}$\,mag; \citealt{Goulding23-UNCOVER}). However, we note that the \texttt{CIGALE} analysis could be slightly overestimating the dust attenuation, even though the current A$_{\mathrm{V}}$ values derived from emission lines and SED-fitting are still consistent within uncertainties (see Section \ref{sect:sed}), and lower than the measured nebular upper-limit. Additionally, the current dust attenuation value of \texttt{CIGALE} is in agreement with the upper-limit measured in the dust continuum emission at 90\,$\mu$m (see Figure \ref{fig:SED}).

Recent deep ALMA Band 7 observations have placed a 3$\sigma$ upper-limit of 5.5\,$\times$\,10$^5$\,M$_{\odot}$ to the cold dust mass present in U26185 \citep{Algera25-ALMA}. This upper-limit represents a small dust-to-stellar mass ratio (M$_{\mathrm{dust}}$/M$_{\mathrm{stellar}}$\,$<$\,3.2$\times$10$^{-3}$), suggesting very low or negligible dust content in this galaxy, and compatible with no attenuation as derived from the optical emission lines and SED-fitting analysis. Therefore, we assume an A$_{\mathrm{V}}$\,=\,0\,mag throughout this paper, and a H$\beta$ flux derived assuming Case B recombination and the H$\alpha$ flux (H$\beta_{\mathrm{\,Case\,B}}$; see Table \ref{tab:fluxes}).

\subsection{Star-formation history and ionizing photon production efficiency}\label{Sect:sfr}

The instantaneous SFR derived from the total H$\alpha$ luminosity and considering no-dust attenuation, SFR$_{\mathrm{H}\alpha}$, is 1.4\,$\pm$\,0.2 and 1.3\,$\pm$\,0.2\,$M_{\odot}$\,yr$^{-1}$ using the relations by \cite{Theios+19} and  \cite{Reddy+22}, calibrated for galaxies with metallicities of 0.1\,$Z_\odot$ and 0.05\,$Z_\odot$, respectively. These values are consistent within 1$\sigma$ uncertainties with those obtained from SED-fitting analyses using \texttt{CIGALE} (SFR$_{\mathrm{SED-10\,Myr}}$\,$=$\,1.7$\pm$0.4\,M$_{\odot}$\,yr$^{-1}$) and \texttt{Bagpipes} SFR\,$=$\,1.25$^{+0.18}_{-0.20}$\,M$_{\odot}$\,yr$^{-1}$; \citealt{Goulding23-UNCOVER}) both averaged over the last 10\,Myr. Additionally, the SFR derived from the UV luminosity (SFR$_{\mathrm{UV}}$) average over the past 100\,Myr, is 1.0\,M$_{\odot}$\,yr$^{-1}$. We assume a metallicity of 0.1\,$Z_\odot$ \citep{Theios+19} and a UV luminosity density of 1.5\,$\times$\,10$^{28}$\,erg\,s$^{-1}$\,Hz$^{-1}$ at 1500\,$\AA$, computed from the average fluxes of the F150W and F200W NIRCam bands. 

The resulting burstiness parameter (SFR$_{\mathrm{H}\alpha}$-to-SFR$_{\mathrm{UV}}$ ratio) is 1.4\,$\pm$\,0.2 for a metallicity of 0.1\,$Z_\odot$, where SFR$_{\mathrm{H}\alpha}$ and SFR$_{\mathrm{UV}}$ trace star formation over 10\,Myr and 100\,Myr timescales, respectively, assuming constant SFH. The different timescales probed by these SFR indicators allow us to investigate the burstiness of the star formation. The inferred burstiness parameter consistent with unity in 2$\sigma$, indicating that the galaxy is not dominated by a recent strong burst. This is consistent with the SFH derived from \texttt{CIGALE}, which suggests that the stellar population has a mass-weighted age of 65\,$\pm$\,20\,Myr, also consistent with the half-mass ages derived from \texttt{Bagpipes} (65$_{-20}^{+31}$\,Myr; \citealt{Goulding23-UNCOVER}). This is also supported by the Balmer break strength higher than 1, indicating that a more mature stellar population should be invoked. Although the error makes the measurement in full agreement with unity, the slightly elevated value than 1 (1.4) suggests the presence of an additional young stellar population coexisting with the bulk of the stellar component, in agreement with the young burst identified by \texttt{CIGALE}, which contributes $\sim$10\% of the total stellar mass.

Following the methodology outlined by \cite{Alvarez-Marquez+23-MACS}, we derive an ionizing photon production efficiency of $\log$($\xi_\mathrm{ion}$\,[Hz\,erg$^{-1}$])\,=\,25.50\,$\pm$\,0.06 under the assumption of zero LyC escape fraction and non-dust attenuation. This value is higher than the commonly adopted canonical value of 25.2\,$\pm$\,0.1\,Hz\,erg$^{-1}$ \citep{Robertson+23}, yet remains consistent with values inferred from the redshift extrapolation of intermediate-redshift galaxies (\citealt{Matthee+17}; see also Figure 3 of \citealt{Alvarez-Marquez+23-MACS}). The inferred $\log(\xi_{\mathrm{ion}})$ is typical of sources in the epoch of reionization (e.g., \citealt{Atek+24, Fujimoto+23, Tang+23, Morishita+23, Rinaldi+2024, Alvarez-Marquez+23-MACS, Morishita2024,Simmonds2024,Komarova2025,Llerena2025,Prieto-Jimenez2025,Zavala+2024}), and of galaxies at intermediate redshifts ($2<z<5$) exhibiting the highest specific star-formation rates, $\log$(sSFR[yr$^{-1}$])\,$\sim$\,$-$7.5  \citep{Castellano+23}, similar ones than that of U26185 ($\log$(sSFR[yr$^{-1}$])\,$\sim$\,$-$7.6\,$\pm$\,1.2). In the context of galaxies at $z>10$, U26185 exhibits an intermediate $\xi_{\mathrm{ion}}$ compared to GHZ2 (25.7$_{-0.1}^{+0.2}$\,Hz\,erg$^{-1}$, \citealt{Calabro2024}), GN-z11 (25.66\,$\pm$\,0.06\,Hz\,erg$^{-1}$, \citealt{Alvarez-Marquez_2025}), and MACS0647-JD and JADES-GS-z14-0 (25.3\,$\pm$\,0.1\,Hz\,erg$^{-1}$, \citealt{Hsiao+2024_MIRI,Helton2025}). 

U26185 is also characterized by a very high H$\alpha$ equivalent width, EW(H$\alpha$)\,$=$\,1399\,$\pm$\,271\AA. This value is consistent with the extrapolation of the EW(H$\alpha$)\,$\propto$\,(1+z)$^{2.1}$ relation derived from a combination of pre-JWST measurements and the results of the JADES and MIDIS JWST surveys (\citealt{Rinaldi+23}; see also Figure 4 of \citealt{Alvarez-Marquez+23-MACS}). Binary Population and Spectral Synthesis (BPASS; \citealt{Eldridge+Stanway+20}) models indicate that EW(H$\alpha$) values exceeding 1000\,$\AA$ can only be produced by stellar populations younger than 10\,Myr. Additionally, ionizing photon production efficiencies of $\log(\xi_{\mathrm{ion}})\gtrsim25.5$\,Hz\,erg$^{-1}$ require young, massive star-forming bursts with stellar masses of at least $\sim10^{7}\,M_{\odot}$, owing to the stochastic nature of massive star formation \citep{Stanway+Eldridge-23}. Furthermore, the location of U26185 in the $\log(\xi_{\mathrm{ion}})$–EW(H$\alpha$) plane (see Figure 5 of \citealt{Prieto-Jimenez2025}) suggests stellar population ages of $\sim$4–6\,Myr, for an instantaneous burst. All these independent constraints, together with the slightly elevated value relative to unity of the burstiness parameter, are consistent with the star-forming burst inferred from \texttt{CIGALE}, which indicates the presence of a young stellar component with an age of 7\,$\pm$\,3\,Myr and a stellar mass of (1.7\,$\pm$\,0.4)\,$\times$\,10$^7$\,M$_{\odot}$.

In summary, the galaxy U26185 is characterized by a SFH that can be decomposed into two distinct stellar populations. The dominant, more mature component$-$which contains the bulk of the stellar mass$-$has a mass-weighted age of 65\,$\pm$\,20\,Myr and a stellar mass of (1.5\,$\pm$\,0.3)\,$\times$\,10$^{8}$\,M$_{\odot}$, indicating that the majority of the galaxy’s stars formed around redshift $z\sim11$. The young stellar population contributes $\sim$10\% of the total stellar mass, with a mass of (1.7\,$\pm$\,0.4)\,$\times$\,10$^7$\,M$_{\odot}$ and an age of 7\,$\pm$\,3\,Myr. The ongoing burst is characterized by a star-formation rate of 1.3\,$\pm$\,0.2\,M$_{\odot}$\,yr$^{-1}$, assuming a metallicity of 0.05\,$Z_\odot$.

\begin{figure*}
\centering
   \includegraphics[width=0.9\linewidth]{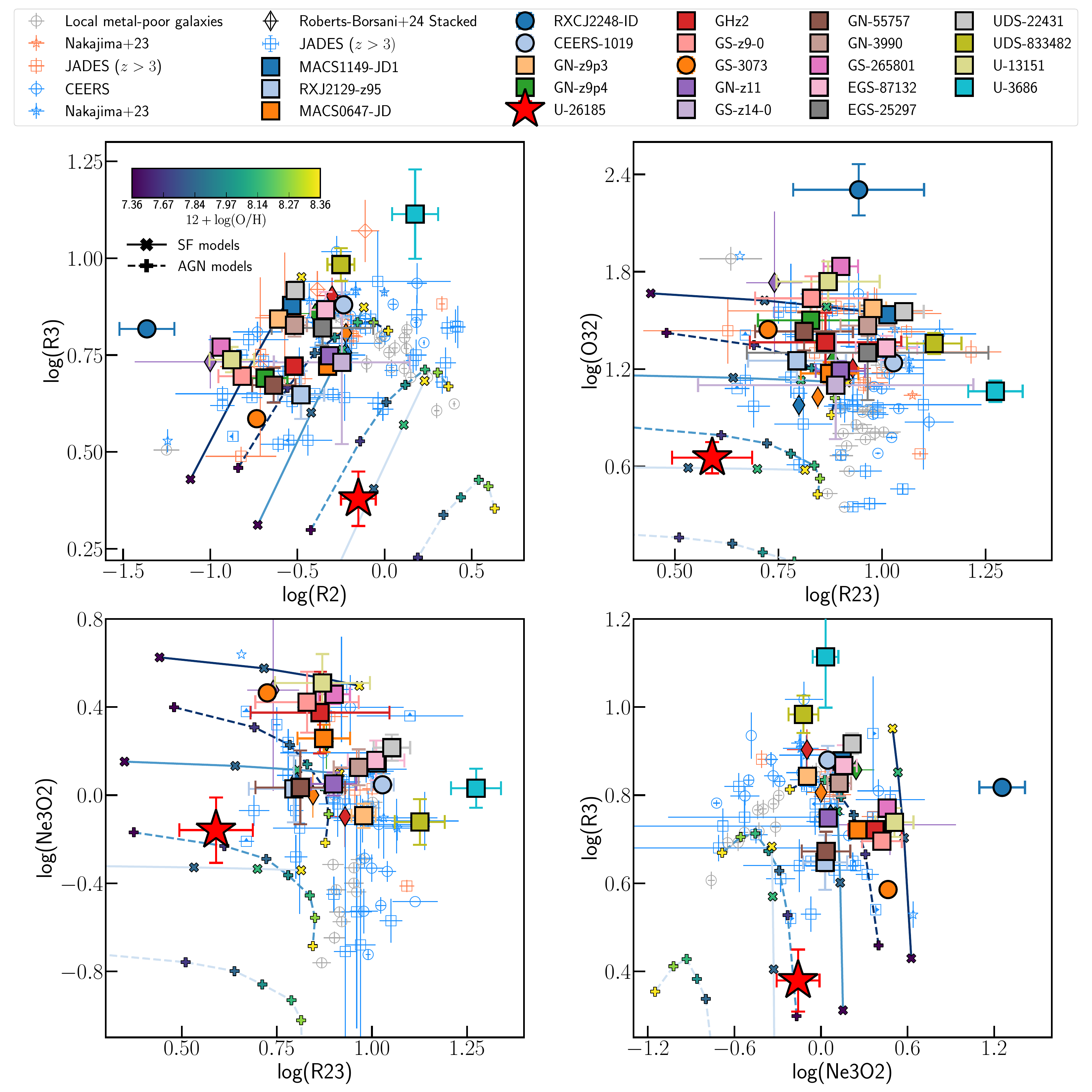}
      \caption{Line ratio diagnostic diagrams. Red star and colored squares and circles represent U26185 and the high-z and N-rich samples, as presented in Sect.~\ref{Sect:lineratios}. Gray, blue and red points display a sample of local metal-poor galaxies \citep{Izotov+06} and high-z SF and AGN galaxies from the main JWST surveys (Sect.~\ref{Sect:lineratios} for further details), respectively. Colored squares and circles represent recent JWST-observed $z>9$ galaxies and $z>5$ N-rich galaxies, respectively. Blue, orange, red, green and purple diamonds display the stacked values for the $z>$\,[5, 6, 7, 8, 9] galaxy subsamples from \citet{Roberts-Borsani+24}, respectively. Solid and dashed lines represent the CLOUDY photo-ionization models presented in \citet{Calabro+23} for the AGN and SF scenarios, respectively. Increasingly darker blue lines represent an increasing log($U$)=[-3, -2.5, -2] values while the metallicities are displayed as colored markers. }
         \label{fig:line_ratios}
\end{figure*}

\subsection{Ionization and emission line ratios}\label{Sect:lineratios}

We combine the MIRI LRS and NIRSpec R100 line fluxes (see Table~\ref{tab:fluxes}) to derive the following emission line ratios, which are then used in the diagnostic diagrams:

\begin{eqnarray*}
    \mathrm{R2} &=& [\mathrm{O\,II}]3727, 3730\AA/\mathrm{H}\beta \\
    \mathrm{R3} &=&  [\mathrm{O\,III}]5008\AA/\mathrm{H}\beta \\
    \mathrm{R23} &=& ([\mathrm{O\,III}]5008\AA + [\mathrm{O\,II}]3727, 3730\AA)/\mathrm{H}\beta \\
    \mathrm{O32}  &=& [\mathrm{O\,III}]4960,5008\AA/[\mathrm{O\,II}]3727, 3730\AA \\
    \mathrm{Ne3O2}  &=& [\mathrm{Ne\,III}]3870\AA/[\mathrm{O\,II}]3727, 3730\AA \\
    \mathrm{\hat{R}} &=& 
0.47  \times \mathrm{R2} + 0.88 \times \mathrm{R3} \\
\end{eqnarray*}
where H$\beta$ is derived assuming H$\alpha$/H$\beta$\,=\,2.80 (see $\S$\,\ref{Sect:LRS_fluxes}). The line ratios along with their associated uncertainties, are listed on Table~\ref{tab:metallicities}.

We use the combination of these line ratios to investigate the nature of the ionization source in U26185. Figure~\ref{fig:line_ratios} displays, as a red star, the position of U26185 in these diagnostic diagrams. For comparison, we also include data (colored squares) for $z$>9 galaxies (MACS1149-JD1, \citealt{Stiavelli+23}; RXJ2129-z95, \citealt{Williams+23}; MACS0647-JD, \citealt{Hsiao+23-NIRCam}; GN-z9p3, \citealt{Boyett+23}; GN-z9p4, \citealt{Schaerer-Rui2024}; GHz2, \citealt{Castellano2024,Zavala+2024};  GS-z9-0, \citealt{Curti+25}; GN-z11, \citealt{Bunker+23,Alvarez-Marquez_2025}; GS-z14-0, \citealt{Helton2025}; EGS-25297, \citealt{Donnan+25}; GN-55757, GN-3990, GS-265801, EGS-87132, UDS-22431, UDS-833482, U-13151 and U-3686, \citealt{Pollock+25} and references therein) and other $z$>5 N-rich galaxies (colored circles: CEERS-1019, \citealt{Larson+23,Zamora+25}; GS-3073, \citealt{Ji+Ubler2024}; RXCJ2248-ID, \citealt{Topping+24}). Blue and red markers display a sample of $z>3$ star-forming and AGN galaxies, respectively, obtained from JWST surveys (JADES, \citealt{Cameron2023,Bunker+24,Curti+24}; CEERS, \citealt{Sanders+24}; ERO+GLASS+CEERS, \citealt{Nakajima+23}). Colored diamonds represent the stacked value for the $z>5$ subsamples drawn from the 500 high redshift galaxies presented in \citet{Roberts-Borsani+24}. In addition, we include a sample of local metal-poor galaxies (gray circles,  \citealt{Izotov+06}) and the CLOUDY photoionization models for SFGs and AGN as presented in \citet{Calabro+23} and \citet{Calabro2024}.

U26185 appears as an outlier from the population of high-$z$ star-forming and AGN galaxies, presenting the lowest R3, O32 and R23 observed ratios up to date (see Figure~\ref{fig:line_ratios}). U26185 also clearly deviates from the ratios observed in local metal-poor galaxies. The line ratios in U26185 are 
compatible with low ionization parameter ($\log$($U$)\,$\sim$\,$-$2.5) and metal-poor (12+log(O/H)\,$\sim$\,7.5) CLOUDY photoionization models of SFGs. This low ionization parameter is also consistent with the value independently derived by other tracers. We derive ionization parameters $\log$($U$) of $-$2.5\,$\pm$\,0.1 and $-$2.4\,$\pm$\,0.1 based on the O32\,--\,$\log$($U$) relations from  \citet{Diaz+00}, derived using single-star photoionization models, and \citet{Papovich+22} for local and 1\,$<$\,$z$\,$<$\,2 galaxies, respectively. These values are also in agreement with the results obtained from \texttt{CIGALE} (Table~\ref{tab:properties}). If we consider the AGN CLOUDY models, the line ratios would predict a larger ionization parameter ($\log$($U$)\,$\sim$\,$-$2) and a higher metallicity (12+log(O/H)\,$\sim$\,8.1). See next Section for metallicity estimates and discussions. 

\subsection{Metallicity}\label{sect:metallicity}

\begin{figure*}
\centering
   \includegraphics[width=\linewidth]{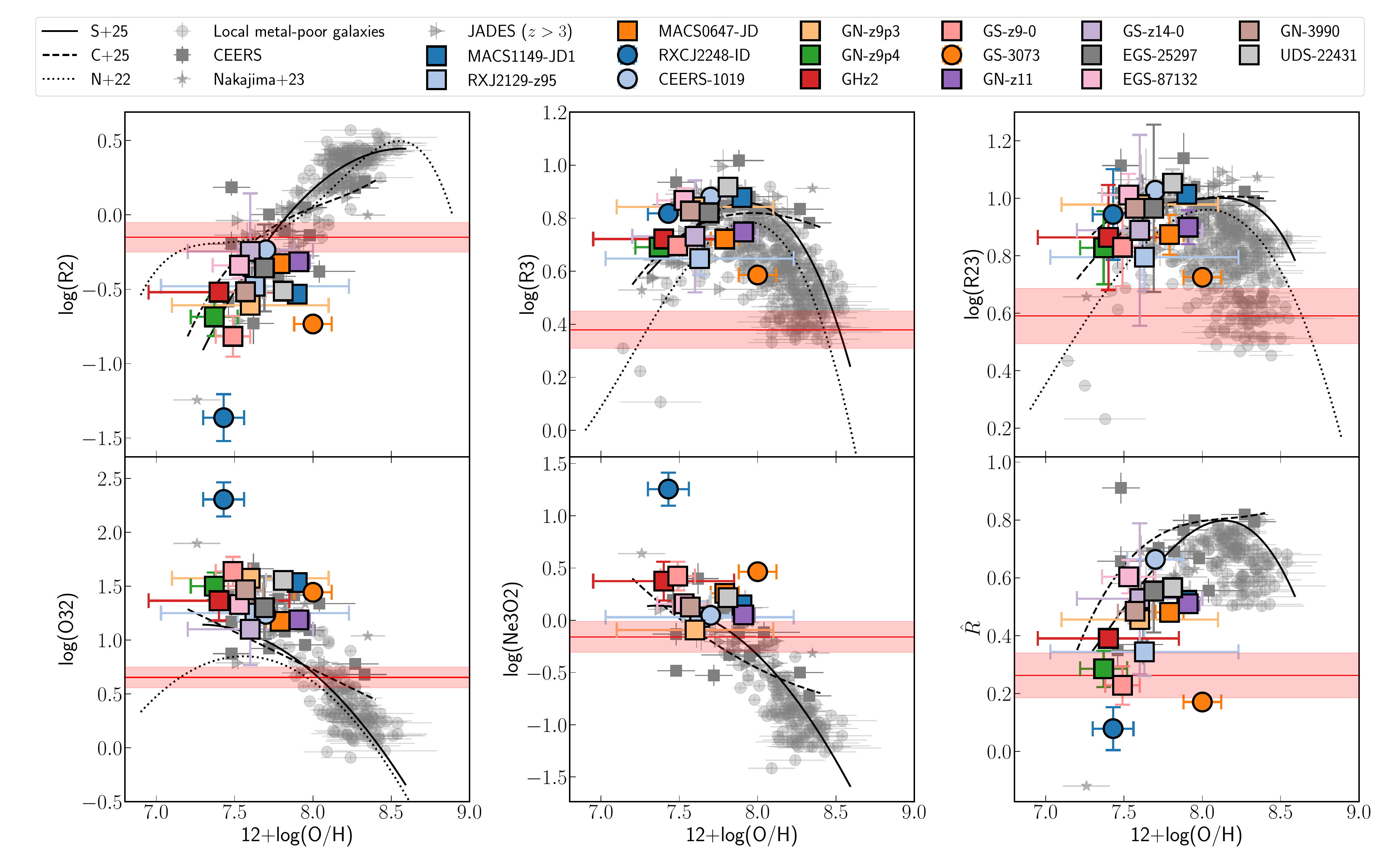}
      \caption{Metallicity relations based on the different line ratios. Grey points  and colored squares and circles display the samples of local metal-poor and high-$z$ galaxies, as in Fig.~\ref{fig:line_ratios}. Solid, dashed and dotted line represent the best-fit expressions derived for each line ratio by \citet{Sanders+25}, \citet{Chakraborty+25} and \citet{Nakajima+22}, respectively. Red horizontal line and shade mark the U26185 value and its uncertainty, respectively.}
         \label{fig:Z_ratios}
\end{figure*}

The dependence of the optical line ratios with the metallicity has been studied in detail 
using the $T_\mathrm{e}$-direct method and large samples of galaxies at redshifts up to 10. In this work, we use some of the most recent studies to infer the metallicity based on these optical line ratios (see Section \ref{Sect:lineratios}). Concretely, we consider the \citet{Sanders+25} and \citet{Chakraborty+25} relations (henceforth, S+25 and C+25, respectively) derived from high-$z$ (2\,$<$\,$z$\,$<$\,10) sample of galaxies with metallicities in the range of 12+$\log$(O/H)\,$\sim$\,7.2--8.6 observed with JWST. For completeness, we also consider the relations from \citet{Nakajima+22} (henceforth, N+22), derived from a SDSS sample of local galaxies covering a wider metallicity range (12+$\log$(O/H)\,$\sim$\,6.9--8.9). Figure~\ref{fig:Z_ratios} displays the empirical relations presented in N+22, S+25 and C+25 for each line ratio as dotted, solid and dashed lines, respectively. Gray points represent general samples of local metal-poor galaxies \citep{Izotov+06}, and $z$\,$>$\,3 galaxies from CEERS \citep{Sanders+25}, JADES \citep{Bunker+23,Cameron2023} and ERO+GLASS+CEERS \citep{Nakajima+23}. For comparison, we included the high-$z$ and N-rich galaxies as presented in Figure~\ref{fig:line_ratios}. Red horizontal line and shaded area represent the line ratio values and uncertainties for U26185, respectively. Table~\ref{tab:metallicities} summarizes the U26185 metallicities derived from the empirical relations presented in S+25, C+25 and N+22. In addition to the values derived from the general samples, we include the metallicities obtained from the medium 
EW(H$\beta$) bin (i.e., 100\,$<$\,EW(H$\beta$)\,[$\AA$]\,$<$\,200) in N+22, to explore potential dependence on EW(H$\beta$). The metallicity values derived from R2, R23 and $\hat{R}$ from S+25 and C+25 are obtained by extrapolating their expressions down to 12+$\log$(O/H)\,$=$\,6.9, to match the N+22 metallicity coverage.

Overall, the emission-line ratios of U26185 place this galaxy in the low-metallicity regime, with all diagnostics consistently indicating 12+$\log$(O/H)\,$<$\,8. The R2, R3, O32, and Ne3O2 indicators are strongly dependent on the ionization conditions, whereas the R23 diagnostic provides the most robust metallicity estimate, with an intrinsic scatter of 0.14\,dex and a weaker dependence on ionization parameters \citep{Nakajima+22,Sanders+25}. We therefore adopt as our fiducial metallicity the R23-based calibration from N+22 (12+log(O/H)=7.31$\pm$0.13), corresponding to the medium EW(H$\beta$) galaxy sample. This choice is motivated by the fact that U26185 exhibits EW(H$\beta$) values within this regime and avoids extrapolation into extremely metal-poor conditions, which would be required when applying the calibrations of S+25 and C+25. We obtain similar metallicities when using the relations derived from local analogs, combining the R23 and O32 ratios to improve the accuracy in the low-metallicity regime (12+log(O/H)=7.3$\pm$0.4; \citealt{Izotov+19}). Using the \texttt{genesis-metallicity} code \citep{genesis-metallicity}, which is based a non-parametric combination of the emission line ratios, we found 12\,+\,log(O/H)\,=\,7.09\,$\pm$\,0.05, fully compatible with the low limit provided by our extrapolations of $\hat{R}$ for S+25 and C+25. 

The adopted metallicity derived from the most reliable R23 line ratio  (12+log(O/H)=7.31$\pm$0.13; 0.04$\pm$0.01\,$Z_\odot$) is in agreement with the one derived from \texttt{CIGALE} (see Table~\ref{tab:properties}). This metallicity represents the lowest metallicity measured so far in galaxies at redshifts above 9. Concretely, other high-$z$ galaxies such as MACS0647, MACSJ1149-JD1, GHz2 or GNz11 present typical values 7.4\,$<$12+$\log$(O/H)\,$<$\,7.9 (see Figure~\ref{fig:Z_ratios}), derived by the $T_\mathrm{e}$-direct method. These higher metallicities could be explained by the larger stellar mass in these galaxies. In Section~\ref{Sect:metal-mass} we will discuss U26185 in the context of the mass-metallicity relation for high-$z$ galaxies.

For completeness, we consider the scenario where the ionization of the nebular emission is produced by an AGN. We use the so-called $T_\mathrm{e}$--AGN method, which considers a different $T_\mathrm{e}$[O$^{++}$]--$T_\mathrm{e}$[O$^{+}$] relation, to derive the metallicity \citep{Dors+20a,Dors+21b}. Assuming a $T_\mathrm{e}$[O$^{++}$]\,$=$\,15000\,K and considering that the higher ionized oxygen species represent $\sim$20$\%$ the total O/H abundance in AGNs \citep{Dors+20b}, we derive a metallicity of 12+log(O/H)=7.9$\pm$0.4. This value is in agreement with the CLOUDY metallicities of AGNs (see  Figure~\ref{fig:line_ratios}).

\begin{table*}
\caption{Nebular metallicity based on different line-metallicity relations}\label{tab:metallicities}
\centering
\begin{tabular}{lc|cccc}
\hline
Ratio & Value & S+25 (z=2--10) & C+25 (z=3--10)& N+22 (z$<$0.2) & N+22 med-EW  \\
 & & [7.3, 8.6] & [7.2, 8.4] & [6.9, 8.9]& [6.9, 8.9]  \\ 
\hline
log(R2) & -0.15 $\pm$ 0.10 & 7.73 $\pm$ 0.07 & 7.65 $\pm$ 0.13 & 7.66 $\pm$ 0.31 & 7.40 $\pm$ 0.17 \\
log(R3) & 0.38 $\pm$ 0.07 & 7.00 $\pm$ 0.05$^*$ & 7.01 $\pm$ 0.05$^*$ & 7.30 $\pm$ 0.07 & 7.30 $\pm$ 0.09 \\
log(R23) & 0.59 $\pm$ 0.10 & 7.04 $\pm$ 0.05$^*$ & 7.04 $\pm$ 0.09$^*$ & 7.27 $\pm$ 0.12 & 7.31 $\pm$ 0.13 \\
log(O32) & 0.65 $\pm$ 0.10 & 8.03 $\pm$ 0.08 & 8.10 $\pm$ 0.11 &  7.15 $\pm$ 0.12 & 7.64 $\pm$ 0.25 \\
log(Ne3O2) & -0.16 $\pm$ 0.15 & 7.87 $\pm$ 0.15 & 7.66 $\pm$ 0.15 & -& - \\
$\hat{R}$ & 0.26 $\pm$ 0.08 & 7.18 $\pm$ 0.11$^*$ & 7.14 $\pm$ 0.05$^*$ & - & - \\
\hline
\end{tabular}
\tablefoot{S+25, C+25 and N+22 stands for \citet{Sanders+25}, \citet{Chakraborty+25} and \citet{Nakajima+22}, respectively. Med-EW values are derived using the relation for the 100<EW(H$\beta$)[$\AA$]<200 bin in \citet{Nakajima+22}.$^*$: Values derived by extrapolating the expressions from S+25 and C+25 down to 12+log(O/H)=6.9.}
\end{table*}

\section{Discussion}\label{Sect4:disc}

\subsection{U26185: no evidence of an elusive X-ray luminous AGN}\label{Sect:ion_source}

U26185 has been identified as one of the highest redshift AGNs known so far based on just its detection as a luminous X-ray source \citep{Bogdan24-UNCOVER}. 
The detection at rest-frame 23-79 keV but not at 11-23 keV is interpreted as due to a heavy absorption of the energy source. Assuming a conservative column density of 2\,$\times$\,10$^{24}$\,cm$^{-2}$, the X-ray emission is consistent with the presence of a Compton-thick AGN characterized by an intrinsic 2-10\,keV luminosity of 1.9\,$\times$\,10$^{44}$\,erg\,s$^{-1}$.  This luminosity corresponds to a black hole mass of about 4\,$\times$\,10$^7$\,M$_{\odot}$ for a X-ray-to-bolometric luminosities ratio of 21 and an Eddington accretion rate \citep{Bogdan24-UNCOVER}. This high X-ray luminosity is similar to that of galaxy GHz9 (1.8-3.8\,$\times$\,10$^{44}$\,erg\,s$^{-1}$), a galaxy at a redshift of 10.145, recently detected by Chandra 
and also showing hybrid properties of coexisting star-formation and AGN from the JWST NIRSpec spectrum \citep{Napolitano25-GHz9, Kovacs24-GHz9}. The Compton-thick scenario predicts the existence of an obscuring material with a very large column density, and the presence of a luminous AGN. This material can be associated with dust and gas in the host galaxy and/or in a dusty torus around the accreting BH, or with the high density gas in the Broad Line Region (BLR), at distances closer to the BH and accretion disk than the torus. However, as explained below, no trace of internal dust and of a luminous AGN is present in the UV and optical spectrum of U26185 when combining the NIRSpec and MIRI LRS data. In the following, based on the combination of the NIRSpec and LRS data, we evaluate whether there is an evidence for obscuring dust and for a luminous Compton-thick AGN,  or whether the alternative scenario of a lower luminosity AGN should be further considered requiring additional X-ray data. 

If U26185 were hosting a classical type 1 AGN, its X-ray luminosity could be derived from the (extinction-corrected) H$\alpha$ luminosity assuming the well known relation with the intrinsic 2$-$10\,keV emission for low-$z$ quasars and type 1 AGNs \citep{Ho2001, Jin2012} holds at redshift 10. Following this relation, the derived 2-10 keV luminosity would be in the 8$-$11\,$\times$\,10$^{42}$\,erg\,s$^{-1}$ range, conservatively assuming the entire H$\alpha$ luminosity is due to a type 1 AGN. This X-ray luminosity is a factor of at least ten lower than the estimated from the X-ray flux under the scenario of a Compton-thick type 1 AGN. Our H$\alpha$-derived X-ray luminosity is even factors $\sim$2-6 lower than the 2-10 keV luminosity derived with the alternative lower column density solution (see discussion above and \citealt{Bogdan24-UNCOVER}). 

A standard type 2 AGN could also be invoked as the source of the X-ray emission. A majority (75\%) of low-$z$ Seyfert 2 galaxies have column densities above 10$^{23}$\,cm$^{-2}$, with half of them having densities higher than 10$^{24}$\,cm$^{-2}$ \citep{Risaliti99-xray}, similar to those derived for U26185 \citep{Bogdan24-UNCOVER}. However, although with some scatter, the median 2-10 keV-to-H$\alpha$ luminosity ratio is about 2 for Seyfert 2 galaxies \citep{Ho2001} and therefore the H$\alpha$-derived X-ray luminosity (1.2$\times$10$^{42}$\,erg\,s$^{-1}$) would be much lower (factors of 20$-$40) than the lowest 2-10 keV luminosity estimated from \textit{Chandra} (>2$-$4\,$\times$\,10$^{43}$\,erg\,s$^{-1}$, \citealt{Bogdan24-UNCOVER}). Moreover, the extension of the line-free continuum measurements up to the rest-frame 0.7\,$\mu$m with the LRS (see Table \ref{tab:fluxes} and Figure \ref{fig:SED}) does not show any indication of an increased red continuum due to hot-dust emission. That  red continuum is present in the spectrum of high-$z$ little red dots (LRDs; e.g. \citealt{Akins25}), and has been detected in GN-z11 \citep{Crespo25-GNz11}, a galaxy at redshift 10.6 where the presence of a supermassive BH has been invoked \citep{Maiolino2024_BH} while the UV-optical spectrum appears to be dominated by a nuclear starburst \citep{Bunker+23, Alvarez-Marquez_2025}. Finally, note that, there is no detection of high excitation UV/optical narrow lines typical of Seyfert 2 AGNs, thus reinforcing the idea, even if present, a typical Seyfert 2 is not contributing to the UV and optical spectrum of U26185. 

U26185 shows extremely low excitation conditions as indicated by the R3 and O32 line ratios (see Table \ref{tab:metallicities} and figure \ref{fig:line_ratios}).  The ionization parameter (logU) corresponds to a low value of $-$2.6$\pm$0.2 and $-$2.5$\pm$0.1, as derived from the SED-fitting and the O32-$\log$U relation \citep{Diaz+00}, respectively. The different optical line diagnostics place U26185 as an outlier relative to EoR ($z$\,$\sim$\,6$-$9) and other pre-EoR ($z$\,$>$\,9) sources, including star-forming galaxies, AGNs, composite systems, and (broad-line) little red dots (LRDs; see Figure \ref{fig:line_ratios}). In addition, it is worth pointing out the large difference in the optical line ratios relative to GHz9 \citep{Napolitano25-GHz9} at redshift of 10.145. U26185 presents much lower ratios (e.g., O32 and R23 ratios in Figure \ref{fig:line_ratios}) than measured in these two galaxies, identified as having a coexistent AGN and nuclear starburst. GHz9 has a similar 2-10\,keV luminosity ($\sim$1.8-3.8\,$\times$\,10$^{44}$\,erg\,s$^{-1}$) than U26185. However, GHz9 shows the presence of strong, high equivalent width, UV lines (CIV]$\lambda$$\lambda$1548,1551, OIII]$\lambda$$\lambda$1661,1666, CIII]$\lambda$$\lambda$1908) observed in AGNs, and no detected in the U26185 spectrum. Therefore, this does not support the presence of a luminous AGN in U26185 as in GHz9.

The combined UV and optical emission lines and spectral energy distribution in U26185 has no trace of the X-ray emitting AGN and appears consistent with a pure stellar origin. Hard X-ray emitting AGNs with no trace of accretion in their optical spectrum have been identified at low-redshifts \citep{Rigby2006}. In these galaxies, the obscuration in the optical of the AGN characteristics is abscribed to obscuration by dust in the host galaxy. However, there is no evidence of large amounts of dust in the host of U26185. As already mentioned in Section \ref{Sect:extinction}, deep ALMA observations have put a strong upper limit of 5.5$\times$ 10$^5$ M$_{\odot}$ to the mass of cold dust in U26185 \citep{Algera25-ALMA}. This upper limit is consistent with the very low dust attenuation (A$_{\mathrm{V-lines}}$\,$=$\,0$^{+0.4}_{-0.0}$\,mag) measured in the stellar continuum and ISM ionized gas, and also consistent with the lack of need for the DLA component to explain the Lyman break in this galaxy \citep{Goulding23-UNCOVER} . This also indicates that we are dealing with a system with a very low dust content in the host galaxy, with a dust to stellar mass ratio of less than 0.0032. However, while extremely unlikely, this amount of cold dust could still be compatible with the existence in the nucleus of U26185 of a type 2 AGN.  This dust mass could be located in the circumnuclear region (i.e. inner several-tens parsecs) or in the outer regions of a dusty torus around the black hole. The median total gas mass in the dusty torus of hard X-ray selected low-$z$ Seyfert 2 is (3.9$\pm$5.1)\,$\times$\,10$^5$\,M$_{\odot}$ with column densities (N$_H$) between 10$^{23}$ and 10$^{24}$\,cm$^{-2}$ \citep{Garcia-Bernete19-torus}. A very specific size and geometry with a large covering factor would be however required in order to block completely the UV-optical radiation coming from the accretion disk and BLR, but not the observed optical emission lines. In addition, if the dust were close to the AGN, hot dust emitting at rest-frame near-infrared wavelengths would be expected and could be searched with deep MIRI imaging at longer wavelengths (10$-$15\,$\mu$m).

Other than dust in the host galaxy and/or in a dusty torus, the high column-density Compton-thick obscuring material could be located in the dust-free BLR gas clouds as demonstrated by X-ray variability in low-redshift AGNs (e.g. \cite{Risaliti2002, Risaliti2011}). A similar conclusion has been derived from an X-ray study of a large sample of high-redshift (2<z<11) broad- and narrow-line AGNs recently detected by JWST \citep{Maiolino2025-Xray}. According to this study, absorption by the BLR gas clouds could explain the X-ray weakness of these sources relative to the standard AGNs. This scenario predicts a covering factor of the BH and accretion disk by the BLR clouds much larger than in low-redshift AGNs. The scenario is supported by an equivalent width of the H$\alpha$ line in these high-z AGNs  factor 2.6 larger than in lower redshift AGNs. However, the median H$\alpha$ equivalent width of the JWST AGNs is 570$\AA$, a factor about 2.5 smaller than the value measured for U26185 (1399$\pm$271$\AA$). Therefore, if the obscuration of the AGN in U26185 were due to the BLR clouds, U26185 would represent an extreme case of this scenario since only about 
2$\%$ of the JWST AGNs have equivalent widths above 1000$\AA$ \citep{Maiolino2025-Xray}. 

In summary, all the evidence gathered so far with all the available NIRSpec and MIRI LRS spectroscopy, combined with the upper limit to the mass of cold dust in U26185, supports the idea that this galaxy is a dust-free, star-forming galaxy, not showing any evidence in its UV and optical spectrum for the presence of a Compton-thick X-ray luminous. However, a non standard,  extremely young, deeply embedded, dust-poor accretion phase can not be fully excluded without further exploration.
Additional data with X-ray satellites and JWST would be required to further investigate the nature of this intriguing source. Deep NIRSpec R1000 spectra would be required to search for the presence of the coronal [NeV]3426 line. High spectral resolution with MIRI MRS could help to set stronger limits on the presence of a broad H$\alpha$ emission line associated with a type 1 AGN. In addition, deep imaging at longer wavelengths with MIRI Imager would extend the wavelength coverage into the rest-frame near-infrared range, searching for the presence of hot dust associated with an obscured AGN like recently identified in the high-$z$ source {\it Virgil} \citep{Rinaldi2025-Virgil}. 

The low significance (4.2$\sigma$) of the  X-ray detection in U26185 is also consistent with a source with a lower 2-10 keV luminosity ($>$ 2-4 $\times$ 10$^{43}$ erg s$^{-1}$) and lower column density  ($>$ 1e+22 cm$^{-2}$; \citealt{Bogdan24-UNCOVER}). Taking the AGN scenario aside, even this hard X-ray luminosity is orders of magnitude higher than the expected value due to high-mass binary stars for a SFR of 1.3\,M$_{\odot}$\,yr$^{-1}$ with a standard IMF (see X-ray-to-SFR relations in, e.g., \citealt{Grimm2003-SFR-Xray}; \citealt{Mineo2014-SFR-Xray}), and to the predictions of population III X-ray binaries \citep{Sartorio2023-SFR-Xray}. Thus, the star formation that dominates the UV and optical emission is not able to explain even the lower X-ray luminosity scenario. In addition to JWST data at longer wavelengths, deeper X-ray observations would be required to confirm the detection of the enigmatic X-ray source with a significance higher than the actual detection. U26185 is just 1 arcmin apart from the center of the galaxy cluster and the detection could be affected by the fluctuation of the X-ray background signal from the intracluster medium.

\subsection{U26185: A metal-poor star-forming galaxy}
\label{Sect:metal-mass}

Based on the results presented in the previous section, we rule out the presence of a classical Type 1 AGN and an obscured Seyfert 2 galaxy, as well as a LRD, challenging the presence of the supermassive active black hole (M$_{\mathrm{BH}}$\,$\sim$\,10$^{7-8}$\,M$_{\odot}$) in the nucleus of U26185, as previously inferred by \cite{Goulding23-UNCOVER}. Therefore, U26185 exhibits the UV-to-optical spectral properties consistent with a purely star-forming galaxy. In the following, we discuss U26185 in the context of the stellar mass to metallicity relation in high-$z$ galaxies.

U26185 is the faintest UV known star-forming galaxy (M$_{\mathrm{UV}}$\,$=$\,$-$18.8\,mag) observed with MIRI spectroscopy at $z$\,$\sim$\,10, enabled by its relatively high gravitational magnification factor ($\mu$\,$=$\,4.07$^{+0.03}_{-0.11}$). Emission-line diagnostics and SED-fitting analyses indicate intrinsic SFRs of 1.3\,$\pm$\,0.2\,M$_{\odot}$\,yr$^{-1}$ (1.0\,M$_{\odot}$\,yr$^{-1}$) derived from H$\alpha$ (UV continuum) corresponding to star-formation timescales of 10\,Myr (100\,Myr). The inferred stellar mass is (1.7\,$\pm$\,0.3)\,$\times$\,10$^{8}$\,$\mathrm{M}_{\odot}$, yielding a specific star-formation rate of 7.6\,$\pm$\,1.2\,Gyr$^{-1}$ (5.9\,Gyr$^{-1}$). The SFR and stellar mass estimates identifies U26185 as a normal star-forming galaxy, placing it within a factor of less than two above the empirical main sequence of star-forming galaxies at z = 7--9 \citep{Merida2025}, and in agreement with predictions from cosmological simulations and semi-analytic models at $z$\,$\sim$\,10 (e.g., \citealt{Kannan2023,Kravtsov2024,Cantarella2025}). U26185 therefore appears to be a representative, normal star-forming galaxy at $z$\,$\sim$\,10, offering an opportunity to probe the typical galaxy population at this epoch, rather than focusing exclusively on extreme, UV-bright systems such as GN-z11 that occupy the bright end of the UV luminosity function (e.g., \citealt{Alvarez-Marquez_2025}).

\begin{figure}
\centering
   \includegraphics[width=\linewidth]{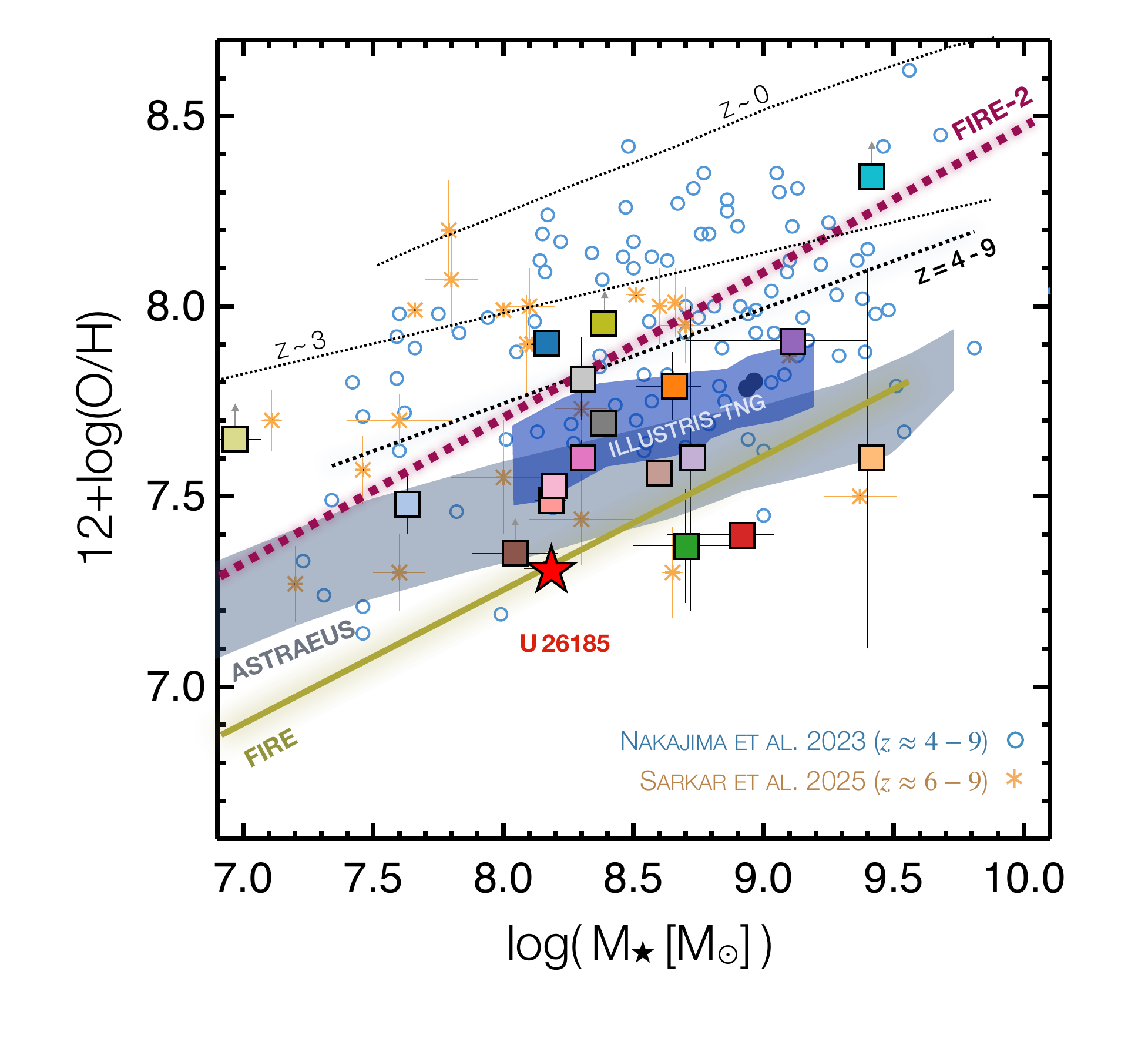}
      \caption{The stellar mass-metallicity plane. The dotted black lines represent the observationally inferred relationships at $z$\,$\approx$\,0.08 \citep{curti+20}, $z$\,$\approx$\,3 \citep{Li2023_MZR}, and $z\approx4-9$ \citep{Nakajima+23, Curti+24, Sarkar+25}. Squares represent $z$\,$>$\,9 galaxies with NIRSpec and/or MIRI detections (color coded following Figure \ref{fig:line_ratios}). The red star is, U\,26185, as constrained in this work. For comparison, we also include $z\sim9-10$ predictions from simulations as follows: FIRE \citep{Ma2016}, ILLUSTRIS-TNG \citep{Torrey2019}, ASTRAEUS \citep{Ucci2023}, FIRE-2 \citep{Marszewski2024}.}
         \label{fig:mass_Z_relation}
\end{figure}

Figure \ref{fig:mass_Z_relation} shows the position of U26185 in the stellar mass-metallicity plane, compared with galaxies at $z$\,$\approx$\,4$-$9 and with the known $z>9$ systems with metallicity measurements. It also includes empirical stellar mass-metallicity relations (MZR) at $z$\,$<$\,9, as well as MZRs predicted by cosmological simulations at $z$\,$\approx$\,9-10. This figure shows that approximately half of the $z>9$ systems fall within the scatter of lower-redshift samples,  in agreement with the weak (or lack of ) redshift evolution of the MZR predicted by simulations within $z\approx5-10$ (e.g. \citealt{Ucci2023,Wilkins2023,Marszewski2024}). Nevertheless, the remaining galaxies are systematically offset toward lower metallicities, potentially indicating more primeval (metal-poor) galaxies or a larger scatter in the MZR.
Actually, U26185 lies among the most metal-poor galaxies observed at $z$\,$\gtrsim$\,10 to date. Its low metallicity contrasts sharply with that of UV-bright systems such as GN-z11, GS-z14-0, and MACS0647-JD at $z$\,$>$\,10, which exhibit also significantly higher stellar masses. This difference may indicate a more advanced chemical enrichment history in these luminous systems, consistent with a more evolved nature, compared to a more representative population of normal star-forming galaxies exemplified by U26185.

Compared with simulation predictions, U26185 is in agreement with MZRs from FIRE (\citealt{Ma2016}) and ASTRAEUS (\citealt{Ucci2023}) simulations, but lies below (with a lower metallicity) of the predictions from ILLUSTRIS-TNG (\citealt{Torrey2019}) and FIRE-2 (\citealt{Marszewski2024}). It is worth noting, however, that the simulations predict a high scatter at relatively low stellar masses ($M_\star\lesssim10^{8.5}\,M_\odot$; \citealt{Marszewski2024}), which could explain the differences between the inferred metallicity and the simulations' predictions. Larger sample of galaxies at $z\gtrsim10$ with different stellar masses and precise metallicity measurements would be needed to put strong constraints on the redshift evolution of the MZR.

U26185 seems thus to be representative of a population of relatively low-mass ($M_\star\sim10^{8}\,M_\odot$) galaxies and, while it seems to lie on the expected main-sequence of star-forming galaxies, its low metallicity (being one of the most metal-poor galaxy identified at $z>10$), indicates that it might be in an earlier evolutionary phase than other brighter $z\gtrsim10$ systems. It represents one of the first cases in which JWST has constrained the rest-frame optical spectrum of this otherwise hidden population of faint galaxies, made accessible by gravitational lensing despite their intrinsically low luminosities. The PRISMS program will increase the number of these faint sources and provide a valuable dataset to test the redshift evolution of the MZR in the earliest epochs of galaxy formation.

\section{Summary and conclusions}\label{Sect:conclusion_Summary}

This work presents the first results of the JWST cycle 4 PRISMS (PRImordial galaxy Survey with MIRI Spectroscopy; ID\,8051) program, aimed at the characterization of the rest-frame optical spectra of a large sample of ten sources representative of intermediate UV-luminosity galaxies  (-20.5\,$<$\,M$_{\mathrm{UV}}$\,[mag]\,$<$\,-17.6)  at redshifts of about 10, i.e. at 500\,Myr after the Big Bang.  

We report deep (13.9\,hours) MIRI Low-Resolution Spectroscopy of the gravitationally lensed ($\mu$\,$=$\,4.07) galaxy U26185 at a redshift of $z$\,$=$\,10.054\,$\pm$\,0.011. It is a faint UV galaxy (M$_{\mathrm{UV}}$\,$=$\,-18.83$\pm$0.07\,mag) previously identified as a X-ray luminous AGN. We probe its rest-frame optical spectrum, detecting the H$\beta$+[O\,III]$\lambda\lambda$4960,5008$\AA$ complex and H$\alpha$ emission-line with signal-to-noise ratios of 10 and 8, respectively. In addition, we detect continuum emission at rest-frame wavelengths of 0.45 and 0.57\,$\mu$m with signal-to-noise ratios of 6–7 when stacking multiple spectral elements, and we place strong upper-limits on the continuum at 0.7\,$\mu$m. 

The combination of the MIRI LRS spectrum with previously reported NIRCam photometry, NIRSpec spectroscopy, and ALMA [OIII]88\,$\mu$m line and 90\,$\mu$m continuum observations allows us to investigate the star-formation history, the physical properties of the interstellar medium, and the nature of the ionization source. The main results are summarized below: 

\begin{itemize}

\item U26185 is characterized by a star-formation history composed of two distinct stellar populations. The more mature component has a mass-weighted age of 65\,$\pm$\,20\,Myr and a stellar mass of (1.5\,$\pm$\,0.3)\,$\times$\,10$^{8}$\,M$_{\odot}$, indicating that the majority of the galaxy’s stars formed around redshift $z\sim11$. The young stellar population contributes $\sim$10\% of the total stellar mass, with a mass of (1.7\,$\pm$\,0.4)\,$\times$\,10$^7$\,M$_{\odot}$ and an age of 7\,$\pm$\,3\,Myr. The ongoing burst is characterized by a star-formation rate of 1.3\,$\pm$\,0.2\,M$_{\odot}$\,yr$^{-1}$ and a specific star-formation of 7.6\,$\pm$\,1.2\,Gyr$^{-1}$. The star-formation rate and total stellar mass locate U26185 in the main sequence of star-forming galaxies at $z$\,$\sim$\,10, as predicted by simulations. 

\item The ionizing photon production efficiency is $\log$($\xi_\mathrm{ion}$)\,=\,25.50\,$\pm$\,0.06\,Hz\,erg$^{-1}$, in agreement with sources identified in the Epoch of Reionization and beyond. U26185 is also characterized by a very high H$\alpha$ equivalent width, EW(H$\alpha$)\,$=$\,1399\,$\pm$\,271\AA, compatible with the presence of stellar population ages of $\sim$4–6\,Myr, for an instantaneous burst. Both, high ionizing photon production efficiency and EW(H$\alpha$), reinforce the presence of a massive ($\sim$\,10$^{7}$\,M$_{\odot}$) star-forming burst with ages at their maximum photon production efficiency and ionization.  

\item U26185 has negligible or zero dust attenuation (A$_{\mathrm{V-lines}}$\,$=$\,0$^{+0.4}_{-0.0}$\,mag). Consistently, SED-fitting analysis yield stellar\,$+$\,nebular dust attenuations compatible with the Balmer decrements, with a best value of A$_{\mathrm{V-SED}}$\,$=$\,0.20\,$\pm$\,0.09\,mag.

\item The available emission line ratios (R2, R3, R23, O32, Ne2O2, and $\hat{R}$) locate U26185 as the lowest excitation ($\log$(U)\,$=$\,$-$2.5\,$\pm$\,0.1) source, as well as the most metal-poor (Z\,$=$\,0.04\,$\pm$\,0.02\,Z$_{\odot}$) galaxy, identified so far at redshifts above 9. 

\item The emission line ratios and UV-to-optical continuum, traced by the new MIRI LRS and the ancillary NIRSpec spectra, are consistent with a stellar origin. The extremely low excitation/ionization traced by the line ratios and the high equivalent width of the H$\alpha$ line do not show any evidence for the presence of the X-ray luminous AGN previously identified in this galaxy. This, together with the measured low nebular and stellar extinction, consistent with zero, and the low upper limit to the cold dust mass measured by ALMA observations, does not support the presence of a  massive, Compton-thick, standard active black hole, either obscured by a dusty torus, or by the dust-free BLR gas clouds. Additional MIRI data are needed to further prove or discard the presence of a luminous Compton-thick AGN in this galaxy. Deep MIRI high resolution spectroscopy is required to search for a faint BLR traced by a broad H$\alpha$ emission line, and deep MIRI imaging at wavelengths above 10 $\mu$m would trace hot dust emission in the (circum)nuclear regions around an AGN. In addition, deeper X-ray imaging is required to increase the significance of the X-ray detection (currently at a significance of 4.2$\sigma$) and therefore confirm, or discard, the presence of the X-ray luminous AGN.

\item The metallicity and stellar mass of U26185 is consistent with the stellar mass$-$metallicty predictions from some cosmological simulations (FIRE, ASTRAEUS) while clearly deviating from others (ILLUSTRIS-TNG and FIRE-2). U26185 lies among the most metal-poor galaxies known at $z$\,$>$\,9. Their metallicity and stellar mass are significantly lower than those of UV-bright systems such as GN-z11 and GS-z14-0, therefore, probing a different (unexplored) regime in the stellar mass$-$metallicty relation at redshift of about 10. 

\end{itemize}

Conclusively, U26185 seems thus to be representative of a population of relatively low-mass ($M_\star\sim10^{8}\,M_\odot$), normal star-forming galaxies and, while it seems to lie on the expected main-sequence of star-forming galaxies, its low metallicity indicates that it might be in an earlier evolutionary phase than other UV-bright systems at $z$\,$\gtrsim$\,10. It represents one of the first cases in which JWST has constrained the rest-frame optical spectrum of this otherwise hidden population of faint galaxies, made accessible by gravitational lensing despite their intrinsically low-luminosities. The PRISMS program will increase the number of these faint sources and provide a valuable dataset to investigate the stellar populations, physical properties of the interstellar medium, and nature of the ionized source of primordial galaxies, just 500\,Myr after the Big Bang.

\begin{acknowledgements}

The authors would like to thank Almudena Alonso-Herrero and Ismael Garc\'ia-Bernete for discussions about AGNs and mechanisms for obscuring their radiation. J.A.-M., C.P.-J., C.B.-P., B.R.P. acknowledge support by grant PID2024-158856NA-I00, J.A.-M., L.C., C.P.-J., C.B.-P., B.R.P. acknowledge support by grant PIB2021-127718NB-100, P.G.P.-G. acknowledges support from grant PID2022-139567NB-I00 from the Spanish Ministry of Science and Innovation/State Agency of Research MCIN/AEI/10.13039/501100011033 and by “ERDF A way of making Europe”. J.A.-M., L.C., C.P.-J., C.B.-P., B.R.P., P.G.P.-G. acknowledge support by grant CSIC/BILATERALES2025/BIJSP25022. L.A.B. acknowledges support from the Dutch Research Council (NWO) under grant VI.Veni.242.055 (\url{https://doi.org/10.61686/LAJVP77714}). M.C. acknowledges INAF GO Grant 2024 "Revealing the nature of bright galaxies at cosmic dawn with deep JWST spectroscopy". T.H. was supported by JSPS KAKENHI 25K00020. Y.H. acknowledges support from the Japan Society for the Promotion of Science (JSPS) Grant-in-Aid for Scientific Research (24H00245), the JSPS Core-to-Core Program (JPJSCCA20210003), and the JSPS International Leading Research (22K21349). Y.F. is supported by JSPS KAKENHI Grant Numbers JP22K21349 and JP23K13149. D.L. was supported by research grants (VIL16599,VIL54489) from VILLUM FONDEN.

The data were obtained from the Mikulski Archive for Space Telescopes at the Space Telescope Science Institute, which is operated by the Association of Universities for Research in Astronomy, Inc., under NASA contract NAS 5-03127 for \textit{JWST}; and from the \href{https://jwst.esac.esa.int/archive/}{European \textit{JWST} archive (e\textit{JWST})} operated by the ESDC. 

This research made use of Photutils, an Astropy package for detection and photometry of astronomical sources \citep{larry_bradley_2022_6825092}.

\end{acknowledgements}

\bibliographystyle{aa} 
\bibliography{bibliography.bib} 

@ARTICLE{Sartorio2023-SFR-Xray,
       author = {{Sartorio}, Nina S. and {Fialkov}, A. and {Hartwig}, T. and {Mirouh}, G.~M. and {Izzard}, R.~G. and {Magg}, M. and {Klessen}, R.~S. and {Glover}, S.~C.~O. and {Chen}, L. and {Tarumi}, Y. and {Hendriks}, D.~D.},
        title = "{Population III X-ray binaries and their impact on the early universe}",
      journal = {\mnras},
         year = 2023,
        month = may,
       volume = {521},
       number = {3},
        pages = {4039-4055},
          doi = {10.1093/mnras/stad697},
archivePrefix = {arXiv},
       eprint = {2303.03435},
 primaryClass = {astro-ph.HE},
       adsurl = {https://ui.adsabs.harvard.edu/abs/2023MNRAS.521.4039S}
}

@ARTICLE{Grimm2003-SFR-Xray,
       author = {{Grimm}, H.-J. and {Gilfanov}, M. and {Sunyaev}, R.},
        title = "{High-mass X-ray binaries as a star formation rate indicator in distant galaxies}",
      journal = {\mnras},
         year = 2003,
        month = mar,
       volume = {339},
       number = {3},
        pages = {793-809},
          doi = {10.1046/j.1365-8711.2003.06224.x},
archivePrefix = {arXiv},
       eprint = {astro-ph/0205371},
 primaryClass = {astro-ph},
       adsurl = {https://ui.adsabs.harvard.edu/abs/2003MNRAS.339..793G}
}

@ARTICLE{Mineo2014-SFR-Xray,
       author = {{Mineo}, S. and {Gilfanov}, M. and {Lehmer}, B.~D. and {Morrison}, G.~E. and {Sunyaev}, R.},
        title = "{X-ray emission from star-forming galaxies - III. Calibration of the L$_{X}$-SFR relation up to redshift z {\ensuremath{\approx}} 1.3}",
      journal = {\mnras},
         year = 2014,
        month = jan,
       volume = {437},
       number = {2},
        pages = {1698-1707},
          doi = {10.1093/mnras/stt1999},
archivePrefix = {arXiv},
       eprint = {1207.2157},
 primaryClass = {astro-ph.HE},
       adsurl = {https://ui.adsabs.harvard.edu/abs/2014MNRAS.437.1698M}
}

@ARTICLE{Balogh1999,
       author = {{Balogh}, Michael L. and {Morris}, Simon L. and {Yee}, H.~K.~C. and {Carlberg}, R.~G. and {Ellingson}, Erica},
        title = "{Differential Galaxy Evolution in Cluster and Field Galaxies at z\raisebox{-0.5ex}\textasciitilde0.3}",
      journal = {\apj},
         year = 1999,
        month = dec,
       volume = {527},
       number = {1},
        pages = {54-79},
          doi = {10.1086/308056},
archivePrefix = {arXiv},
       eprint = {astro-ph/9906470},
 primaryClass = {astro-ph},
       adsurl = {https://ui.adsabs.harvard.edu/abs/1999ApJ...527...54B}
}

@ARTICLE{Ciesla2024,
       author = {{Ciesla}, L. and {Elbaz}, D. and {Ilbert}, O. and {Buat}, V. and {Magnelli}, B. and {Narayanan}, D. and {Daddi}, E. and {G{\'o}mez-Guijarro}, C. and {Arango-Toro}, R.},
        title = "{Identification of a transition from stochastic to secular star formation around z = 9 with JWST}",
      journal = {\aap},
         year = 2024,
        month = jun,
       volume = {686},
          eid = {A128},
        pages = {A128},
          doi = {10.1051/0004-6361/202348091},
archivePrefix = {arXiv},
       eprint = {2309.15720},
 primaryClass = {astro-ph.GA},
       adsurl = {https://ui.adsabs.harvard.edu/abs/2024A&A...686A.128C}
}

@ARTICLE{Hutter2025,
       author = {{Hutter}, Anne and {Cueto}, Elie R. and {Dayal}, Pratika and {Gottl{\"o}ber}, Stefan and {Trebitsch}, Maxime and {Yepes}, Gustavo},
        title = "{ASTRAEUS: X. Indications of a top-heavy initial mass function in highly star-forming galaxies from JWST observations at z > 10}",
      journal = {\aap},
         year = 2025,
        month = feb,
       volume = {694},
          eid = {A254},
        pages = {A254},
          doi = {10.1051/0004-6361/202452460},
archivePrefix = {arXiv},
       eprint = {2410.00730},
 primaryClass = {astro-ph.GA},
       adsurl = {https://ui.adsabs.harvard.edu/abs/2025A&A...694A.254H}
}

@ARTICLE{Tang2025_z9-14,
       author = {{Tang}, Mengtao and {Stark}, Daniel P. and {Mason}, Charlotte A. and {Gelli}, Viola and {Chen}, Zuyi and {Topping}, Michael W.},
        title = "{The JWST Spectroscopic Properties of Galaxies at $z=9-14$}",
      journal = {arXiv e-prints},
         year = 2025,
        month = jul,
          eid = {arXiv:2507.08245},
        pages = {arXiv:2507.08245},
          doi = {10.48550/arXiv.2507.08245},
archivePrefix = {arXiv},
       eprint = {2507.08245},
 primaryClass = {astro-ph.GA},
       adsurl = {https://ui.adsabs.harvard.edu/abs/2025arXiv250708245T}
}

@ARTICLE{Witstok2025Natur_z13,
       author = {{Witstok}, Joris and {Jakobsen}, Peter and {Maiolino}, Roberto and {Helton}, Jakob M. and {Johnson}, Benjamin D. and {Robertson}, Brant E. and {Tacchella}, Sandro and {Cameron}, Alex J. and {Smit}, Renske and {Bunker}, Andrew J. and {Saxena}, Aayush and {Sun}, Fengwu and {Alberts}, Stacey and {Arribas}, Santiago and {Baker}, William M. and {Bhatawdekar}, Rachana and {Boyett}, Kristan and {Cargile}, Phillip A. and {Carniani}, Stefano and {Charlot}, St{\'e}phane and {Chevallard}, Jacopo and {Curti}, Mirko and {Curtis-Lake}, Emma and {D'Eugenio}, Francesco and {Eisenstein}, Daniel J. and {Hainline}, Kevin N. and {Jones}, Gareth C. and {Kumari}, Nimisha and {Maseda}, Michael V. and {P{\'e}rez-Gonz{\'a}lez}, Pablo G. and {Rinaldi}, Pierluigi and {Scholtz}, Jan and {{\"U}bler}, Hannah and {Williams}, Christina C. and {Willmer}, Christopher N.~A. and {Willott}, Chris and {Zhu}, Yongda},
        title = "{Witnessing the onset of reionization through Lyman-{\ensuremath{\alpha}} emission at redshift 13}",
      journal = {\nat},
         year = 2025,
        month = mar,
       volume = {639},
       number = {8056},
        pages = {897-901},
          doi = {10.1038/s41586-025-08779-5},
archivePrefix = {arXiv},
       eprint = {2408.16608},
 primaryClass = {astro-ph.GA},
       adsurl = {https://ui.adsabs.harvard.edu/abs/2025Natur.639..897W}
}

@ARTICLE{Roberts-Borsani+23Natur,
       author = {{Roberts-Borsani}, Guido and {Treu}, Tommaso and {Chen}, Wenlei and {Morishita}, Takahiro and {Vanzella}, Eros and {Zitrin}, Adi and {Bergamini}, Pietro and {Castellano}, Marco and {Fontana}, Adriano and {Glazebrook}, Karl and {Grillo}, Claudio and {Kelly}, Patrick L. and {Merlin}, Emiliano and {Nanayakkara}, Themiya and {Paris}, Diego and {Rosati}, Piero and {Yang}, Lilan and {Acebron}, Ana and {Bonchi}, Andrea and {Boyett}, Kit and {Brada{\v{c}}}, Maru{\v{s}}a and {Brammer}, Gabriel and {Broadhurst}, Tom and {Calabr{\'o}}, Antonello and {Diego}, Jose M. and {Dressler}, Alan and {Furtak}, Lukas J. and {Filippenko}, Alexei V. and {Henry}, Alaina and {Koekemoer}, Anton M. and {Leethochawalit}, Nicha and {Malkan}, Matthew A. and {Mason}, Charlotte and {Mercurio}, Amata and {Metha}, Benjamin and {Pentericci}, Laura and {Pierel}, Justin and {Rieck}, Steven and {Roy}, Namrata and {Santini}, Paola and {Strait}, Victoria and {Strausbaugh}, Robert and {Trenti}, Michele and {Vulcani}, Benedetta and {Wang}, Lifan and {Wang}, Xin and {Windhorst}, Rogier A.},
        title = "{The nature of an ultra-faint galaxy in the cosmic dark ages seen with JWST}",
      journal = {\nat},
         year = 2023,
        month = jun,
       volume = {618},
       number = {7965},
        pages = {480-483},
          doi = {10.1038/s41586-023-05994-w},
archivePrefix = {arXiv},
       eprint = {2210.15639},
 primaryClass = {astro-ph.GA},
       adsurl = {https://ui.adsabs.harvard.edu/abs/2023Natur.618..480R}
}

@ARTICLE{Castellano2025_GHz2,
       author = {{Castellano}, M. and {Napolitano}, L. and {Moreschini}, B. and {Calabr{\`o}}, A. and {Christensen}, L. and {Llerena}, M. and {Bakx}, T.~J.~L.~C. and {Belfiore}, F. and {Bevacqua}, D. and {Dickinson}, M. and {Fontana}, A. and {Gandolfi}, G. and {Gasparetto}, T. and {Marconi}, A. and {Mascia}, S. and {Merlin}, E. and {Morishita}, T. and {Nanayakkara}, T. and {Paris}, D. and {Pentericci}, L. and {P{\'e}rez-D{\'\i}az}, B. and {Roberts-Borsani}, G. and {Rojas Ruiz}, S. and {Santini}, P. and {Treu}, T. and {Vanzella}, E. and {Vulcani}, B. and {Wang}, X. and {Yoon}, I. and {Zavala}, J.},
        title = "{Investigating ionising sources and the complex interstellar medium of GHZ2 at $z=12.3$}",
      journal = {arXiv e-prints},
         year = 2025,
        month = dec,
          eid = {arXiv:2512.08490},
        pages = {arXiv:2512.08490},
          doi = {10.48550/arXiv.2512.08490},
archivePrefix = {arXiv},
       eprint = {2512.08490},
 primaryClass = {astro-ph.GA},
       adsurl = {https://ui.adsabs.harvard.edu/abs/2025arXiv251208490C}
}

@ARTICLE{Castellano2025,
       author = {{Castellano}, M. and {Fontana}, A. and {Merlin}, E. and {Santini}, P. and {Napolitano}, L. and {Menci}, N. and {P{\'e}rez-Gonz{\'a}lez}, P.~G. and {Calabr{\`o}}, A. and {Paris}, D. and {Pentericci}, L. and {Zavala}, J.~A. and {Dickinson}, M. and {Finkelstein}, S.~L. and {Treu}, T. and {Amorin}, R.~O. and {Arrabal Haro}, P. and {Bergamini}, P. and {Bisigello}, L. and {Catone}, M. and {Daddi}, E. and {Dayal}, P. and {Dekel}, A. and {Ferrara}, A. and {Fortuni}, F. and {Gandolfi}, G. and {Giavalisco}, M. and {Grillo}, C. and {Guida}, S.~T. and {Hathi}, N.~P. and {Holwerda}, B.~W. and {Koekemoer}, A.~M. and {Kokorev}, V. and {Li}, Z. and {Llerena}, M. and {Lucas}, R.~A. and {Mascia}, S. and {Metha}, B. and {Morishita}, T. and {Nanayakkara}, T. and {Pacucci}, F. and {Roberts-Borsani}, G. and {Rodighiero}, G. and {Rosati}, P. and {Salazar}, V. and {Schneider}, R. and {Somerville}, R.~S. and {Taylor}, A. and {Trenti}, M. and {Trinca}, A. and {Wang}, X. and {Watson}, P.~J. and {Yang}, L. and {Yung}, L.~Y.~A.},
        title = "{Pushing JWST to the extremes: Search and scrutiny of bright galaxy candidates at z ≃ 15─30}",
      journal = {\aap},
         year = 2025,
        month = dec,
       volume = {704},
          eid = {A158},
        pages = {A158},
          doi = {10.1051/0004-6361/202555082},
archivePrefix = {arXiv},
       eprint = {2504.05893},
 primaryClass = {astro-ph.GA},
       adsurl = {https://ui.adsabs.harvard.edu/abs/2025A&A...704A.158C}
}

@ARTICLE{Fudamoto2020,
       author = {{Fudamoto}, Y. and {Oesch}, P.~A. and {Faisst}, A. and {B{\'e}thermin}, M. and {Ginolfi}, M. and {Khusanova}, Y. and {Loiacono}, F. and {Le F{\`e}vre}, O. and {Capak}, P. and {Schaerer}, D. and {Silverman}, J.~D. and {Cassata}, P. and {Yan}, L. and {Amorin}, R. and {Bardelli}, S. and {Boquien}, M. and {Cimatti}, A. and {Dessauges-Zavadsky}, M. and {Fujimoto}, S. and {Gruppioni}, C. and {Hathi}, N.~P. and {Ibar}, E. and {Jones}, G.~C. and {Koekemoer}, A.~M. and {Lagache}, G. and {Lemaux}, B.~C. and {Maiolino}, R. and {Narayanan}, D. and {Pozzi}, F. and {Riechers}, D.~A. and {Rodighiero}, G. and {Talia}, M. and {Toft}, S. and {Vallini}, L. and {Vergani}, D. and {Zamorani}, G. and {Zucca}, E.},
        title = "{The ALPINE-ALMA [CII] survey. Dust attenuation properties and obscured star formation at z {\ensuremath{\sim}} 4.4-5.8}",
      journal = {\aap},
         year = 2020,
        month = nov,
       volume = {643},
          eid = {A4},
        pages = {A4},
          doi = {10.1051/0004-6361/202038163},
archivePrefix = {arXiv},
       eprint = {2004.10760},
 primaryClass = {astro-ph.GA},
       adsurl = {https://ui.adsabs.harvard.edu/abs/2020A&A...643A...4F}
}

@ARTICLE{Storey2000,
       author = {{Storey}, P.~J. and {Zeippen}, C.~J.},
        title = "{Theoretical values for the [OIII] 5007/4959 line-intensity ratio and homologous cases}",
      journal = {\mnras},
         year = 2000,
        month = mar,
       volume = {312},
       number = {4},
        pages = {813-816},
          doi = {10.1046/j.1365-8711.2000.03184.x},
       adsurl = {https://ui.adsabs.harvard.edu/abs/2000MNRAS.312..813S}
}

@ARTICLE{Burgarella2025,
       author = {{Burgarella}, Denis and {Buat}, V{\'e}ronique and {Theul{\'e}}, Patrice and {Zavala}, Jorge and {Dickinson}, Mark and {Arrabal Haro}, Pablo and {Bagley}, Micaela B. and {Boquien}, M{\'e}d{\'e}ric and {Cleri}, Nikko and {Dewachter}, Tim and {Ferguson}, Henry C. and {Fern{\`a}ndez}, Vital and {Finkelstein}, Steven L. and {Gawiser}, Eric and {Grazian}, Andrea and {Grogin}, Norman and {Holwerda}, Benne W. and {Kartaltepe}, Jeyhan S. and {Kewley}, Lisa and {Kirkpatrick}, Allison and {Kocevski}, Dale and {Koekemoer}, Anton M. and {Long}, Arianna and {Lotz}, Jennifer and {Lucas}, Ray A. and {Mobasher}, Bahram and {Papovich}, Casey and {P{\'e}rez-Gonz{\`a}lez}, Pablo G. and {Pirzkal}, Nor and {Ravindranath}, Swara and {Rodighiero}, Giulia and {Roehlly}, Yannick and {Rose}, Caitlin and {Seill{\'e}}, Lise-Marie and {Somerville}, Rachel and {Wilkins}, Steve and {Yang}, Guang and {Yung}, L.~Y. Aaron},
        title = "{CEERS: Possibly forging the first dust grains in the universe: A population of galaxies with spectroscopically derived extremely low dust attenuation (GELDA) at 4.0 < z {\ensuremath{\lesssim}} 11.4}",
      journal = {\aap},
         year = 2025,
        month = jul,
       volume = {699},
          eid = {A336},
        pages = {A336},
          doi = {10.1051/0004-6361/202554231},
archivePrefix = {arXiv},
       eprint = {2504.13118},
 primaryClass = {astro-ph.GA},
       adsurl = {https://ui.adsabs.harvard.edu/abs/2025A&A...699A.336B}
}

@ARTICLE{Alvarez-Marquez_2025,
       author = {{{\'A}lvarez-M{\'a}rquez}, J. and {Crespo G{\'o}mez}, A. and {Colina}, L. and {Langeroodi}, D. and {Marques-Chaves}, R. and {Prieto-Jim{\'e}nez}, C. and {Bik}, A. and {Alonso-Herrero}, A. and {Boogaard}, L. and {Costantin}, L. and {Garc{\'\i}a-Mar{\'\i}n}, M. and {Gillman}, S. and {Hjorth}, J. and {Iani}, E. and {Jermann}, I. and {Labiano}, A. and {Melinder}, J. and {Meyer}, R. and {{\"O}stlin}, G. and {P{\'e}rez-Gonz{\'a}lez}, P.~G. and {Rinaldi}, P. and {Walter}, F. and {van der Werf}, P. and {Wright}, G.},
        title = "{Insight into the starburst nature of Galaxy GN-z11 with JWST MIRI spectroscopy}",
      journal = {\aap},
         year = 2025,
        month = mar,
       volume = {695},
          eid = {A250},
        pages = {A250},
          doi = {10.1051/0004-6361/202451731},
archivePrefix = {arXiv},
       eprint = {2412.12826},
 primaryClass = {astro-ph.GA},
       adsurl = {https://ui.adsabs.harvard.edu/abs/2025A&A...695A.250A}
}

@ARTICLE{Rinaldi2025-Virgil,
       author = {{Rinaldi}, Pierluigi and {P{\'e}rez-Gonz{\'a}lez}, Pablo G. and {Rieke}, George H. and {Lyu}, Jianwei and {D'Eugenio}, Francesco and {Wu}, Zihao and {Carniani}, Stefano and {Looser}, Tobias J. and {Shivaei}, Irene and {Boogaard}, Leindert A. and {Diaz-Santos}, Tanio and {Colina}, Luis and {{\"O}stlin}, G{\"o}ran and {Alberts}, Stacey and {{\'A}lvarez-M{\'a}rquez}, Javier and {Annuziatella}, Marianna and {Aravena}, Manuel and {Bhatawdekar}, Rachana and {Bunker}, Andrew J. and {Caputi}, Karina I. and {Charlot}, St{\'e}phane and {Crespo G{\'o}mez}, Alejandro and {Curti}, Mirko and {Eckart}, Andreas and {Gillman}, Steven and {Hainline}, Kevin and {Kumari}, Nimisha and {Hjorth}, Jens and {Iani}, Edoardo and {Inami}, Hanae and {Ji}, Zhiyuan and {Johnson}, Benjamin D. and {Jones}, Gareth C. and {Labiano}, {\'A}lvaro and {Maiolino}, Roberto and {Melinder}, Jens and {Moutard}, Thibaud and {Peissker}, Florian and {Rieke}, Marcia and {Robertson}, Brant and {Scholtz}, Jan and {Tacchella}, Sandro and {van der Werf}, Paul P. and {Walter}, Fabian and {Williams}, Christina C. and {Willott}, Chris and {Witstok}, Joris and {{\"U}bler}, Hannah and {Zhu}, Yongda},
        title = "{Deciphering the Nature of Virgil: An Obscured Active Galactic Nucleus Lurking within an Apparently Normal Ly{\ensuremath{\alpha}} Emitter during Cosmic Reionization}",
      journal = {\apj},
         year = 2025,
        month = nov,
       volume = {994},
       number = {1},
          eid = {86},
        pages = {86},
          doi = {10.3847/1538-4357/ae089c},
archivePrefix = {arXiv},
       eprint = {2504.01852},
 primaryClass = {astro-ph.GA},
       adsurl = {https://ui.adsabs.harvard.edu/abs/2025ApJ...994...86R}
}

@ARTICLE{Maiolino2025-Xray,
       author = {{Maiolino}, Roberto and {Risaliti}, Guido and {Signorini}, Matilde and {Trefoloni}, Bartolomeo and {Juod{\v{z}}balis}, Ignas and {Scholtz}, Jan and {{\"U}bler}, Hannah and {D'Eugenio}, Francesco and {Carniani}, Stefano and {Fabian}, Andy and {Ji}, Xihan and {Mazzolari}, Giovanni and {Bertola}, Elena and {Brusa}, Marcella and {Bunker}, Andrew J. and {Charlot}, Stephane and {Comastri}, Andrea and {Cresci}, Giovanni and {DeCoursey}, Christa Noel and {Egami}, Eiichi and {Fiore}, Fabrizio and {Gilli}, Roberto and {Perna}, Michele and {Tacchella}, Sandro and {Venturi}, Giacomo},
        title = "{JWST meets Chandra: a large population of Compton thick, feedback-free, and intrinsically X-ray weak AGN, with a sprinkle of SNe}",
      journal = {\mnras},
         year = 2025,
        month = apr,
       volume = {538},
       number = {3},
        pages = {1921-1943},
          doi = {10.1093/mnras/staf359},
archivePrefix = {arXiv},
       eprint = {2405.00504},
 primaryClass = {astro-ph.GA},
       adsurl = {https://ui.adsabs.harvard.edu/abs/2025MNRAS.538.1921M}
}

@ARTICLE{Risaliti2011,
       author = {{Risaliti}, G. and {Nardini}, E. and {Salvati}, M. and {Elvis}, M. and {Fabbiano}, G. and {Maiolino}, R. and {Pietrini}, P. and {Torricelli-Ciamponi}, G.},
        title = "{X-ray absorption by broad-line region clouds in Mrk 766}",
      journal = {\mnras},
         year = 2011,
        month = jan,
       volume = {410},
       number = {2},
        pages = {1027-1035},
          doi = {10.1111/j.1365-2966.2010.17503.x},
archivePrefix = {arXiv},
       eprint = {1008.5067},
 primaryClass = {astro-ph.CO},
       adsurl = {https://ui.adsabs.harvard.edu/abs/2011MNRAS.410.1027R}
}

@ARTICLE{Risaliti2002,
       author = {{Risaliti}, G. and {Elvis}, M. and {Nicastro}, F.},
        title = "{Ubiquitous Variability of X-Ray-absorbing Column Densities in Seyfert 2 Galaxies}",
      journal = {\apj},
         year = 2002,
        month = may,
       volume = {571},
       number = {1},
        pages = {234-246},
          doi = {10.1086/324146},
archivePrefix = {arXiv},
       eprint = {astro-ph/0107510},
 primaryClass = {astro-ph},
       adsurl = {https://ui.adsabs.harvard.edu/abs/2002ApJ...571..234R}
}

@ARTICLE{Rigby2006,
       author = {{Rigby}, J.~R. and {Rieke}, G.~H. and {Donley}, J.~L. and {Alonso-Herrero}, A. and {P{\'e}rez-Gonz{\'a}lez}, P.~G.},
        title = "{Why X-Ray-selected Active Galactic Nuclei Appear Optically Dull}",
      journal = {\apj},
         year = 2006,
        month = jul,
       volume = {645},
       number = {1},
        pages = {115-133},
          doi = {10.1086/504067},
archivePrefix = {arXiv},
       eprint = {astro-ph/0603313},
 primaryClass = {astro-ph},
       adsurl = {https://ui.adsabs.harvard.edu/abs/2006ApJ...645..115R}
}

@ARTICLE{Kravtsov2024,
       author = {{Kravtsov}, Andrey and {Belokurov}, Vasily},
        title = "{Stochastic star formation and the abundance of $z>10$ UV-bright galaxies}",
      journal = {arXiv e-prints},
         year = 2024,
        month = may,
          eid = {arXiv:2405.04578},
        pages = {arXiv:2405.04578},
          doi = {10.48550/arXiv.2405.04578},
archivePrefix = {arXiv},
       eprint = {2405.04578},
 primaryClass = {astro-ph.GA},
       adsurl = {https://ui.adsabs.harvard.edu/abs/2024arXiv240504578K}
}

@ARTICLE{Wilkins2023,
       author = {{Wilkins}, Stephen M. and {Vijayan}, Aswin P. and {Lovell}, Christopher C. and {Roper}, William J. and {Zackrisson}, Erik and {Irodotou}, Dimitrios and {Seeyave}, Louise T.~C. and {Kuusisto}, Jussi K. and {Thomas}, Peter A. and {Caruana}, Joseph and {Conselice}, Christopher J.},
        title = "{First Light And Reionization Epoch Simulations (FLARES) VII: The star formation and metal enrichment histories of galaxies in the early Universe}",
      journal = {\mnras},
         year = 2023,
        month = jan,
       volume = {518},
       number = {3},
        pages = {3935-3948},
          doi = {10.1093/mnras/stac3281},
archivePrefix = {arXiv},
       eprint = {2208.00976},
 primaryClass = {astro-ph.GA},
       adsurl = {https://ui.adsabs.harvard.edu/abs/2023MNRAS.518.3935W}
}

@ARTICLE{Cantarella2025,
       author = {{Cantarella}, Sebastiano and {De Lucia}, Gabriella and {Fontanot}, Fabio and {Hirschmann}, Michaela and {Xie}, Lizhi and {Franco}, Maximilien and {Plat}, Ad{\`e}le},
        title = "{Probing the dawn of galaxies: star formation and feedback in the JWST era through the GAEA model}",
      journal = {arXiv e-prints},
         year = 2025,
        month = nov,
          eid = {arXiv:2511.03787},
        pages = {arXiv:2511.03787},
          doi = {10.48550/arXiv.2511.03787},
archivePrefix = {arXiv},
       eprint = {2511.03787},
 primaryClass = {astro-ph.GA},
       adsurl = {https://ui.adsabs.harvard.edu/abs/2025arXiv251103787C}
}

@ARTICLE{Kannan2023,
       author = {{Kannan}, Rahul and {Springel}, Volker and {Hernquist}, Lars and {Pakmor}, R{\"u}diger and {Delgado}, Ana Maria and {Hadzhiyska}, Boryana and {Hern{\'a}ndez-Aguayo}, C{\'e}sar and {Barrera}, Monica and {Ferlito}, Fulvio and {Bose}, Sownak and {White}, Simon D.~M. and {Frenk}, Carlos and {Smith}, Aaron and {Garaldi}, Enrico},
        title = "{The MillenniumTNG project: the galaxy population at z {\ensuremath{\geq}} 8}",
      journal = {\mnras},
         year = 2023,
        month = sep,
       volume = {524},
       number = {2},
        pages = {2594-2605},
          doi = {10.1093/mnras/stac3743},
archivePrefix = {arXiv},
       eprint = {2210.10066},
 primaryClass = {astro-ph.GA},
       adsurl = {https://ui.adsabs.harvard.edu/abs/2023MNRAS.524.2594K}
}

@ARTICLE{Ucci2023,
       author = {{Ucci}, Graziano and {Dayal}, Pratika and {Hutter}, Anne and {Kobayashi}, Chiaki and {Gottl{\"o}ber}, Stefan and {Yepes}, Gustavo and {Hunt}, Leslie and {Legrand}, Laurent and {Tortora}, Crescenzo},
        title = "{Astraeus V: the emergence and evolution of metallicity scaling relations during the epoch of reionization}",
      journal = {\mnras},
         year = 2023,
        month = jan,
       volume = {518},
       number = {3},
        pages = {3557-3575},
          doi = {10.1093/mnras/stac2654},
archivePrefix = {arXiv},
       eprint = {2112.02115},
 primaryClass = {astro-ph.GA},
       adsurl = {https://ui.adsabs.harvard.edu/abs/2023MNRAS.518.3557U}
}

@ARTICLE{Torrey2019,
       author = {{Torrey}, Paul and {Vogelsberger}, Mark and {Marinacci}, Federico and {Pakmor}, R{\"u}diger and {Springel}, Volker and {Nelson}, Dylan and {Naiman}, Jill and {Pillepich}, Annalisa and {Genel}, Shy and {Weinberger}, Rainer and {Hernquist}, Lars},
        title = "{The evolution of the mass-metallicity relation and its scatter in IllustrisTNG}",
      journal = {\mnras},
         year = 2019,
        month = apr,
       volume = {484},
       number = {4},
        pages = {5587-5607},
          doi = {10.1093/mnras/stz243},
archivePrefix = {arXiv},
       eprint = {1711.05261},
 primaryClass = {astro-ph.GA},
       adsurl = {https://ui.adsabs.harvard.edu/abs/2019MNRAS.484.5587T}
}

@ARTICLE{Marszewski2024,
       author = {{Marszewski}, Andrew and {Sun}, Guochao and {Faucher-Gigu{\`e}re}, Claude-Andr{\'e} and {Hayward}, Christopher C. and {Feldmann}, Robert},
        title = "{The High-Redshift Gas-Phase Mass{\textendash}Metallicity Relation in FIRE-2}",
      journal = {\apjl},
         year = 2024,
        month = jun,
       volume = {967},
       number = {2},
          eid = {L41},
        pages = {L41},
          doi = {10.3847/2041-8213/ad4cee},
archivePrefix = {arXiv},
       eprint = {2403.08853},
 primaryClass = {astro-ph.GA},
       adsurl = {https://ui.adsabs.harvard.edu/abs/2024ApJ...967L..41M}
}

@ARTICLE{Ma2016,
       author = {{Ma}, Xiangcheng and {Hopkins}, Philip F. and {Faucher-Gigu{\`e}re}, Claude-Andr{\'e} and {Zolman}, Nick and {Muratov}, Alexander L. and {Kere{\v{s}}}, Du{\v{s}}an and {Quataert}, Eliot},
        title = "{The origin and evolution of the galaxy mass-metallicity relation}",
      journal = {\mnras},
         year = 2016,
        month = feb,
       volume = {456},
       number = {2},
        pages = {2140-2156},
          doi = {10.1093/mnras/stv2659},
archivePrefix = {arXiv},
       eprint = {1504.02097},
 primaryClass = {astro-ph.GA},
       adsurl = {https://ui.adsabs.harvard.edu/abs/2016MNRAS.456.2140M}
}

@ARTICLE{Sarkar+25,
       author = {{Sarkar}, Arnab and {Chakraborty}, Priyanka and {Vogelsberger}, Mark and {McDonald}, Michael and {Torrey}, Paul and {Garcia}, Alex M. and {Khullar}, Gourav and {Ferland}, Gary J. and {Forman}, William and {Wolk}, Scott and {Schneider}, Benjamin and {Bautz}, Mark and {Miller}, Eric and {Grant}, Catherine and {ZuHone}, John},
        title = "{Unveiling the Cosmic Chemistry: Revisiting the Mass{\textendash}Metallicity Relation with JWST/NIRSpec at 4 < z < 10}",
      journal = {\apj},
         year = 2025,
        month = jan,
       volume = {978},
       number = {2},
          eid = {136},
        pages = {136},
          doi = {10.3847/1538-4357/ad8f32},
archivePrefix = {arXiv},
       eprint = {2408.07974},
 primaryClass = {astro-ph.GA},
       adsurl = {https://ui.adsabs.harvard.edu/abs/2025ApJ...978..136S}
}

@ARTICLE{Li2023_MZR,
       author = {{Li}, Mingyu and {Cai}, Zheng and {Bian}, Fuyan and {Lin}, Xiaojing and {Li}, Zihao and {Wu}, Yunjing and {Sun}, Fengwu and {Zhang}, Shiwu and {Golden-Marx}, Emmet and {Sun}, Zechang and {Zou}, Siwei and {Fan}, Xiaohui and {Egami}, Eiichi and {Charlot}, Stephane and {Bruzual}, Gustavo and {Chevallard}, Jacopo},
        title = "{The Mass-Metallicity Relation of Dwarf Galaxies at Cosmic Noon from JWST Observations}",
      journal = {\apjl},
         year = 2023,
        month = sep,
       volume = {955},
       number = {1},
          eid = {L18},
        pages = {L18},
          doi = {10.3847/2041-8213/acf470},
archivePrefix = {arXiv},
       eprint = {2211.01382},
 primaryClass = {astro-ph.GA},
       adsurl = {https://ui.adsabs.harvard.edu/abs/2023ApJ...955L..18L}
}

@ARTICLE{Curti+20,
       author = {{Curti}, Mirko and {Mannucci}, Filippo and {Cresci}, Giovanni and {Maiolino}, Roberto},
        title = "{The mass-metallicity and the fundamental metallicity relation revisited on a fully T$_{e}$-based abundance scale for galaxies}",
      journal = {\mnras},
         year = 2020,
        month = jan,
       volume = {491},
       number = {1},
        pages = {944-964},
          doi = {10.1093/mnras/stz2910},
archivePrefix = {arXiv},
       eprint = {1910.00597},
 primaryClass = {astro-ph.GA},
       adsurl = {https://ui.adsabs.harvard.edu/abs/2020MNRAS.491..944C}
}

@ARTICLE{Pollock+25,
       author = {{Pollock}, Clara L. and {Gottumukkala}, Rashmi and {Heintz}, Kasper E. and {Brammer}, Gabriel B. and {Roberts-Borsani}, Guido and {Oesch}, Pascal A. and {Witstok}, Joris and {Arellano-C{\'o}rdova}, Karla Z. and {Cullen}, Fergus and {Scholte}, Dirk and {Terp}, Chamilla and {Rowland}, Lucie and {Sneppen}, Albert and {Ito}, Kei and {Valentino}, Francesco and {Matthee}, Jorryt and {Watson}, Darach and {Toft}, Sune},
        title = "{Novel $z\sim~10$ auroral line measurements extend the gradual offset of the FMR deep into the first Gyr of cosmic time}",
      journal = {arXiv e-prints},
         year = 2025,
        month = jun,
          eid = {arXiv:2506.15779},
        pages = {arXiv:2506.15779},
          doi = {10.48550/arXiv.2506.15779},
archivePrefix = {arXiv},
       eprint = {2506.15779},
 primaryClass = {astro-ph.GA},
       adsurl = {https://ui.adsabs.harvard.edu/abs/2025arXiv250615779P}
}

@ARTICLE{Donnan+25,
       author = {{Donnan}, Callum T. and {Dickinson}, Mark and {Taylor}, Anthony J. and {Arrabal Haro}, Pablo and {Finkelstein}, Steven L. and {Stanton}, Thomas M. and {Jung}, Intae and {Papovich}, Casey and {Akins}, Hollis B. and {Koekemoer}, Anton M. and {McLeod}, Derek J. and {Napolitano}, Lorenzo and {Amor{\'\i}n}, Ricardo O. and {Begley}, Ryan and {Burgarella}, Denis and {Carnall}, Adam C. and {Casey}, Caitlin M. and {Calabr{\`o}}, Antonello and {Cullen}, Fergus and {Dunlop}, James S. and {Ellis}, Richard S. and {Fern{\'a}ndez}, Vital and {Giavalisco}, Mauro and {Hirschmann}, Michaela and {Hu}, Weida and {Illingworth}, Garth and {Kartaltepe}, Jeyhan S. and {Kocevski}, Dale D. and {Kokorev}, Vasily and {Leung}, Ho-Hin and {Lucas}, Ray A. and {Morales}, Alexa M. and {McLure}, Ross and {Pentericci}, Laura and {P{\'e}rez-Gonz{\'a}lez}, Pablo G. and {Somerville}, Rachel S. and {Stevenson}, Struan and {Trump}, Jonathan R. and {Yung}, L.~Y. Aaron and {Zavala}, Jorge A.},
        title = "{Very Bright, Very Blue, and Very Red: JWST CAPERS Analysis of Highly Luminous Galaxies with Extreme Ultraviolet Slopes at z = 10}",
      journal = {\apj},
         year = 2025,
        month = nov,
       volume = {993},
       number = {2},
          eid = {224},
        pages = {224},
          doi = {10.3847/1538-4357/ae0a1f},
archivePrefix = {arXiv},
       eprint = {2507.10518},
 primaryClass = {astro-ph.GA},
       adsurl = {https://ui.adsabs.harvard.edu/abs/2025ApJ...993..224D}
}

@ARTICLE{Roberts-Borsani+2025,
       author = {{Roberts-Borsani}, Guido and {Oesch}, Pascal and {Ellis}, Richard and {Weibel}, Andrea and {Giovinazzo}, Emma and {Bouwens}, Rychard and {Dayal}, Pratika and {Fontana}, Adriano and {Heintz}, Kasper and {Matthee}, Jorryt and {Meyer}, Romain and {Pentericci}, Laura and {Shapley}, Alice and {Tacchella}, Sandro and {Treu}, Tommaso and {Walter}, Fabian and {Atek}, Hakim and {Bose}, Sownak and {Castellano}, Marco and {Fudamoto}, Yoshinobu and {Morishita}, Takahiro and {Naidu}, Rohan and {Sanders}, Ryan and {van der Wel}, Arjen},
        title = "{JWST Spectroscopic Insights Into the Diversity of Galaxies in the First 500 Myr: Short-Lived Snapshots Along a Common Evolutionary Pathway}",
      journal = {arXiv e-prints},
         year = 2025,
        month = aug,
          eid = {arXiv:2508.21708},
        pages = {arXiv:2508.21708},
          doi = {10.48550/arXiv.2508.21708},
archivePrefix = {arXiv},
       eprint = {2508.21708},
 primaryClass = {astro-ph.GA},
       adsurl = {https://ui.adsabs.harvard.edu/abs/2025arXiv250821708R}
}

@ARTICLE{Perez-Gonzalez+2025,
       author = {{P{\'e}rez-Gonz{\'a}lez}, Pablo G. and {{\"O}stlin}, G{\"o}ran and {Costantin}, Luca and {Melinder}, Jens and {Finkelstein}, Steven L. and {Somerville}, Rachel S. and {Annunziatella}, Marianna and {{\'A}lvarez-M{\'a}rquez}, Javier and {Colina}, Luis and {Dekel}, Avishai and {Ferguson}, Henry C. and {Li}, Zhaozhou and {Yung}, L.~Y. Aaron and {Bagley}, Micaela B. and {Boogaard}, Leindert A. and {Burgarella}, Denis and {Calabr{\`o}}, Antonello and {Caputi}, Karina I. and {Cheng}, Yingjie and {Dickinson}, Mark and {Eckart}, Andreas and {Giavalisco}, Mauro and {Gillman}, Steven and {Greve}, Thomas R. and {Hamed}, Mahmoud and {Hathi}, Nimish P. and {Hjorth}, Jens and {Huertas-Company}, Marc and {Kartaltepe}, Jeyhan S. and {Koekemoer}, Anton M. and {Kokorev}, Vasily and {Labiano}, {\'A}lvaro and {Langeroodi}, Danial and {Leung}, Gene C.~K. and {Natarajan}, Priyamvada and {Papovich}, Casey and {Peissker}, Florian and {Pentericci}, Laura and {Pirzkal}, Nor and {Rinaldi}, Pierluigi and {van der Werf}, Paul and {Walter}, Fabian},
        title = "{The Rise of the Galactic Empire: Ultraviolet Luminosity Functions at z {\ensuremath{\sim}} 17 and z {\ensuremath{\sim}} 25 Estimated with the MIDIS+NGDEEP Ultra-deep JWST/NIRCam Data Set}",
      journal = {\apj},
         year = 2025,
        month = oct,
       volume = {991},
       number = {2},
          eid = {179},
        pages = {179},
          doi = {10.3847/1538-4357/adf8c9},
archivePrefix = {arXiv},
       eprint = {2503.15594},
 primaryClass = {astro-ph.GA},
       adsurl = {https://ui.adsabs.harvard.edu/abs/2025ApJ...991..179P}
}

@ARTICLE{Schouws+2025,
       author = {{Schouws}, Sander and {Bouwens}, Rychard J. and {Ormerod}, Katherine and {Smit}, Renske and {Algera}, Hiddo and {Sommovigo}, Laura and {Hodge}, Jacqueline and {Ferrara}, Andrea and {Oesch}, Pascal A. and {Rowland}, Lucie E. and {van Leeuwen}, Ivana and {Stefanon}, Mauro and {Herard-Demanche}, Thomas and {Fudamoto}, Yoshinobu and {R{\"o}ttgering}, Huub and {van der Werf}, Paul},
        title = "{Detection of [O III]$_{88 {\ensuremath{\mu}}m}$ in JADES-GS-z14-0 at z = 14.1793}",
      journal = {\apj},
         year = 2025,
        month = jul,
       volume = {988},
       number = {1},
          eid = {19},
        pages = {19},
          doi = {10.3847/1538-4357/adbf1b},
archivePrefix = {arXiv},
       eprint = {2409.20549},
 primaryClass = {astro-ph.GA},
       adsurl = {https://ui.adsabs.harvard.edu/abs/2025ApJ...988...19S}
}

@ARTICLE{Naidu+2025,
       author = {{Naidu}, Rohan P. and {Oesch}, Pascal A. and {Brammer}, Gabriel and {Weibel}, Andrea and {Li}, Yijia and {Matthee}, Jorryt and {Chisholm}, John and {Pollock}, Clara L. and {Heintz}, Kasper E. and {Johnson}, Benjamin D. and {Shen}, Xuejian and {Hviding}, Raphael E. and {Leja}, Joel and {Tacchella}, Sandro and {Ganguly}, Arpita and {Witten}, Callum and {Atek}, Hakim and {Belli}, Sirio and {Bose}, Sownak and {Bouwens}, Rychard and {Dayal}, Pratika and {Decarli}, Roberto and {de Graaff}, Anna and {Fudamoto}, Yoshinobu and {Giovinazzo}, Emma and {Greene}, Jenny E. and {Illingworth}, Garth and {Inoue}, Akio K. and {Kane}, Sarah G. and {Labbe}, Ivo and {Leonova}, Ecaterina and {Marques-Chaves}, Rui and {Meyer}, Romain A. and {Nelson}, Erica J. and {Roberts-Borsani}, Guido and {Schaerer}, Daniel and {Simcoe}, Robert A. and {Stefanon}, Mauro and {Sugahara}, Yuma and {Toft}, Sune and {van der Wel}, Arjen and {van Dokkum}, Pieter and {Walter}, Fabian and {Watson}, Darach and {Weaver}, John R. and {Whitaker}, Katherine E.},
        title = "{A Cosmic Miracle: A Remarkably Luminous Galaxy at $z_{\rm{spec}}=14.44$ Confirmed with JWST}",
      journal = {arXiv e-prints},
         year = 2025,
        month = may,
          eid = {arXiv:2505.11263},
        pages = {arXiv:2505.11263},
          doi = {10.48550/arXiv.2505.11263},
archivePrefix = {arXiv},
       eprint = {2505.11263},
 primaryClass = {astro-ph.GA},
       adsurl = {https://ui.adsabs.harvard.edu/abs/2025arXiv250511263N}
}

@ARTICLE{Merida2025,
       author = {{M{\'e}rida}, Rosa M. and {Sawicki}, Marcin and {Iyer}, Kartheik G. and {Noirot}, Ga{\"e}l and {Willott}, Chris J. and {Brada{\v{c}}}, Maru{\v{s}}a and {Desprez}, Guillaume and {Martis}, Nicholas S. and {Muzzin}, Adam and {Rihtar{\v{s}}i{\v{c}}}, Gregor and {Sarrouh}, Ghassan T.~E. and {Favaro}, Jeremy and {Gaspar}, Gaia and {Harshan}, Anishya and {Jude{\v{z}}}, Jon},
        title = "{Probing the Star Formation Main Sequence down to 10$^{7} M_\odot$ at $1 < z < 9$}",
      journal = {arXiv e-prints},
         year = 2025,
        month = sep,
          eid = {arXiv:2509.22871},
        pages = {arXiv:2509.22871},
          doi = {10.48550/arXiv.2509.22871},
archivePrefix = {arXiv},
       eprint = {2509.22871},
 primaryClass = {astro-ph.GA},
       adsurl = {https://ui.adsabs.harvard.edu/abs/2025arXiv250922871M}
}

@ARTICLE{Llerena2025,
       author = {{Llerena}, M. and {Pentericci}, L. and {Napolitano}, L. and {Mascia}, S. and {Amor{\'\i}n}, R. and {Calabr{\`o}}, A. and {Castellano}, M. and {Cleri}, N.~J. and {Giavalisco}, M. and {Grogin}, N.~A. and {Hathi}, N.~P. and {Hirschmann}, M. and {Koekemoer}, A.~M. and {Nanayakkara}, T. and {Pacucci}, F. and {Shen}, L. and {Wilkins}, S.~M. and {Yoon}, I. and {Yung}, L.~Y.~A. and {Bhatawdekar}, R. and {Lucas}, R.~A. and {Wang}, X. and {Arrabal Haro}, P. and {Bagley}, M.~B. and {Finkelstein}, S.~L. and {Kartaltepe}, J.~S. and {Merlin}, E. and {Papovich}, C. and {Pirzkal}, N. and {Santini}, P.},
        title = "{The ionizing photon production efficiency of star-forming galaxies at z {\ensuremath{\sim}} 4─10}",
      journal = {\aap},
         year = 2025,
        month = jun,
       volume = {698},
          eid = {A302},
        pages = {A302},
          doi = {10.1051/0004-6361/202453251},
archivePrefix = {arXiv},
       eprint = {2412.01358},
 primaryClass = {astro-ph.GA},
       adsurl = {https://ui.adsabs.harvard.edu/abs/2025A&A...698A.302L}
}

@ARTICLE{Prieto-Jimenez2025,
       author = {{Prieto-Jim{\'e}nez}, C. and {{\'A}lvarez-M{\'a}rquez}, J. and {Colina}, L. and {Crespo G{\'o}mez}, A. and {Bik}, A. and {{\"O}stlin}, G. and {Alonso-Herrero}, A. and {Boogaard}, L. and {Caputi}, K.~I. and {Costantin}, L. and {Eckart}, A. and {Garc{\'\i}a-Mar{\'\i}n}, M. and {Gillman}, S. and {Hjorth}, J. and {Iani}, E. and {Jermann}, I. and {Labiano}, A. and {Langeroodi}, D. and {Melinder}, J. and {Moutard}, T. and {Pei{\ss}ker}, F. and {P{\'e}rez-Gonz{\'a}lez}, P.~G. and {Pye}, J.~P. and {Rinaldi}, P. and {Tikkanen}, T.~V. and {van der Werf}, P. and {Walter}, F. and {Hashimoto}, T. and {Sugahara}, Y. and {G{\"u}del}, M. and {Henning}, T.},
        title = "{Spatially resolved H{\ensuremath{\alpha}} emission in B14-65666: Compact starbursts, ionizing efficiency, and gas kinematics in an advanced merger at the Epoch of Reionization}",
      journal = {\aap},
         year = 2025,
        month = sep,
       volume = {701},
          eid = {A31},
        pages = {A31},
          doi = {10.1051/0004-6361/202555057},
archivePrefix = {arXiv},
       eprint = {2507.06793},
 primaryClass = {astro-ph.GA},
       adsurl = {https://ui.adsabs.harvard.edu/abs/2025A&A...701A..31P}
}

@ARTICLE{Komarova2025,
       author = {{Komarova}, Lena and {Stefanon}, Mauro and {Laza-Ramos}, Andres and {Algera}, Hiddo S. and {Aravena}, Manuel and {Bouwens}, Rychard J. and {Bowler}, Rebecca and {da Cunha}, Elisabete and {Dayal}, Pratika and {Ferrara}, Andrea and {Fisher}, Rebecca and {Nanayakkara}, Themiya and {Rowland}, Lucie E. and {Schouws}, Sander and {Smit}, Renske and {Sommovigo}, Laura and {Stark}, Daniel P. and {van der Werf}, Paul},
        title = "{REBELS-IFU: Spatially Resolved Ionizing Photon Production Efficiencies of 12 Bright Galaxies in the Epoch of Reionization}",
      journal = {arXiv e-prints},
         year = 2025,
        month = nov,
          eid = {arXiv:2511.10743},
        pages = {arXiv:2511.10743},
          doi = {10.48550/arXiv.2511.10743},
archivePrefix = {arXiv},
       eprint = {2511.10743},
 primaryClass = {astro-ph.GA},
       adsurl = {https://ui.adsabs.harvard.edu/abs/2025arXiv251110743K}
}

@ARTICLE{Helton2025,
       author = {{Helton}, Jakob M. and {Morrison}, Jane E. and {Hainline}, Kevin N. and {D'Eugenio}, Francesco and {Rieke}, George H. and {Alberts}, Stacey and {Carniani}, Stefano and {Leja}, Joel and {Li}, Yijia and {Rinaldi}, Pierluigi and {Scholtz}, Jan and {Stone}, Meredith and {Willmer}, Christopher N.~A. and {Wu}, Zihao and {Baker}, William M. and {Bunker}, Andrew J. and {Charlot}, Stephane and {Chevallard}, Jacopo and {Cleri}, Nikko J. and {Curti}, Mirko and {Curtis-Lake}, Emma and {Egami}, Eiichi and {Eisenstein}, Daniel J. and {Jakobsen}, Peter and {Ji}, Zhiyuan and {Johnson}, Benjamin D. and {Kumari}, Nimisha and {Lin}, Xiaojing and {Lyu}, Jianwei and {Maiolino}, Roberto and {Maseda}, Michael and {P{\'e}rez-Gonz{\'a}lez}, Pablo G. and {Rieke}, Marcia J. and {Robertson}, Brant and {Saxena}, Aayush and {Sun}, Fengwu and {Tacchella}, Sandro and {{\"U}bler}, Hannah and {Venturi}, Giacomo and {Williams}, Christina C. and {Willott}, Chris and {Witstok}, Joris and {Zhu}, Yongda},
        title = "{Ionizing Photon Production Efficiencies and Chemical Abundances at Cosmic Dawn Revealed by Ultra-Deep Rest-Frame Optical Spectroscopy of JADES-GS-z14-0}",
      journal = {arXiv e-prints},
         year = 2025,
        month = dec,
          eid = {arXiv:2512.19695},
        pages = {arXiv:2512.19695},
          doi = {10.48550/arXiv.2512.19695},
archivePrefix = {arXiv},
       eprint = {2512.19695},
 primaryClass = {astro-ph.GA},
       adsurl = {https://ui.adsabs.harvard.edu/abs/2025arXiv251219695H}
}

@ARTICLE{Roberts-Borsani+24,
       author = {{Roberts-Borsani}, Guido and {Treu}, Tommaso and {Shapley}, Alice and {Fontana}, Adriano and {Pentericci}, Laura and {Castellano}, Marco and {Morishita}, Takahiro and {Bergamini}, Pietro and {Rosati}, Piero},
        title = "{Between the Extremes: A JWST Spectroscopic Benchmark for High-redshift Galaxies Using {\ensuremath{\sim}}500 Confirmed Sources at z {\ensuremath{\geq}} 5}",
      journal = {\apj},
         year = 2024,
        month = dec,
       volume = {976},
       number = {2},
          eid = {193},
        pages = {193},
          doi = {10.3847/1538-4357/ad85d3},
archivePrefix = {arXiv},
       eprint = {2403.07103},
 primaryClass = {astro-ph.GA},
       adsurl = {https://ui.adsabs.harvard.edu/abs/2024ApJ...976..193R}
}

@ARTICLE{Shivaei2020,
       author = {{Shivaei}, Irene and {Reddy}, Naveen and {Rieke}, George and {Shapley}, Alice and {Kriek}, Mariska and {Battisti}, Andrew and {Mobasher}, Bahram and {Sanders}, Ryan and {Fetherolf}, Tara and {Azadi}, Mojegan and {Coil}, Alison L. and {Freeman}, William R. and {de Groot}, Laura and {Leung}, Gene and {Price}, Sedona H. and {Siana}, Brian and {Zick}, Tom},
        title = "{The MOSDEF Survey: The Variation of the Dust Attenuation Curve with Metallicity}",
      journal = {\apj},
         year = 2020,
        month = aug,
       volume = {899},
       number = {2},
          eid = {117},
        pages = {117},
          doi = {10.3847/1538-4357/aba35e},
archivePrefix = {arXiv},
       eprint = {2005.01742},
 primaryClass = {astro-ph.GA},
       adsurl = {https://ui.adsabs.harvard.edu/abs/2020ApJ...899..117S}
}

@ARTICLE{Pei1992,
       author = {{Pei}, Yichuan C.},
        title = "{Interstellar Dust from the Milky Way to the Magellanic Clouds}",
      journal = {\apj},
         year = 1992,
        month = aug,
       volume = {395},
        pages = {130},
          doi = {10.1086/171637},
       adsurl = {https://ui.adsabs.harvard.edu/abs/1992ApJ...395..130P}
}

@ARTICLE{Garcia-Bernete19-torus,
       author = {{Garc{\'\i}a-Bernete}, I. and {Ramos Almeida}, C. and {Alonso-Herrero}, A. and {Ward}, M.~J. and {Acosta-Pulido}, J.~A. and {Pereira-Santaella}, M. and {Hern{\'a}n-Caballero}, A. and {Asensio Ramos}, A. and {Gonz{\'a}lez-Mart{\'\i}n}, O. and {Levenson}, N.~A. and {Mateos}, S. and {Carrera}, F.~J. and {Ricci}, C. and {Roche}, P. and {Marquez}, I. and {Packham}, C. and {Masegosa}, J. and {Fuller}, L.},
        title = "{Torus model properties of an ultra-hard X-ray selected sample of Seyfert galaxies}",
      journal = {\mnras},
         year = 2019,
        month = jul,
       volume = {486},
       number = {4},
        pages = {4917-4935},
          doi = {10.1093/mnras/stz1003},
archivePrefix = {arXiv},
       eprint = {1904.03694},
 primaryClass = {astro-ph.GA},
       adsurl = {https://ui.adsabs.harvard.edu/abs/2019MNRAS.486.4917G}
}

@ARTICLE{Algera25-ALMA,
       author = {{Algera}, Hiddo S.~B. and {Weaver}, John R. and {Bakx}, Tom J.~L.~C. and {Aravena}, Manuel and {Bouwens}, Rychard J. and {Cescon}, Karin and {Chen}, Chian-Chou and {da Cunha}, Elisabete and {Dayal}, Pratika and {Faisst}, Andreas and {Ferrara}, Andrea and {Fujimoto}, Seiji and {Hashimoto}, Takuya and {Heintz}, Kasper and {Herrera-Camus}, Rodrigo and {Hodge}, Jacqueline and {Inami}, Hanae and {Inoue}, Akio K. and {Matthee}, Jorryt and {Meyer}, Romain and {Mizukoshi}, Shoichiro and {Mondal}, Chayan and {Nanayakkara}, Themiya and {Oesch}, Pascal A. and {Pallottini}, Andrea and {R{\"o}ttgering}, Huub and {Rowland}, Lucie E. and {Schouws}, Sander and {Smit}, Renske and {Sommovigo}, Laura and {Stark}, Daniel P. and {Sugahara}, Yuma and {Vallini}, Livia and {Vijarnwannaluk}, Bovornpratch and {van der Werf}, Paul and {Werner}, Norbert and {Witstok}, Joris and {Xiao}, Mengyuan},
        title = "{A first systematic study of [OIII] 88$μ$m at $z>8$: two luminous oxygen lines and a powerful ionized outflow in the first 600 million years}",
      journal = {arXiv e-prints},
         year = 2025,
        month = dec,
          eid = {arXiv:2512.14486},
        pages = {arXiv:2512.14486},
archivePrefix = {arXiv},
       eprint = {2512.14486},
 primaryClass = {astro-ph.GA},
       adsurl = {https://ui.adsabs.harvard.edu/abs/2025arXiv251214486A}
}

@ARTICLE{Risaliti99-xray,
       author = {{Risaliti}, G. and {Maiolino}, R. and {Salvati}, M.},
        title = "{The Distribution of Absorbing Column Densities among Seyfert 2 Galaxies}",
      journal = {\apj},
         year = 1999,
        month = sep,
       volume = {522},
       number = {1},
        pages = {157-164},
          doi = {10.1086/307623},
archivePrefix = {arXiv},
       eprint = {astro-ph/9902377},
 primaryClass = {astro-ph},
       adsurl = {https://ui.adsabs.harvard.edu/abs/1999ApJ...522..157R}
}

@ARTICLE{Kovacs24-GHz9,
       author = {{Kov{\'a}cs}, Orsolya E. and {Bogd{\'a}n}, {\'A}kos and {Natarajan}, Priyamvada and {Werner}, Norbert and {Azadi}, Mojegan and {Volonteri}, Marta and {Tremblay}, Grant R. and {Chadayammuri}, Urmila and {Forman}, William R. and {Jones}, Christine and {Kraft}, Ralph P.},
        title = "{A Candidate Supermassive Black Hole in a Gravitationally Lensed Galaxy at Z {\ensuremath{\approx}} 10}",
      journal = {\apjl},
         year = 2024,
        month = apr,
       volume = {965},
       number = {2},
          eid = {L21},
        pages = {L21},
          doi = {10.3847/2041-8213/ad391f},
archivePrefix = {arXiv},
       eprint = {2403.14745},
 primaryClass = {astro-ph.GA},
       adsurl = {https://ui.adsabs.harvard.edu/abs/2024ApJ...965L..21K}
}

@ARTICLE{Napolitano2025_glass_sample,
       author = {{Napolitano}, L. and {Castellano}, M. and {Pentericci}, L. and {Arrabal Haro}, P. and {Fontana}, A. and {Treu}, T. and {Bergamini}, P. and {Calabr{\`o}}, A. and {Mascia}, S. and {Morishita}, T. and {Roberts-Borsani}, G. and {Santini}, P. and {Vanzella}, E. and {Vulcani}, B. and {Zakharova}, D. and {Bakx}, T. and {Dickinson}, M. and {Grillo}, C. and {Leethochawalit}, N. and {Llerena}, M. and {Merlin}, E. and {Paris}, D. and {Rojas-Ruiz}, S. and {Rosati}, P. and {Wang}, X. and {Yoon}, I. and {Zavala}, J.},
        title = "{Seven wonders of Cosmic Dawn: JWST confirms a high abundance of galaxies and AGN at z ≃ 9{\textendash}11 in the GLASS field}",
      journal = {\aap},
         year = 2025,
        month = jan,
       volume = {693},
          eid = {A50},
        pages = {A50},
          doi = {10.1051/0004-6361/202452090},
archivePrefix = {arXiv},
       eprint = {2410.10967},
 primaryClass = {astro-ph.GA},
       adsurl = {https://ui.adsabs.harvard.edu/abs/2025A&A...693A..50N}
}

@ARTICLE{Napolitano25-GHz9,
       author = {{Napolitano}, Lorenzo and {Castellano}, Marco and {Pentericci}, Laura and {Vignali}, Cristian and {Gilli}, Roberto and {Fontana}, Adriano and {Santini}, Paola and {Treu}, Tommaso and {Calabr{\`o}}, Antonello and {Llerena}, Mario and {Piconcelli}, Enrico and {Zappacosta}, Luca and {Mascia}, Sara and {Tripodi}, Roberta and {Arrabal Haro}, Pablo and {Bergamini}, Pietro and {Bakx}, Tom J.~L.~C. and {Dickinson}, Mark and {Glazebrook}, Karl and {Henry}, Alaina and {Leethochawalit}, Nicha and {Mazzolari}, Giovanni and {Merlin}, Emiliano and {Morishita}, Takahiro and {Nanayakkara}, Themiya and {Paris}, Diego and {Puccetti}, Simonetta and {Roberts-Borsani}, Guido and {Rojas Ruiz}, Sofia and {Rosati}, Piero and {Vanzella}, Eros and {Vito}, Fabio and {Vulcani}, Benedetta and {Wang}, Xin and {Yoon}, Ilsang and {Zavala}, Jorge A.},
        title = "{The Dual Nature of GHZ9: Coexisting Active Galactic Nuclei and Star Formation Activity in a Remote X-Ray Source at z = 10.145}",
      journal = {\apj},
         year = 2025,
        month = aug,
       volume = {989},
       number = {1},
          eid = {75},
        pages = {75},
          doi = {10.3847/1538-4357/ade706},
archivePrefix = {arXiv},
       eprint = {2410.18763},
 primaryClass = {astro-ph.GA},
       adsurl = {https://ui.adsabs.harvard.edu/abs/2025ApJ...989...75N}
}

@ARTICLE{Crespo25-GNz11,
       author = {{Crespo G{\'o}mez}, A. and {Colina}, L. and {P{\'e}rez-Gonz{\'a}lez}, P.~G. and {{\'A}lvarez-M{\'a}rquez}, J. and {Garc{\'\i}a-Mar{\'\i}n}, M. and {Alonso-Herrero}, A. and {Annunziatella}, M. and {Bik}, A. and {Bosman}, S. and {Bunker}, A.~J. and {Labiano}, A. and {Langeroodi}, D. and {Rinaldi}, P. and {{\"O}stlin}, G. and {Boogaard}, L. and {Gillman}, S. and {Barro}, G. and {Finkelstein}, S.~L. and {Leung}, G.~C.~K.},
        title = "{MIRI spectrophotometry of GN-z11: Detection and nature of an optical red continuum component}",
      journal = {arXiv e-prints},
         year = 2025,
        month = dec,
          eid = {arXiv:2512.02997},
        pages = {arXiv:2512.02997},
          doi = {10.48550/arXiv.2512.02997},
archivePrefix = {arXiv},
       eprint = {2512.02997},
 primaryClass = {astro-ph.GA},
       adsurl = {https://ui.adsabs.harvard.edu/abs/2025arXiv251202997C}
}

@ARTICLE{Akins25,
       author = {{Akins}, Hollis B. and {Casey}, Caitlin M. and {Lambrides}, Erini and {Allen}, Natalie and {Andika}, Irham T. and {Brinch}, Malte and {Champagne}, Jaclyn B. and {Cooper}, Olivia and {Ding}, Xuheng and {Drakos}, Nicole E. and {Faisst}, Andreas and {Finkelstein}, Steven L. and {Franco}, Maximilien and {Fujimoto}, Seiji and {Gentile}, Fabrizio and {Gillman}, Steven and {Gozaliasl}, Ghassem and {Harish}, Santosh and {Hayward}, Christopher C. and {Hirschmann}, Michaela and {Ilbert}, Olivier and {Kartaltepe}, Jeyhan S. and {Kocevski}, Dale D. and {Koekemoer}, Anton M. and {Kokorev}, Vasily and {Liu}, Daizhong and {Long}, Arianna S. and {McCracken}, Henry Joy and {McKinney}, Jed and {Onoue}, Masafusa and {Paquereau}, Louise and {Renzini}, Alvio and {Rhodes}, Jason and {Robertson}, Brant E. and {Shuntov}, Marko and {Silverman}, John D. and {Tanaka}, Takumi S. and {Toft}, Sune and {Trakhtenbrot}, Benny and {Valentino}, Francesco and {Zavala}, Jorge},
        title = "{COSMOS-Web: The Overabundance and Physical Nature of ``Little Red Dots''{\textemdash}Implications for Early Galaxy and SMBH Assembly}",
      journal = {\apj},
         year = 2025,
        month = sep,
       volume = {991},
       number = {1},
          eid = {37},
        pages = {37},
          doi = {10.3847/1538-4357/ade984},
archivePrefix = {arXiv},
       eprint = {2406.10341},
 primaryClass = {astro-ph.GA},
       adsurl = {https://ui.adsabs.harvard.edu/abs/2025ApJ...991...37A}
}

@ARTICLE{Bogdan24-UNCOVER,
       author = {{Bogd{\'a}n}, {\'A}kos and {Goulding}, Andy D. and {Natarajan}, Priyamvada and {Kov{\'a}cs}, Orsolya E. and {Tremblay}, Grant R. and {Chadayammuri}, Urmila and {Volonteri}, Marta and {Kraft}, Ralph P. and {Forman}, William R. and {Jones}, Christine and {Churazov}, Eugene and {Zhuravleva}, Irina},
        title = "{Evidence for heavy-seed origin of early supermassive black holes from a z {\ensuremath{\approx}} 10 X-ray quasar}",
      journal = {Nature Astronomy},
         year = 2024,
        month = jan,
       volume = {8},
       number = {1},
        pages = {126-133},
          doi = {10.1038/s41550-023-02111-9},
archivePrefix = {arXiv},
       eprint = {2305.15458},
 primaryClass = {astro-ph.GA},
       adsurl = {https://ui.adsabs.harvard.edu/abs/2024NatAs...8..126B}
}

@ARTICLE{Goulding23-UNCOVER,
       author = {{Goulding}, Andy D. and {Greene}, Jenny E. and {Setton}, David J. and {Labbe}, Ivo and {Bezanson}, Rachel and {Miller}, Tim B. and {Atek}, Hakim and {Bogd{\'a}n}, {\'A}kos and {Brammer}, Gabriel and {Chemerynska}, Iryna and {Cutler}, Sam E. and {Dayal}, Pratika and {Fudamoto}, Yoshinobu and {Fujimoto}, Seiji and {Furtak}, Lukas J. and {Kokorev}, Vasily and {Khullar}, Gourav and {Leja}, Joel and {Marchesini}, Danilo and {Natarajan}, Priyamvada and {Nelson}, Erica and {Oesch}, Pascal A. and {Pan}, Richard and {Papovich}, Casey and {Price}, Sedona H. and {van Dokkum}, Pieter and {Wang}, Bingjie and {Weaver}, John R. and {Whitaker}, Katherine E. and {Zitrin}, Adi},
        title = "{UNCOVER: The Growth of the First Massive Black Holes from JWST/NIRSpec-Spectroscopic Redshift Confirmation of an X-Ray Luminous AGN at z = 10.1}",
      journal = {\apjl},
         year = 2023,
        month = sep,
       volume = {955},
       number = {1},
          eid = {L24},
        pages = {L24},
          doi = {10.3847/2041-8213/acf7c5},
archivePrefix = {arXiv},
       eprint = {2308.02750},
 primaryClass = {astro-ph.GA},
       adsurl = {https://ui.adsabs.harvard.edu/abs/2023ApJ...955L..24G}
}

@ARTICLE{Castellano23-UNCOVER,
       author = {{Castellano}, Marco and {Fontana}, Adriano and {Treu}, Tommaso and {Merlin}, Emiliano and {Santini}, Paola and {Bergamini}, Pietro and {Grillo}, Claudio and {Rosati}, Piero and {Acebron}, Ana and {Leethochawalit}, Nicha and {Paris}, Diego and {Bonchi}, Andrea and {Belfiori}, Davide and {Calabr{\`o}}, Antonello and {Correnti}, Matteo and {Nonino}, Mario and {Polenta}, Gianluca and {Trenti}, Michele and {Boyett}, Kristan and {Brammer}, G. and {Broadhurst}, Tom and {Caminha}, Gabriel B. and {Chen}, Wenlei and {Filippenko}, Alexei V. and {Fortuni}, Flaminia and {Glazebrook}, Karl and {Mascia}, Sara and {Mason}, Charlotte A. and {Menci}, Nicola and {Meneghetti}, Massimo and {Mercurio}, Amata and {Metha}, Benjamin and {Morishita}, Takahiro and {Nanayakkara}, Themiya and {Pentericci}, Laura and {Roberts-Borsani}, Guido and {Roy}, Namrata and {Vanzella}, Eros and {Vulcani}, Benedetta and {Yang}, Lilan and {Wang}, Xin},
        title = "{Early Results from GLASS-JWST. XIX. A High Density of Bright Galaxies at z {\ensuremath{\approx}} 10 in the A2744 Region}",
      journal = {\apjl},
         year = 2023,
        month = may,
       volume = {948},
       number = {2},
          eid = {L14},
        pages = {L14},
          doi = {10.3847/2041-8213/accea5},
archivePrefix = {arXiv},
       eprint = {2212.06666},
 primaryClass = {astro-ph.GA},
       adsurl = {https://ui.adsabs.harvard.edu/abs/2023ApJ...948L..14C}
}

@ARTICLE{Weaver2024,
       author = {{Weaver}, John R. and {Cutler}, Sam E. and {Pan}, Richard and {Whitaker}, Katherine E. and {Labb{\'e}}, Ivo and {Price}, Sedona H. and {Bezanson}, Rachel and {Brammer}, Gabriel and {Marchesini}, Danilo and {Leja}, Joel and {Wang}, Bingjie and {Furtak}, Lukas J. and {Zitrin}, Adi and {Atek}, Hakim and {Chemerynska}, Iryna and {Coe}, Dan and {Dayal}, Pratika and {van Dokkum}, Pieter and {Feldmann}, Robert and {F{\"o}rster Schreiber}, Natascha M. and {Franx}, Marijn and {Fujimoto}, Seiji and {Fudamoto}, Yoshinobu and {Glazebrook}, Karl and {de Graaff}, Anna and {Greene}, Jenny E. and {Juneau}, St{\'e}phanie and {Kassin}, Susan and {Kriek}, Mariska and {Khullar}, Gourav and {Maseda}, Michael V. and {Mowla}, Lamiya A. and {Muzzin}, Adam and {Nanayakkara}, Themiya and {Nelson}, Erica J. and {Oesch}, Pascal A. and {Pacifici}, Camilla and {Papovich}, Casey and {Setton}, David J. and {Shapley}, Alice E. and {Shipley}, Heath V. and {Smit}, Renske and {Stefanon}, Mauro and {Taylor}, Edward N. and {Weibel}, Andrea and {Williams}, Christina C.},
        title = "{The UNCOVER Survey: A First-look HST + JWST Catalog of 60,000 Galaxies near A2744 and beyond}",
      journal = {\apjs},
         year = 2024,
        month = jan,
       volume = {270},
       number = {1},
          eid = {7},
        pages = {7},
          doi = {10.3847/1538-4365/ad07e0},
archivePrefix = {arXiv},
       eprint = {2301.02671},
 primaryClass = {astro-ph.GA},
       adsurl = {https://ui.adsabs.harvard.edu/abs/2024ApJS..270....7W}
}

@ARTICLE{Furtak2023,
       author = {{Furtak}, Lukas J. and {Zitrin}, Adi and {Weaver}, John R. and {Atek}, Hakim and {Bezanson}, Rachel and {Labb{\'e}}, Ivo and {Whitaker}, Katherine E. and {Leja}, Joel and {Price}, Sedona H. and {Brammer}, Gabriel B. and {Wang}, Bingjie and {Marchesini}, Danilo and {Pan}, Richard and {Dayal}, Pratika and {van Dokkum}, Pieter and {Feldmann}, Robert and {Fujimoto}, Seiji and {Franx}, Marijn and {Khullar}, Gourav and {Nelson}, Erica J. and {Mowla}, Lamiya A.},
        title = "{UNCOVERing the extended strong lensing structures of Abell 2744 with the deepest JWST imaging}",
      journal = {\mnras},
         year = 2023,
        month = aug,
       volume = {523},
       number = {3},
        pages = {4568-4582},
          doi = {10.1093/mnras/stad1627},
archivePrefix = {arXiv},
       eprint = {2212.04381},
 primaryClass = {astro-ph.GA},
       adsurl = {https://ui.adsabs.harvard.edu/abs/2023MNRAS.523.4568F}
}

@ARTICLE{Price2025,
       author = {{Price}, Sedona H. and {Bezanson}, Rachel and {Labbe}, Ivo and {Furtak}, Lukas J. and {de Graaff}, Anna and {Greene}, Jenny E. and {Kokorev}, Vasily and {Setton}, David J. and {Suess}, Katherine A. and {Brammer}, Gabriel and {Cutler}, Sam E. and {Leja}, Joel and {Pan}, Richard and {Wang}, Bingjie and {Weaver}, John R. and {Whitaker}, Katherine E. and {Atek}, Hakim and {Burgasser}, Adam J. and {Chemerynska}, Iryna and {Dayal}, Pratika and {Feldmann}, Robert and {F{\"o}rster Schreiber}, Natascha M. and {Fudamoto}, Yoshinobu and {Fujimoto}, Seiji and {Glazebrook}, Karl and {Goulding}, Andy D. and {Khullar}, Gourav and {Kriek}, Mariska and {Marchesini}, Danilo and {Maseda}, Michael V. and {Miller}, Tim B. and {Muzzin}, Adam and {Nanayakkara}, Themiya and {Nelson}, Erica and {Oesch}, Pascal A. and {Shipley}, Heath and {Smit}, Renske and {Taylor}, Edward N. and {Dokkum}, Pieter van and {Williams}, Christina C. and {Zitrin}, Adi},
        title = "{The UNCOVER Survey: First Release of Ultradeep JWST/NIRSpec PRISM Spectra for {\ensuremath{\sim}}700 Galaxies from z {\ensuremath{\sim}} 0.3{\textendash}13 in A2744}",
      journal = {\apj},
         year = 2025,
        month = mar,
       volume = {982},
       number = {1},
          eid = {51},
        pages = {51},
          doi = {10.3847/1538-4357/adaec1},
archivePrefix = {arXiv},
       eprint = {2408.03920},
 primaryClass = {astro-ph.GA},
       adsurl = {https://ui.adsabs.harvard.edu/abs/2025ApJ...982...51P}
}

@ARTICLE{Suess2024,
       author = {{Suess}, Katherine A. and {Weaver}, John R. and {Price}, Sedona H. and {Pan}, Richard and {Wang}, Bingjie and {Bezanson}, Rachel and {Brammer}, Gabriel and {Cutler}, Sam E. and {Labb{\'e}}, Ivo and {Leja}, Joel and {Williams}, Christina C. and {Whitaker}, Katherine E. and {Atek}, Hakim and {Dayal}, Pratika and {de Graaff}, Anna and {Feldmann}, Robert and {Franx}, Marijn and {Fudamoto}, Yoshinobu and {Fujimoto}, Seiji and {Furtak}, Lukas J. and {Goulding}, Andy D. and {Greene}, Jenny E. and {Khullar}, Gourav and {Kokorev}, Vasily and {Kriek}, Mariska and {Lorenz}, Brian and {Marchesini}, Danilo and {Maseda}, Michael V. and {Matthee}, Jorryt and {Miller}, Tim B. and {Mitsuhashi}, Ikki and {Mowla}, Lamiya A. and {Muzzin}, Adam and {Naidu}, Rohan P. and {Nanayakkara}, Themiya and {Nelson}, Erica J. and {Oesch}, Pascal A. and {Setton}, David J. and {Shipley}, Heath and {Smit}, Renske and {Spilker}, Justin S. and {van Dokkum}, Pieter and {Zitrin}, Adi},
        title = "{Medium Bands, Mega Science: A JWST/NIRCam Medium-band Imaging Survey of A2744}",
      journal = {\apj},
         year = 2024,
        month = nov,
       volume = {976},
       number = {1},
          eid = {101},
        pages = {101},
          doi = {10.3847/1538-4357/ad75fe},
archivePrefix = {arXiv},
       eprint = {2404.13132},
 primaryClass = {astro-ph.GA},
       adsurl = {https://ui.adsabs.harvard.edu/abs/2024ApJ...976..101S}
}

@ARTICLE{Bezanson2024,
       author = {{Bezanson}, Rachel and {Labbe}, Ivo and {Whitaker}, Katherine E. and {Leja}, Joel and {Price}, Sedona H. and {Franx}, Marijn and {Brammer}, Gabriel and {Marchesini}, Danilo and {Zitrin}, Adi and {Wang}, Bingjie and {Weaver}, John R. and {Furtak}, Lukas J. and {Atek}, Hakim and {Coe}, Dan and {Cutler}, Sam E. and {Dayal}, Pratika and {van Dokkum}, Pieter and {Feldmann}, Robert and {F{\"o}rster Schreiber}, Natascha M. and {Fujimoto}, Seiji and {Geha}, Marla and {Glazebrook}, Karl and {de Graaff}, Anna and {Greene}, Jenny E. and {Juneau}, St{\'e}phanie and {Kassin}, Susan and {Kriek}, Mariska and {Khullar}, Gourav and {Maseda}, Michael and {Mowla}, Lamiya A. and {Muzzin}, Adam and {Nanayakkara}, Themiya and {Nelson}, Erica J. and {Oesch}, Pascal A. and {Pacifici}, Camilla and {Pan}, Richard and {Papovich}, Casey and {Setton}, David J. and {Shapley}, Alice E. and {Smit}, Renske and {Stefanon}, Mauro and {Taylor}, Edward N. and {Williams}, Christina C.},
        title = "{The JWST UNCOVER Treasury Survey: Ultradeep NIRSpec and NIRCam Observations before the Epoch of Reionization}",
      journal = {\apj},
         year = 2024,
        month = oct,
       volume = {974},
       number = {1},
          eid = {92},
        pages = {92},
          doi = {10.3847/1538-4357/ad66cf},
archivePrefix = {arXiv},
       eprint = {2212.04026},
 primaryClass = {astro-ph.GA},
       adsurl = {https://ui.adsabs.harvard.edu/abs/2024ApJ...974...92B}
}

@ARTICLE{Kendrew2015,
       author = {{Kendrew}, Sarah and {Scheithauer}, Silvia and {Bouchet}, Patrice and {Amiaux}, Jerome and {Azzollini}, Ruym{\'a}n and {Bouwman}, Jeroen and {Chen}, C.~H. and {Dubreuil}, D. and {Fischer}, Sebastian and {Glasse}, Alistair and {Greene}, T.~P. and {Lagage}, P.-O. and {Lahuis}, Fred and {Ronayette}, Samuel and {Wright}, David and {Wright}, G.~S.},
        title = "{The Mid-Infrared Instrument for the James Webb Space Telescope, IV: The Low-Resolution Spectrometer}",
      journal = {\pasp},
         year = 2015,
        month = jul,
       volume = {127},
       number = {953},
        pages = {623},
          doi = {10.1086/682255},
archivePrefix = {arXiv},
       eprint = {1512.03000},
 primaryClass = {astro-ph.IM},
       adsurl = {https://ui.adsabs.harvard.edu/abs/2015PASP..127..623K}
}

@ARTICLE{Carniani+24_ALMA,
       author = {{Carniani}, Stefano and {D'Eugenio}, Francesco and {Ji}, Xihan and {Parlanti}, Eleonora and {Scholtz}, Jan and {Sun}, Fengwu and {Venturi}, Giacomo and {Bakx}, Tom J.~L.~C. and {Curti}, Mirko and {Maiolino}, Roberto and {Tacchella}, Sandro and {Zavala}, Jorge A. and {Hainline}, Kevin and {Witstok}, Joris and {Johnson}, Benjamin D. and {Alberts}, Stacey and {Bunker}, Andrew J. and {Charlot}, St{\'e}phane and {Eisenstein}, Daniel J. and {Helton}, Jakob M. and {Jakobsen}, Peter and {Kumari}, Nimisha and {Robertson}, Brant and {Saxena}, Aayush and {{\"U}bler}, Hannah and {Williams}, Christina C. and {Willmer}, Christopher N.~A. and {Willott}, Chris},
        title = "{The eventful life of a luminous galaxy at z = 14: metal enrichment, feedback, and low gas fraction?}",
      journal = {arXiv e-prints},
         year = 2024,
        month = sep,
          eid = {arXiv:2409.20533},
        pages = {arXiv:2409.20533},
          doi = {10.48550/arXiv.2409.20533},
archivePrefix = {arXiv},
       eprint = {2409.20533},
 primaryClass = {astro-ph.GA},
       adsurl = {https://ui.adsabs.harvard.edu/abs/2024arXiv240920533C}
}

@ARTICLE{Ferrara23,
       author = {{Ferrara}, Andrea and {Pallottini}, Andrea and {Dayal}, Pratika},
        title = "{On the stunning abundance of super-early, luminous galaxies revealed by JWST}",
      journal = {\mnras},
         year = 2023,
        month = jul,
       volume = {522},
       number = {3},
        pages = {3986-3991},
          doi = {10.1093/mnras/stad1095},
archivePrefix = {arXiv},
       eprint = {2208.00720},
 primaryClass = {astro-ph.GA},
       adsurl = {https://ui.adsabs.harvard.edu/abs/2023MNRAS.522.3986F}
}

@ARTICLE{Curti+25,
       author = {{Curti}, Mirko and {Witstok}, Joris and {Jakobsen}, Peter and {Kobayashi}, Chiaki and {Curtis-Lake}, Emma and {Hainline}, Kevin and {Ji}, Xihan and {D'Eugenio}, Francesco and {Chevallard}, Jacopo and {Maiolino}, Roberto and {Scholtz}, Jan and {Carniani}, Stefano and {Arribas}, Santiago and {Baker}, William M. and {Bhatawdekar}, Rachana and {Boyett}, Kristan and {Bunker}, Andrew J. and {Cameron}, Alex and {Cargile}, Phillip A. and {Charlot}, St{\'e}phane and {Eisenstein}, Daniel J. and {Ji}, Zhiyuan and {Johnson}, Benjamin D. and {Kumari}, Nimisha and {Maseda}, Michael V. and {Robertson}, Brant and {Silcock}, Maddie S. and {Tacchella}, Sandro and {{\"U}bler}, Hannah and {Venturi}, Giacomo and {Williams}, Christina C. and {Willmer}, Christopher N.~A. and {Willott}, Chris},
        title = "{JADES: The star formation and chemical enrichment history of a luminous galaxy at z {\ensuremath{\sim}} 9.43 probed by ultra-deep JWST/NIRSpec spectroscopy}",
      journal = {\aap},
         year = 2025,
        month = may,
       volume = {697},
          eid = {A89},
        pages = {A89},
          doi = {10.1051/0004-6361/202451410},
archivePrefix = {arXiv},
       eprint = {2407.02575},
 primaryClass = {astro-ph.GA},
       adsurl = {https://ui.adsabs.harvard.edu/abs/2025A&A...697A..89C}
}

@ARTICLE{Curti+24,
       author = {{Curti}, Mirko and {Maiolino}, Roberto and {Curtis-Lake}, Emma and {Chevallard}, Jacopo and {Carniani}, Stefano and {D'Eugenio}, Francesco and {Looser}, Tobias J. and {Scholtz}, Jan and {Charlot}, Stephane and {Cameron}, Alex and {{\"U}bler}, Hannah and {Witstok}, Joris and {Boyett}, Kristian and {Laseter}, Isaac and {Sandles}, Lester and {Arribas}, Santiago and {Bunker}, Andrew and {Giardino}, Giovanna and {Maseda}, Michael V. and {Rawle}, Tim and {Rodr{\'\i}guez Del Pino}, Bruno and {Smit}, Renske and {Willott}, Chris J. and {Eisenstein}, Daniel J. and {Hausen}, Ryan and {Johnson}, Benjamin and {Rieke}, Marcia and {Robertson}, Brant and {Tacchella}, Sandro and {Williams}, Christina C. and {Willmer}, Christopher and {Baker}, William M. and {Bhatawdekar}, Rachana and {Egami}, Eiichi and {Helton}, Jakob M. and {Ji}, Zhiyuan and {Kumari}, Nimisha and {Perna}, Michele and {Shivaei}, Irene and {Sun}, Fengwu},
        title = "{JADES: Insights into the low-mass end of the mass-metallicity-SFR relation at 3 < z < 10 from deep JWST/NIRSpec spectroscopy}",
      journal = {\aap},
         year = 2024,
        month = apr,
       volume = {684},
          eid = {A75},
        pages = {A75},
          doi = {10.1051/0004-6361/202346698},
archivePrefix = {arXiv},
       eprint = {2304.08516},
 primaryClass = {astro-ph.GA},
       adsurl = {https://ui.adsabs.harvard.edu/abs/2024A&A...684A..75C}
}

@ARTICLE{Simmonds2024,
       author = {{Simmonds}, C. and {Tacchella}, S. and {Hainline}, K. and {Johnson}, B.~D. and {McClymont}, W. and {Robertson}, B. and {Saxena}, A. and {Sun}, F. and {Witten}, C. and {Baker}, W.~M. and {Bhatawdekar}, R. and {Boyett}, K. and {Bunker}, A.~J. and {Charlot}, S. and {Curtis-Lake}, E. and {Egami}, E. and {Eisenstein}, D.~J. and {Hausen}, R. and {Maiolino}, R. and {Maseda}, M.~V. and {Scholtz}, J. and {Williams}, C.~C. and {Willott}, C. and {Witstok}, J.},
        title = "{Low-mass bursty galaxies in JADES efficiently produce ionizing photons and could represent the main drivers of reionization}",
      journal = {\mnras},
         year = 2024,
        month = jan,
       volume = {527},
       number = {3},
        pages = {6139-6157},
          doi = {10.1093/mnras/stad3605},
archivePrefix = {arXiv},
       eprint = {2310.01112},
 primaryClass = {astro-ph.GA},
       adsurl = {https://ui.adsabs.harvard.edu/abs/2024MNRAS.527.6139S}
}

@ARTICLE{Bunker+24,
       author = {{Bunker}, Andrew J. and {Cameron}, Alex J. and {Curtis-Lake}, Emma and {Jakobsen}, Peter and {Carniani}, Stefano and {Curti}, Mirko and {Witstok}, Joris and {Maiolino}, Roberto and {D'Eugenio}, Francesco and {Looser}, Tobias J. and {Willott}, Chris and {Bonaventura}, Nina and {Hainline}, Kevin and {{\"U}bler}, Hannah and {Willmer}, Christopher N.~A. and {Saxena}, Aayush and {Smit}, Renske and {Alberts}, Stacey and {Arribas}, Santiago and {Baker}, William M. and {Baum}, Stefi and {Bhatawdekar}, Rachana and {Bowler}, Rebecca A.~A. and {Boyett}, Kristan and {Charlot}, Stephane and {Chen}, Zuyi and {Chevallard}, Jacopo and {Circosta}, Chiara and {DeCoursey}, Christa and {de Graaff}, Anna and {Egami}, Eiichi and {Eisenstein}, Daniel J. and {Endsley}, Ryan and {Ferruit}, Pierre and {Giardino}, Giovanna and {Hausen}, Ryan and {Helton}, Jakob M. and {Hviding}, Raphael E. and {Ji}, Zhiyuan and {Johnson}, Benjamin D. and {Jones}, Gareth C. and {Kumari}, Nimisha and {Laseter}, Isaac and {L{\"u}tzgendorf}, Nora and {Maseda}, Michael V. and {Nelson}, Erica and {Parlanti}, Eleonora and {Perna}, Michele and {Rauscher}, Bernard J. and {Rawle}, Tim and {Rix}, Hans-Walter and {Rieke}, Marcia and {Robertson}, Brant and {Rodr{\'\i}guez Del Pino}, Bruno and {Sandles}, Lester and {Scholtz}, Jan and {Sharpe}, Katherine and {Skarbinski}, Maya and {Stark}, Daniel P. and {Sun}, Fengwu and {Tacchella}, Sandro and {Topping}, Michael W. and {Villanueva}, Natalia C. and {Wallace}, Imaan E.~B. and {Williams}, Christina C. and {Woodrum}, Charity},
        title = "{JADES NIRSpec initial data release for the Hubble Ultra Deep Field: Redshifts and line fluxes of distant galaxies from the deepest JWST Cycle 1 NIRSpec multi-object spectroscopy}",
      journal = {\aap},
         year = 2024,
        month = oct,
       volume = {690},
          eid = {A288},
        pages = {A288},
          doi = {10.1051/0004-6361/202347094},
archivePrefix = {arXiv},
       eprint = {2306.02467},
 primaryClass = {astro-ph.GA},
       adsurl = {https://ui.adsabs.harvard.edu/abs/2024A&A...690A.288B}
}

@ARTICLE{Zamora+25,
       author = {{Zamora}, Sandra and {Carniani}, Stefano and {Bertola}, Elena and {Parlanti}, Eleonora and {P{\'e}rez-Gonz{\'a}lez}, Pablo G. and {Arribas}, Santiago and {B{\"o}ker}, Torsten and {Bunker}, Andrew J. and {D'Eugenio}, Francesco and {Maiolino}, Roberto and {Perna}, Michele and {Rodr{\'\i}guez Del Pino}, Bruno and {{\"U}bler}, Hannah and {Cresci}, Giovanni and {Jones}, Gareth C. and {Lamperti}, Isabella and {Scholtz}, Jan and {Trefoloni}, Bartolomeo and {Venturi}, Giacomo},
        title = "{GA-NIFS: Understanding the ionization nature of EGSY8p7/CEERS-1019. Evidence for a star formation-driven outflow at z = 8.6}",
      journal = {arXiv e-prints},
         year = 2025,
        month = dec,
          eid = {arXiv:2512.09022},
        pages = {arXiv:2512.09022},
          doi = {10.48550/arXiv.2512.09022},
archivePrefix = {arXiv},
       eprint = {2512.09022},
 primaryClass = {astro-ph.GA},
       adsurl = {https://ui.adsabs.harvard.edu/abs/2025arXiv251209022Z}
}

@ARTICLE{Ji+Ubler2024,
       author = {{Ji}, Xihan and {{\"U}bler}, Hannah and {Maiolino}, Roberto and {D'Eugenio}, Francesco and {Arribas}, Santiago and {Bunker}, Andrew J. and {Charlot}, St{\'e}phane and {Perna}, Michele and {Rodr{\'\i}guez Del Pino}, Bruno and {B{\"o}ker}, Torsten and {Cresci}, Giovanni and {Curti}, Mirko and {Kumari}, Nimisha and {Lamperti}, Isabella},
        title = "{GA-NIFS: An extremely nitrogen-loud and chemically stratified galaxy at $z\sim 5.55$}",
      journal = {arXiv e-prints},
         year = 2024,
        month = apr,
          eid = {arXiv:2404.04148},
        pages = {arXiv:2404.04148},
          doi = {10.48550/arXiv.2404.04148},
archivePrefix = {arXiv},
       eprint = {2404.04148},
 primaryClass = {astro-ph.GA},
       adsurl = {https://ui.adsabs.harvard.edu/abs/2024arXiv240404148J}
}

@misc{Schaerer-Rui2024,
      title={Discovery of a new N-emitter in the epoch of reionization}, 
      author={D. Schaerer and R. Marques-Chaves and M. Xiao and D. Korber},
      year={2024},
      eprint={2406.08408},
      archivePrefix={arXiv},
      primaryClass={astro-ph.GA}
}

@ARTICLE{Carniani+24,
       author = {{Carniani}, Stefano and {Hainline}, Kevin and {D'Eugenio}, Francesco and {Eisenstein}, Daniel J. and {Jakobsen}, Peter and {Witstok}, Joris and {Johnson}, Benjamin D. and {Chevallard}, Jacopo and {Maiolino}, Roberto and {Helton}, Jakob M. and {Willott}, Chris and {Robertson}, Brant and {Alberts}, Stacey and {Arribas}, Santiago and {Baker}, William M. and {Bhatawdekar}, Rachana and {Boyett}, Kristan and {Bunker}, Andrew J. and {Cameron}, Alex J. and {Cargile}, Phillip A. and {Charlot}, St{\'e}phane and {Curti}, Mirko and {Curtis-Lake}, Emma and {Egami}, Eiichi and {Giardino}, Giovanna and {Isaak}, Kate and {Ji}, Zhiyuan and {Jones}, Gareth C. and {Kumari}, Nimisha and {Maseda}, Michael V. and {Parlanti}, Eleonora and {P{\'e}rez-Gonz{\'a}lez}, Pablo G. and {Rawle}, Tim and {Rieke}, George and {Rieke}, Marcia and {Del Pino}, Bruno Rodr{\'\i}guez and {Saxena}, Aayush and {Scholtz}, Jan and {Smit}, Renske and {Sun}, Fengwu and {Tacchella}, Sandro and {{\"U}bler}, Hannah and {Venturi}, Giacomo and {Williams}, Christina C. and {Willmer}, Christopher N.~A.},
        title = "{Spectroscopic confirmation of two luminous galaxies at a redshift of 14}",
      journal = {\nat},
         year = 2024,
        month = sep,
       volume = {633},
       number = {8029},
        pages = {318-322},
          doi = {10.1038/s41586-024-07860-9},
archivePrefix = {arXiv},
       eprint = {2405.18485},
 primaryClass = {astro-ph.GA},
       adsurl = {https://ui.adsabs.harvard.edu/abs/2024Natur.633..318C}
}

@ARTICLE{Cameron2023,
       author = {{Cameron}, Alex J. and {Saxena}, Aayush and {Bunker}, Andrew J. and {D'Eugenio}, Francesco and {Carniani}, Stefano and {Maiolino}, Roberto and {Curtis-Lake}, Emma and {Ferruit}, Pierre and {Jakobsen}, Peter and {Arribas}, Santiago and {Bonaventura}, Nina and {Charlot}, Stephane and {Chevallard}, Jacopo and {Curti}, Mirko and {Looser}, Tobias J. and {Maseda}, Michael V. and {Rawle}, Tim and {Rodr{\'\i}guez Del Pino}, Bruno and {Smit}, Renske and {{\"U}bler}, Hannah and {Willott}, Chris and {Witstok}, Joris and {Egami}, Eiichi and {Eisenstein}, Daniel J. and {Johnson}, Benjamin D. and {Hainline}, Kevin and {Rieke}, Marcia and {Robertson}, Brant E. and {Stark}, Daniel P. and {Tacchella}, Sandro and {Williams}, Christina C. and {Willmer}, Christopher N.~A. and {Bhatawdekar}, Rachana and {Bowler}, Rebecca and {Boyett}, Kristan and {Circosta}, Chiara and {Helton}, Jakob M. and {Jones}, Gareth C. and {Kumari}, Nimisha and {Ji}, Zhiyuan and {Nelson}, Erica and {Parlanti}, Eleonora and {Sandles}, Lester and {Scholtz}, Jan and {Sun}, Fengwu},
        title = "{JADES: Probing interstellar medium conditions at z {\ensuremath{\sim}} 5.5-9.5 with ultra-deep JWST/NIRSpec spectroscopy}",
      journal = {\aap},
         year = 2023,
        month = sep,
       volume = {677},
          eid = {A115},
        pages = {A115},
          doi = {10.1051/0004-6361/202346107},
archivePrefix = {arXiv},
       eprint = {2302.04298},
 primaryClass = {astro-ph.GA},
       adsurl = {https://ui.adsabs.harvard.edu/abs/2023A&A...677A.115C}
}

@ARTICLE{Ji2025_gnz11,
       author = {{Ji}, Xihan and {Maiolino}, Roberto and {Ferland}, Gary and {D'Eugenio}, Francesco and {Bhatawdekar}, Rachana and {Charlot}, St{\'e}phane and {Chevallard}, Jacopo and {Curti}, Mirko and {Curtis-Lake}, Emma and {Hainline}, Kevin and {Ji}, Zhiyuan and {Robertson}, Brant and {Rodr{\'\i}guez Del Pino}, Bruno and {Scholtz}, Jan and {Tacchella}, Sandro and {Williams}, Christina C. and {Witstok}, Joris},
        title = "{JADES {\textendash} the small blue bump in GN-z11: insights into the nuclear region of a galaxy at z = 10.6}",
      journal = {\mnras},
         year = 2025,
        month = aug,
       volume = {541},
       number = {3},
        pages = {2134-2161},
          doi = {10.1093/mnras/staf1083},
archivePrefix = {arXiv},
       eprint = {2405.05772},
 primaryClass = {astro-ph.GA},
       adsurl = {https://ui.adsabs.harvard.edu/abs/2025MNRAS.541.2134J}
}

@ARTICLE{Dekel2023,
       author = {{Dekel}, Avishai and {Sarkar}, Kartick C. and {Birnboim}, Yuval and {Mandelker}, Nir and {Li}, Zhaozhou},
        title = "{Efficient formation of massive galaxies at cosmic dawn by feedback-free starbursts}",
      journal = {\mnras},
         year = 2023,
        month = aug,
       volume = {523},
       number = {3},
        pages = {3201-3218},
          doi = {10.1093/mnras/stad1557},
archivePrefix = {arXiv},
       eprint = {2303.04827},
 primaryClass = {astro-ph.GA},
       adsurl = {https://ui.adsabs.harvard.edu/abs/2023MNRAS.523.3201D}
}

@ARTICLE{Jin2012,
       author = {{Jin}, Chichuan and {Ward}, Martin and {Done}, Chris},
        title = "{A combined optical and X-ray study of unobscured type 1 active galactic nuclei - II. Relation between X-ray emission and optical spectra}",
      journal = {\mnras},
         year = 2012,
        month = jun,
       volume = {422},
       number = {4},
        pages = {3268-3284},
          doi = {10.1111/j.1365-2966.2012.20847.x},
archivePrefix = {arXiv},
       eprint = {1203.0239},
 primaryClass = {astro-ph.HE},
       adsurl = {https://ui.adsabs.harvard.edu/abs/2012MNRAS.422.3268J}
}

@ARTICLE{Ho2001,
       author = {{Ho}, Luis C. and {Feigelson}, Eric D. and {Townsley}, Leisa K. and {Sambruna}, Rita M. and {Garmire}, Gordon P. and {Brandt}, W.~N. and {Filippenko}, Alexei V. and {Griffiths}, Richard E. and {Ptak}, Andrew F. and {Sargent}, Wallace L.~W.},
        title = "{Detection of Nuclear X-Ray Sources in Nearby Galaxies with Chandra}",
      journal = {\apjl},
         year = 2001,
        month = mar,
       volume = {549},
       number = {1},
        pages = {L51-L54},
          doi = {10.1086/319138},
archivePrefix = {arXiv},
       eprint = {astro-ph/0102504},
 primaryClass = {astro-ph},
       adsurl = {https://ui.adsabs.harvard.edu/abs/2001ApJ...549L..51H}
}

@ARTICLE{Maiolino2024_BH,
       author = {{Maiolino}, Roberto and {Scholtz}, Jan and {Witstok}, Joris and {Carniani}, Stefano and {D'Eugenio}, Francesco and {de Graaff}, Anna and {{\"U}bler}, Hannah and {Tacchella}, Sandro and {Curtis-Lake}, Emma and {Arribas}, Santiago and {Bunker}, Andrew and {Charlot}, St{\'e}phane and {Chevallard}, Jacopo and {Curti}, Mirko and {Looser}, Tobias J. and {Maseda}, Michael V. and {Rawle}, Timothy D. and {Rodr{\'\i}guez del Pino}, Bruno and {Willott}, Chris J. and {Egami}, Eiichi and {Eisenstein}, Daniel J. and {Hainline}, Kevin N. and {Robertson}, Brant and {Williams}, Christina C. and {Willmer}, Christopher N.~A. and {Baker}, William M. and {Boyett}, Kristan and {DeCoursey}, Christa and {Fabian}, Andrew C. and {Helton}, Jakob M. and {Ji}, Zhiyuan and {Jones}, Gareth C. and {Kumari}, Nimisha and {Laporte}, Nicolas and {Nelson}, Erica J. and {Perna}, Michele and {Sandles}, Lester and {Shivaei}, Irene and {Sun}, Fengwu},
        title = "{A small and vigorous black hole in the early Universe}",
      journal = {\nat},
         year = 2024,
        month = mar,
       volume = {627},
       number = {8002},
        pages = {59-63},
          doi = {10.1038/s41586-024-07052-5},
archivePrefix = {arXiv},
       eprint = {2305.12492},
 primaryClass = {astro-ph.GA},
       adsurl = {https://ui.adsabs.harvard.edu/abs/2024Natur.627...59M}
}

@ARTICLE{Calabro2024,
       author = {{Calabr{\`o}}, Antonello and {Castellano}, Marco and {Zavala}, Jorge A. and {Pentericci}, Laura and {Arrabal Haro}, Pablo and {Bakx}, Tom J.~L.~C. and {Burgarella}, Denis and {Casey}, Caitlin M. and {Dickinson}, Mark and {Finkelstein}, Steven L. and {Fontana}, Adriano and {Llerena}, Mario and {Mascia}, Sara and {Merlin}, Emiliano and {Mitsuhashi}, Ikki and {Napolitano}, Lorenzo and {Paris}, Diego and {P{\'e}rez-Gonz{\'a}lez}, Pablo G. and {Roberts-Borsani}, Guido and {Santini}, Paola and {Treu}, Tommaso and {Vanzella}, Eros},
        title = "{Evidence of Extreme Ionization Conditions and Low Metallicity in GHZ2/GLASS-Z12 from a Combined Analysis of NIRSpec and MIRI Observations}",
      journal = {\apj},
         year = 2024,
        month = nov,
       volume = {975},
       number = {2},
          eid = {245},
        pages = {245},
          doi = {10.3847/1538-4357/ad7602},
archivePrefix = {arXiv},
       eprint = {2403.12683},
 primaryClass = {astro-ph.GA},
       adsurl = {https://ui.adsabs.harvard.edu/abs/2024ApJ...975..245C}
}

@ARTICLE{Castellano2024,
       author = {{Castellano}, Marco and {Napolitano}, Lorenzo and {Fontana}, Adriano and {Roberts-Borsani}, Guido and {Treu}, Tommaso and {Vanzella}, Eros and {Zavala}, Jorge A. and {Arrabal Haro}, Pablo and {Calabr{\`o}}, Antonello and {Llerena}, Mario and {Mascia}, Sara and {Merlin}, Emiliano and {Paris}, Diego and {Pentericci}, Laura and {Santini}, Paola and {Bakx}, Tom J.~L.~C. and {Bergamini}, Pietro and {Cupani}, Guido and {Dickinson}, Mark and {Filippenko}, Alexei V. and {Glazebrook}, Karl and {Grillo}, Claudio and {Kelly}, Patrick L. and {Malkan}, Matthew A. and {Mason}, Charlotte A. and {Morishita}, Takahiro and {Nanayakkara}, Themiya and {Rosati}, Piero and {Sani}, Eleonora and {Wang}, Xin and {Yoon}, Ilsang},
        title = "{JWST NIRSpec Spectroscopy of the Remarkable Bright Galaxy GHZ2/GLASS-z12 at Redshift 12.34}",
      journal = {\apj},
         year = 2024,
        month = sep,
       volume = {972},
       number = {2},
          eid = {143},
        pages = {143},
          doi = {10.3847/1538-4357/ad5f88},
archivePrefix = {arXiv},
       eprint = {2403.10238},
 primaryClass = {astro-ph.GA},
       adsurl = {https://ui.adsabs.harvard.edu/abs/2024ApJ...972..143C}
}

@ARTICLE{Nakajima+22,
       author = {{Nakajima}, Kimihiko and {Ouchi}, Masami and {Xu}, Yi and {Rauch}, Michael and {Harikane}, Yuichi and {Nishigaki}, Moka and {Isobe}, Yuki and {Kusakabe}, Haruka and {Nagao}, Tohru and {Ono}, Yoshiaki and {Onodera}, Masato and {Sugahara}, Yuma and {Kim}, Ji Hoon and {Komiyama}, Yutaka and {Lee}, Chien-Hsiu and {Zahedy}, Fakhri S.},
        title = "{EMPRESS. V. Metallicity Diagnostics of Galaxies over 12 + log(O/H) ≃ 6.9-8.9 Established by a Local Galaxy Census: Preparing for JWST Spectroscopy}",
      journal = {\apjs},
         year = 2022,
        month = sep,
       volume = {262},
       number = {1},
          eid = {3},
        pages = {3},
          doi = {10.3847/1538-4365/ac7710},
archivePrefix = {arXiv},
       eprint = {2206.02824},
 primaryClass = {astro-ph.GA},
       adsurl = {https://ui.adsabs.harvard.edu/abs/2022ApJS..262....3N}
}

@ARTICLE{Nakajima+23,
       author = {{Nakajima}, Kimihiko and {Ouchi}, Masami and {Isobe}, Yuki and {Harikane}, Yuichi and {Zhang}, Yechi and {Ono}, Yoshiaki and {Umeda}, Hiroya and {Oguri}, Masamune},
        title = "{JWST Census for the Mass-Metallicity Star Formation Relations at z = 4-10 with Self-consistent Flux Calibration and Proper Metallicity Calibrators}",
      journal = {\apjs},
         year = 2023,
        month = dec,
       volume = {269},
       number = {2},
          eid = {33},
        pages = {33},
          doi = {10.3847/1538-4365/acd556},
archivePrefix = {arXiv},
       eprint = {2301.12825},
 primaryClass = {astro-ph.GA},
       adsurl = {https://ui.adsabs.harvard.edu/abs/2023ApJS..269...33N}
}

@ARTICLE{Morishita2024,
       author = {{Morishita}, Takahiro and {Stiavelli}, Massimo and {Grillo}, Claudio and {Rosati}, Piero and {Schuldt}, Stefan and {Trenti}, Michele and {Bergamini}, Pietro and {Boyett}, Kristan N. and {Chary}, Ranga-Ram and {Leethochawalit}, Nicha and {Roberts-Borsani}, Guido and {Treu}, Tommaso and {Vanzella}, Eros},
        title = "{Diverse Oxygen Abundance in Early Galaxies Unveiled by Auroral Line Analysis with JWST}",
      journal = {arXiv e-prints},
         year = 2024,
        month = feb,
          eid = {arXiv:2402.14084},
        pages = {arXiv:2402.14084},
          doi = {10.48550/arXiv.2402.14084},
archivePrefix = {arXiv},
       eprint = {2402.14084},
 primaryClass = {astro-ph.GA},
       adsurl = {https://ui.adsabs.harvard.edu/abs/2024arXiv240214084M}
}

@ARTICLE{Rinaldi+2024,
       author = {{Rinaldi}, P. and {Caputi}, K.~I. and {Iani}, E. and {Costantin}, L. and {Gillman}, S. and {Perez Gonzalez}, P.~G. and {{\"O}stlin}, G. and {Colina}, L. and {Greve}, T.~R. and {N{\o}rgard-Nielsen}, H.~U. and {Wright}, G.~S. and {{\'A}lvarez-M{\'a}rquez}, J. and {Eckart}, A. and {Garc{\'\i}a-Mar{\'\i}n}, M. and {Hjorth}, J. and {Ilbert}, O. and {Kendrew}, S. and {Labiano}, A. and {Le F{\`e}vre}, O. and {Pye}, J. and {Tikkanen}, T. and {Walter}, F. and {van der Werf}, P. and {Ward}, M. and {Annunziatella}, M. and {Azzollini}, R. and {Bik}, A. and {Boogaard}, L. and {Bosman}, S.~E.~I. and {Crespo G{\'o}mez}, A. and {Jermann}, I. and {Langeroodi}, D. and {Melinder}, J. and {Meyer}, R.~A. and {Moutard}, T. and {Peissker}, F. and {van Dishoeck}, E. and {G{\"u}del}, M. and {Henning}, Th. and {Lagage}, P. -O. and {Ray}, T. and {Vandenbussche}, B. and {Waelkens}, C. and {Dayal}, Pratika},
        title = "{MIDIS: Unveiling the Role of Strong H{\ensuremath{\alpha}} Emitters During the Epoch of Reionization with JWST}",
      journal = {\apj},
         year = 2024,
        month = jul,
       volume = {969},
       number = {1},
          eid = {12},
        pages = {12},
          doi = {10.3847/1538-4357/ad4147},
archivePrefix = {arXiv},
       eprint = {2309.15671},
 primaryClass = {astro-ph.GA},
       adsurl = {https://ui.adsabs.harvard.edu/abs/2024ApJ...969...12R}
}

\end{document}